\documentclass[aps,prd,amsmath,floats,floatfix, twocolumn,superscriptaddress,
nofootinbib,showpacs]{revtex4-2}

\usepackage[normalem]{ulem}
\usepackage{graphicx}
\usepackage{dcolumn}
\usepackage{bm}
\usepackage{subfigure}
\usepackage{xcolor}
\usepackage{url}
\usepackage{hyperref}
\usepackage[mathlines]{lineno}

\newcommand\EOBROM{\texttt{SEOBNRv4\_ROM}}
\newcommand\ENIGMA{\texttt{ENIGMA}}
\newcommand\EccentricFD{\texttt{EccentricFD}}
\newcommand\EccentricTD{\texttt{EccentricTD}}
\newcommand\TEOBResum{\texttt{TEOBiResumS\_SM}}
\newcommand\TEOBResumE{\texttt{TEOBResumE}}
\newcommand\Phenom{\texttt{IMRPhenomD}}
\newcommand\BILBY{\texttt{BILBY}}
\newcommand\dynesty{\texttt{dynesty}}
\newcommand\TaylorF{\texttt{TaylorF2}}

\newcommand\SolarMass{M\textsubscript{\( \odot \)}}

\def\apjs{{ApJS}}               

\begin{document}
\preprint{APS/123-QED}
\title{ Correlations in gravitational-wave reconstructions from eccentric binaries: a case study with GW151226 \& GW170608 }
\author{Eamonn O'Shea}%
\email{edo27@cornell.edu}
\affiliation{
	Cornell Center for Astrophysics and Planetary Science, Cornell University, Ithaca, New York 14853, USA
}

\author{Prayush Kumar}%
\email{prayush@icts.res.in}
\affiliation{
    International Centre for Theoretical Sciences, Tata Institute of Fundamental Research, Bangalore 560089, India
}
\affiliation{
	Cornell Center for Astrophysics and Planetary Science, Cornell University, Ithaca, New York 14853, USA
}

\date{\today}

\begin{abstract}
The eccentricity of binary black hole mergers is predicted to be an indicator of the history of their formation. In particular, eccentricity is a strong signature of dynamical formation rather than formation by stellar evolution in isolated stellar systems.
We investigate the efficacy of the existing quasi-circular parameter estimation pipelines to determine the source parameters of such eccentric systems. We create a set of simulated signals with eccentricity up to 0.3 and find that as the eccentricity increases, the
recovered mass parameters are consistent with those of a binary with up to a $\approx 10\%$ higher chirp mass and mass ratio closer to unity. We also employ a full inspiral-merger-ringdown waveform model to perform parameter estimation on two gravitational wave events, GW151226 and GW170608, to investigate this bias on real data. We find that the correlation between the masses and eccentricity persists in real data, but that there is also a correlation between the measured eccentricity and effective spin. In particular, using a non-spinning prior results in a spurious eccentricity measurement for GW151226 as it exhibits signs of non-zero black-hole spin. Performing parameter estimation with an aligned spin, eccentric model, we constrain the eccentricities of GW151226 and GW170608 to be $<0.15$ and $<0.12$ respectively. 
\end{abstract}

\maketitle
\section{Introduction}
The Laser Interferometer Gravitational-wave Observatory (LIGO) and the
Virgo observatory have detected gravitational wave (GW) signals
from dozens of binary black hole (BBH) mergers during their first three observing
runs~\cite{Abbott_2019,abbott2020gwtc2,LIGOScientific:2021djp}. As the field of
gravitational wave astronomy moves forward, the network of Earth-based detectors
will be joined by KAGRA~\cite{kagra} and LIGO India~\cite{ligoIndia} in the coming
few years, and possibly followed up by third-generation instruments such as the
Cosmic Explorer (CE)~\cite{cosmicExplorer} and the Einstein Telescope
(ET)~\cite{Maggiore_2020}. Expanding the network to include more
geographical locations will improve the sky localization of events, and the
increased sensitivity of future detectors (CE and ET) will increase our detection
rates, particularly at lower frequencies.


The mechanisms of formation and merging of these binary systems are
the subject of active research~\cite{LIGOScientific:2016vpg, 
marchant2021role, andrews2020targeted, Zevin_2021}.
One possibility is that of isolated binary formation, either through 
formation of a common envelope phase~\cite{Ivanova_2013, livio, 
kruckow, Dominik:2012kk}, or through chemically homogeneous 
evolution~\cite{de_Mink_2010, de_Mink_2016}. 
This formation channel generally results in black holes with lower 
masses, $\lesssim 50 $ \SolarMass{}, and aligned 
spins~\cite{Zevin_2017,Samsing:2020qqd}.

Another promising formation channel that can account for the rate 
of mergers seen is that of dynamical capture binary mergers in
dense globular clusters~\cite{Portegies_Zwart_2000,Rodriguez_2015,
Rodriguez_2016,Rodriguez_2016_2,O_Leary_2016,Samsing2017, 
Samsing:2020qqd} or in active galactic nuclei (AGN)
disks~\cite{Yang_2019, McKernan_2020, 
Tagawa:2020jnc, Gondan:2020svr, Vajpeyi:2021qsw, Grobner_2020}.
In the case of dense environments, binary black holes merge after
undergoing many interactions with third bodies or other
binaries~\cite{Samsing2017}. Dynamical interactions can produce 
multiple generations of mergers. Such formation channels lead to 
binaries with higher mass~\cite{Gerosa_2017, Rodriguez_2019} that 
can lie in the supernova pair-instability mass  
gap~\cite{Woosley_2019, Woosley_2017, Samsing:2020qqd}, have isotropic 
spins~\cite{Rodriguez_2016,talbot_thrane_2017}, and
eccentricity~\cite{Rodriguez_2018, samsing2018black, Zevin_2019}. 
In such cases the eccentricity of a binary can be driven to
non-zero values through the Lidov-Kozai
effect~\cite{lidov, kozai}. By performing Monte Carlo simulations 
of such dense clusters, it has been 
shown~\cite{PhysRevD.97.103014,PhysRevLett.120.151101} that on the
order of 5\% of mergers in globular clusters can have 
eccentricity \(>0.1\) when the frequency of their orbits enters 
the sensitivity band of terrestrial GW detectors at 10 Hz. 
Interactions within AGN disks can also lead to binaries with 
measurable eccentricity~\cite{Tagawa_2021, samsing2020active}. 
Orbital eccentricity is therefore striking evidence that can 
conclusively establish dense clusters as breeding grounds
for compact binary mergers~\cite{Zevin:2021rtf}.

The presence of eccentricity in any of the gravitational-wave signals 
seen thus far has yet to be conclusively established. Searches
for eccentric compact binary candidates in GW data is an active area of 
research~\cite{Tiwari:2015gal, LIGOScientific:2019dag, 
Cheeseboro:2021rey, Wang:2021qsu, Ravichandran:2023qma}, although the 
first three observing runs of LIGO-Virgo have yielded no extra GW 
detections~\cite{LIGOScientific:2019dag,LIGOScientific:2020ufj,
Nitz_2020,LIGOScientific:2023lpe} beyond those already found by 
non-eccentric searches~\cite{LIGOScientific:2021djp}.
There have been various studies~\cite{Romero_Shaw_2019, Wu_2020,
Romero-Shaw2020, Romero-Shaw:2021ual, Iglesias:2022xfc, Romero-Shaw:2022xko} that attempt
to measure orbital eccentricities of binary black hole merger events, as 
well the neutron star merger events in GW transient 
catalogs~\cite{Lenon_2020}. While they agree that most observed GW 
signals are produced by quasi-circular binary mergers, some of these
studies find evidence of residual orbital eccentricity in a few 
events~\cite{Wu_2020, Romero-Shaw2020, Romero-Shaw:2022xko}.
There is strong statistical evidence that at least one merger in GWTC-2 
is the result of hierarchical mergers~\cite{Zevin_2021}.

Of special note is the event GW190521 that is favoured to have been 
formed from the  merger of second generation black 
holes~\cite{kimball2020evidence}, or possibly through dynamical 
capture~\cite{Gamba:2021gap, gayathri2020gw190521}, although it could be
indistinguishable from a precessing, quasi-circular binary
merger~\cite{Romero-Shaw2020, Romero-Shaw:2022xko} given the short 
length of the signal. 
Our ability to accurately determine the eccentricity of mergers such as
GW190521~\cite{LIGOScientific:2020ufj} will improve as the network of GW 
detectors expands its sensitive frequency band to include deciHertz 
detectors~\cite{Sato_2017, Kawamura:2020pcg, LGWA:2020mma}. It has been shown that 
the addition of detectors such as Cosmic Explorer and the Einstein 
Telescope can substantially increase the signal-to-noise ratio (SNR) of 
eccentric signals, in some cases, to higher values than similar
quasi-circular systems~\cite{chen2020observation}.

There has been previous work to investigate the effect that orbital 
eccentricity can have on the detectability of GW signals when the 
quasi-circular template banks that LIGO-Virgo currently employs are 
used~\cite{Huerta_2013,Tiwari2016,Zevin:2021rtf}. Mildly eccentric 
systems can still be detected by search pipelines, but with a loss of 
SNR. However, moderately and highly eccentric systems (with $e_{15Hz} > 
0.275$) would likely be missed by search pipelines~\cite{Huerta_2013, 
ENIGMA} (although sufficiently loud eccentric events can still be detected by unmodelled transient GW searches that make minimal assumptions about the gravitational waveform~\cite{LIGOScientific:2019dag,LIGOScientific:2020ufj,
Nitz_2020,LIGOScientific:2023lpe}). Once detected, ignoring eccentricity could still lead us to 
infer with biases other intrinsic parameters of the source.

In this paper we pursue two related goals. The first is to understand
how our lack of inclusion of orbital eccentricity in waveform templates
alters our inference on the source parameters of GW signals. To this
end we simulate a set of eccentric inspiral-merger-ringdown (IMR) signals
and perform Bayesian parameter estimation on them using waveform 
templates that represent binary mergers on quasi-circular orbits. We restrict these injections to non-spinning systems in order to keep the dimensionality of the parameter space being sampled small while studying parameter space degeneracies. This is a reasonable approximation as most GW events observed so far (more than half) are consistent with small black hole spins~\cite{gwosc_catalog}.
We find that as we increase the orbital eccentricity of GW sources, our 
quasi-circular templates furnish larger chirp mass values and less 
asymmetric mass ratios than the those simulated.
Although eccentricity alters the rate of inspiral in the same way a
larger chirp mass does~\cite{peters}, it also introduces modulations
in the orbital frequency as a function of time to make the phase
evolution qualitatively very different from what a shift in binary
masses does. This bias is therefore not obvious, especially when
considering the full inspiral-merger-ringdown.
We quantify the bias to find an $\mathcal{O}(10\%)$ drift in chirp mass
measurement alone in the case of low-mass binaries with moderate SNRs 
(20-30), which is significant given that the measurement precision on 
the chirp mass is typically smaller than $\mathcal{O}(1\%)$. We also 
find a similarly significant bias in the measurement of binary mass 
ratio.
Our quantitative findings are consistent with previous
results~\cite{Martel_1999, Lenon_2020}.

Our second goal is to investigate whether this bias appears with real
GW events~\cite{LIGOScientific:2016ebw}. We investigate two events: GW151226 and GW170608, that have been debated in literature for being possibly eccentric~\cite{Wu_2020, Romero_Shaw_2019}. It is useful to note that as Refs~\cite{Wu_2020, Romero_Shaw_2019} use different waveform models to study these events, their eccentricity measurements are not quantitatively comparable~\cite{Mora:2002gf, Knee:2022hth, Shaikh:2023ypz}. However the claim of being eccentric/non-eccentric could still be qualitatively contrasted between both studies, as finite eccentricity measured by one waveform model is unlikely to be measured as nearly zero by another.
%
We find that if we use a non-spinning, eccentric waveform
model to perform parameter estimation, GW151226 is consistent with being
a moderately eccentric binary merger, with a chirp mass that is an underestimate of that found by LIGO-Virgo~\cite{Abbott_2019}. However, if we use a waveform model that includes black hole spins (aligned) in addition to orbital eccentricity, we find that GW151226 is well described as a non-eccentric spinning binary merger with component masses and spins consistent with the original LVC analysis~\cite{Abbott_2019} that generally ignored eccentricity. Inclusion of BH spins significantly weakens the evidence of orbital eccentricity in GW151226.
The implication that the eccentricity measurement can be correlated with
the measured masses \textit{and} spins implies that non-spinning eccentric waveform models may be insufficient for the measurement of eccentricities of binary black holes if any of the binary components possess spin angular momentum.

In Section~\ref{methods} we describe our methods, including a description
of the waveform models we use in~\ref{models}. In Section~\ref{sec:pe} we
describe the setup for our parameter estimation runs using \BILBY{}.
Section~\ref{sec:sim_events} details the eccentric simulated signals and
results of their recovery with quasi-circular templates. 
In section~\ref{sec:gw_events} we employ a full IMR, eccentric waveform to 
perform parameter estimation on the two GW events described above, with and 
without spins included in our prior. We follow up our studies on real events 
in Section~\ref{sec:sim_events} by simulating the best fit quasi-circular 
parameters for these events, and performing inference with the same eccentric 
IMR model to determine how their parameters would be recovered if they were 
truly non-eccentric.

Throughout, we refer to mass as measured in the detector frame. We denote eccentricity as $e_{y}$, meaning eccentricity at a reference frequency of $y$Hz.

\begin{figure*}[t]
	\centering
	\subfigure[ $q=1$, $e=0$ ]{\includegraphics[width=8.6cm]{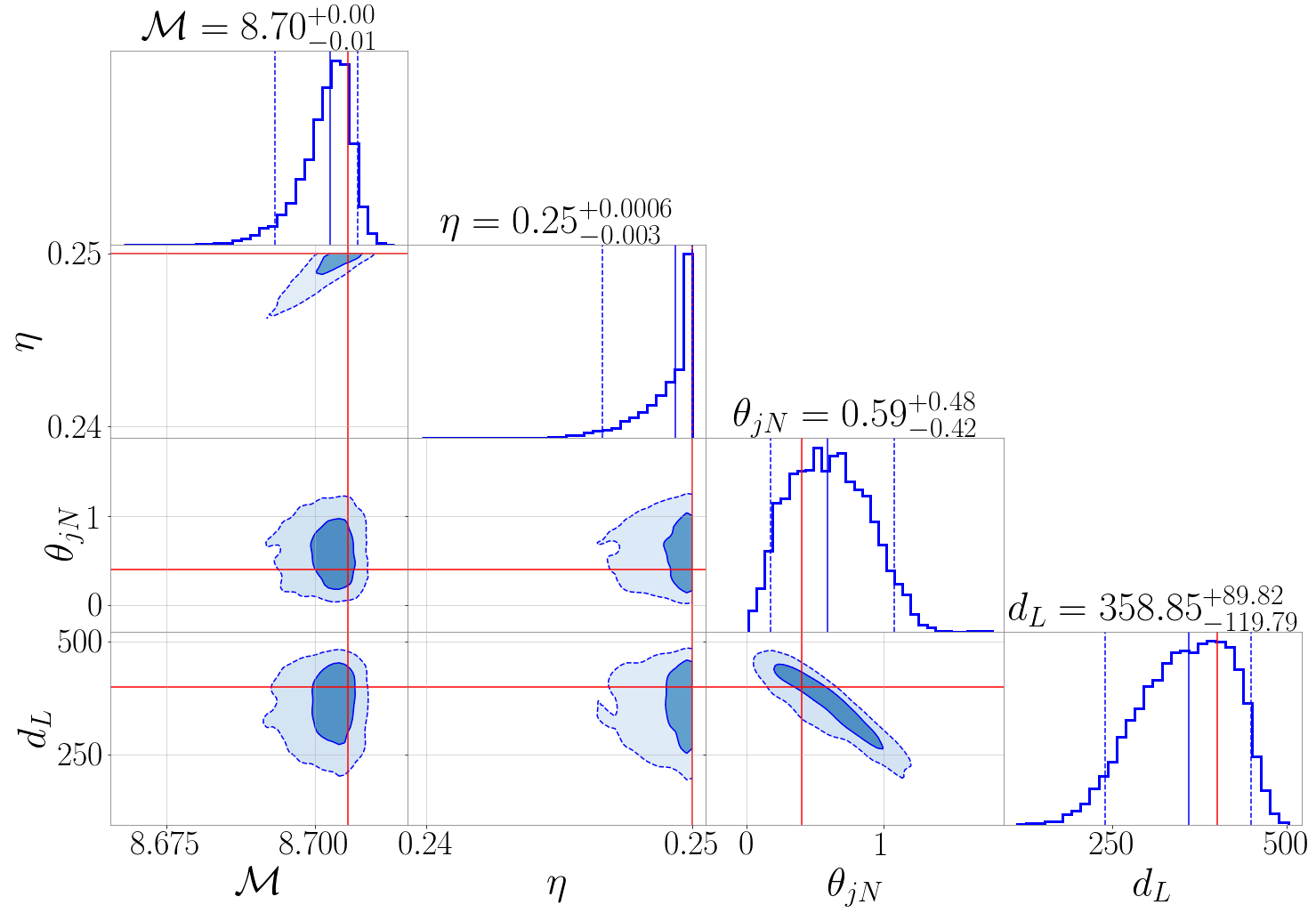}}\quad
	\subfigure[	$q=1$, $e=0.3$]{\includegraphics[width=8.6cm]{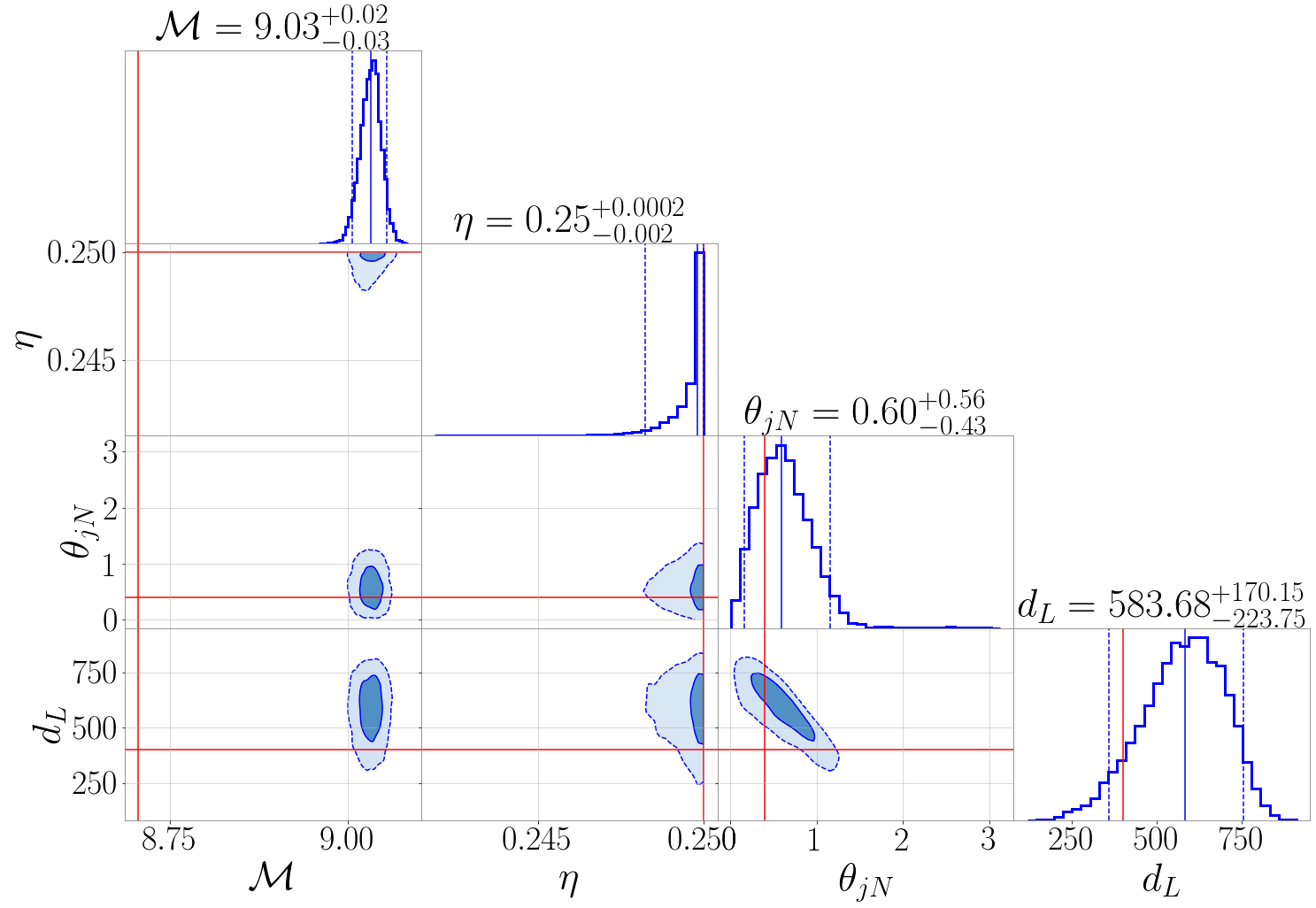}}
	\subfigure[ $q=0.33$, $e=0$ ]{\includegraphics[width=8.6cm]{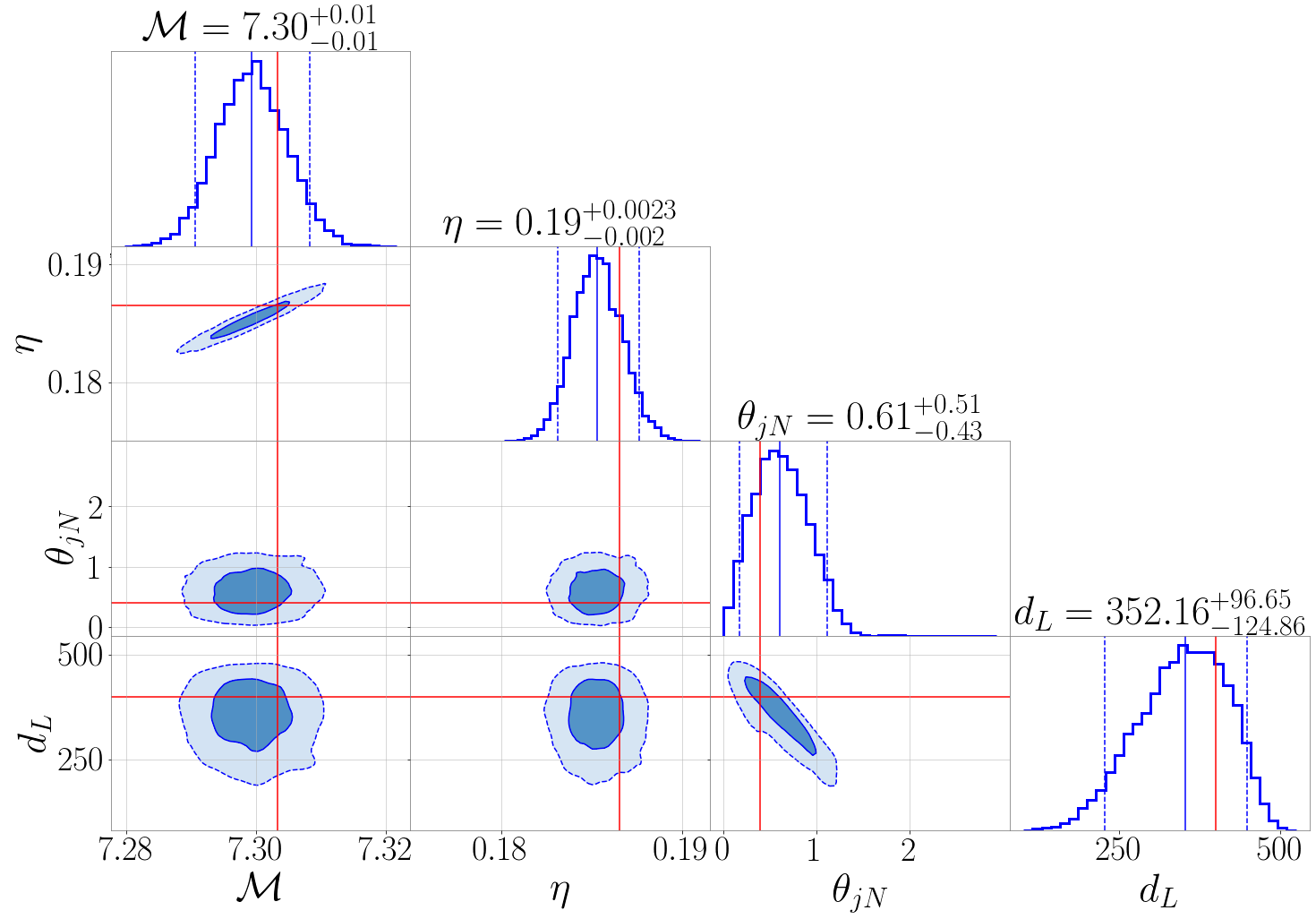}}\quad
	\subfigure[	$q=0.33$, $e=0.3$]{\includegraphics[width=8.6cm]{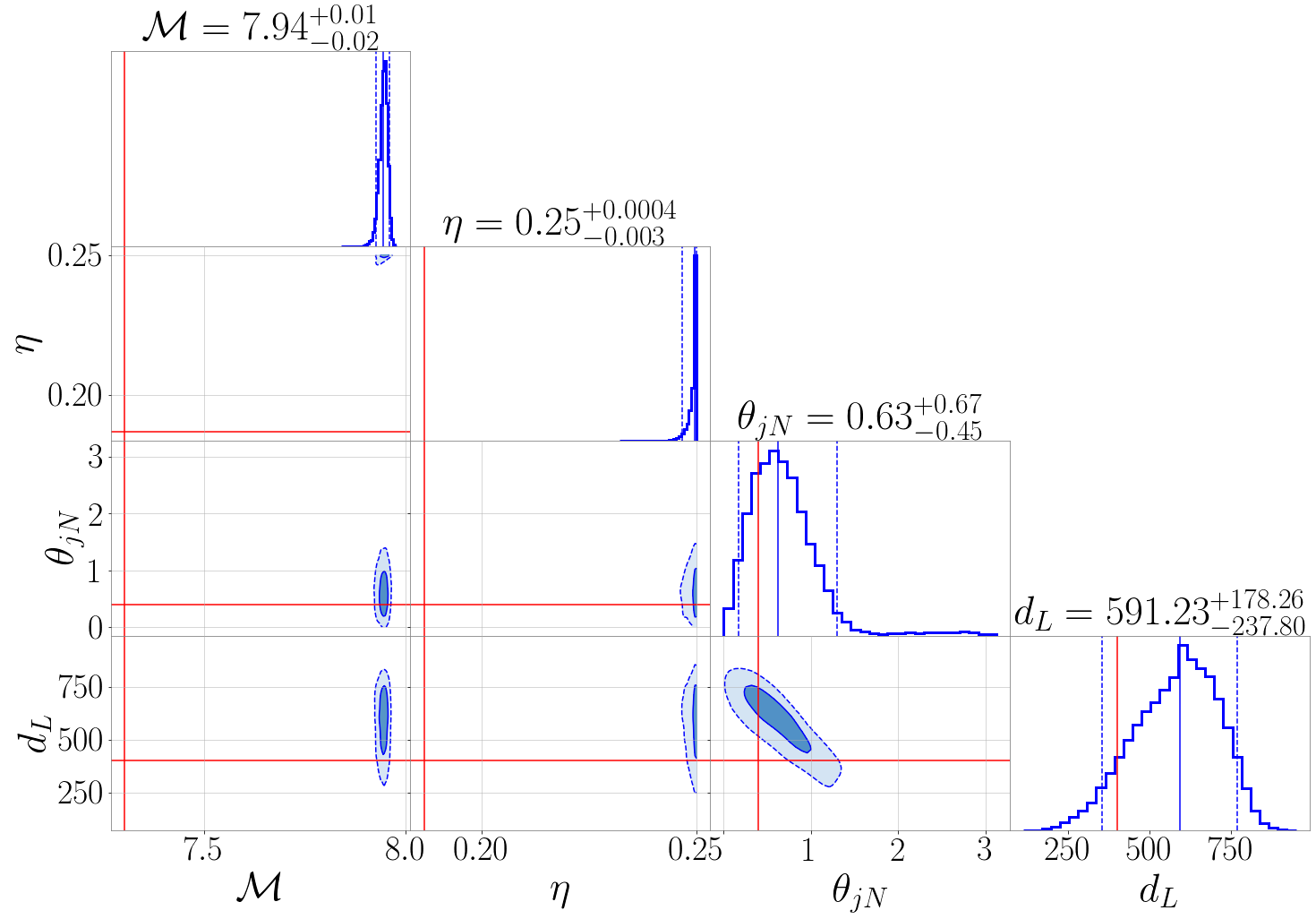}}

	\caption{Comparison of parameter estimation results for simulated signals of zero
		and moderate
		eccentricity using \Phenom{} as the likelihood model and \TEOBResumE{}
		as the signal model. Shown are the posterior
		distributions of chirp mass
		$\mathcal{M}_c$, symmetric mass ratio $\eta$, luminosity distance $D_L$ and
		inclination $\iota$, where the red lines indicate
		signal's values.}

\label{figure:sample_plot}
\end{figure*}

\section{Methods} \label{methods}

\subsection{Waveform models} \label{models}

Detection and characterization of the source properties of 
eccentric signals require accurate and efficient waveform models.
Some of the existing waveform models use a post-Newtonian description and
only predict the inspiral portion of the signal, such as the
\EccentricFD{}~\cite{EccentricFD} and \EccentricTD{}~\cite{EccentricTD},
which are both implemented in LALsuite~\cite{lalsuite}. Recently, models
have been developed to produce full inspiral-merger-ringdown (IMR)
waveform for eccentric binaries, by hybridizing eccentric inspiral
signals with a quasi-circular merger-ringdown portion (\ENIGMA{})~\cite{ENIGMA},
through an effective one-body (EOB) formalism~\cite{seobnre, Chiaramello2020}, 
or by surrogate modelling of numerical relativity waveforms~\cite{Islam_2021}.



The first eccentric model we use is the inspiral portion of \ENIGMA{}~\cite{ENIGMA}.
\ENIGMA{} is a time-domain model designed to
produce full inspiral-merger-ringdown waveforms up to moderate eccentricities.
\ENIGMA{} incorporates
corrections for the energy flux of quasi-circular binaries and gravitational
self-force corrections to the binding energy of compact binaries up to 6PN
order.
Higher-order corrections to the effect of the inclination angle on the
inspiral waveform were introduced in~\cite{chen2020observation}.
The framework coincides with the TaylorT4 approximant at 3PN order
in the zero-eccentricity limit. We opt to use the inspiral portion of the model
alone in our analyses, for reasons elaborated in Appendix~\ref{appendix:enigma}.
In order to produce the inspiral waveform, we integrate the equations of motion
from~\cite{ENIGMA} until the expansion parameter $x = (M\omega)^{2/3}$, reaches the
Schwarzschild innermost stable circular orbit (ISCO), at $x_{\mathrm ISCO} =
\frac{1}{6}$. Here, $M$ is the total mass of the binary and $\omega$ is the mean
orbital frequency.
We refer the reader to~\cite{ENIGMA,chen2020observation} for more details.

The second waveform model we use is an extension of the full IMR quasi-circular model
\TEOBResum{}~\cite{TEOBResum}
to include eccentric effects, which was introduced in~\cite{TEOBResumE}. This
model, which we will refer to as \TEOBResumE, alters the angular momentum flux
portion of \TEOBResum{} with a Newtonian-like prefactor, which generalizes the
quasi-circular model to moderate eccentricities ($e \approx 0.3$). We have used
the publicly available implementation at~\cite{EOBBitbucket}
\footnote{Using the \texttt{eccentric} branch with commit \texttt{10f6110}} without any
changes to the core routines, and use the dominant waveform modes in our analyses.
However, we have loosened the absolute and relative
error tolerances of its ODE integrator from their default values of $10^{-13}$
and $10^{-11}$ to $10^{-8}$ and $10^{-7}$. We find that this allows the likelihood
function to be evaluated on the $\mathcal{O}(10^{-2}\,\, \mathrm{s})$ timescale,
which makes full parameter estimation analyses viable.
To ensure that loosening this tolerance does not significantly degrade the produced
waveforms, we calculate the mismatch, 
\begin{equation}
    \label{eq:mismatch}
    M = 1 - \frac{\langle h_{\mathrm{strict}} | h_{\mathrm{loose}} \rangle}{ \langle h_{\mathrm{strict}} | h_{\mathrm{strict}} \rangle } 
\end{equation}
between the stricter waveform $h_{\mathrm{strict}}$ and the looser waveform $h_{\mathrm{loose}}$,
where we use the straightforward inner product between two timeseries
\begin{equation}
    \langle a | b \rangle = \int a(t) b(t) \mathrm{d} t. 
\end{equation}
We find that waveform generation is robust to such changes with the ODE integrator,
and the mismatches differ by no more than $10^{-3}$ over the parameter space. 

Although this waveform model can produce waveforms with higher-order modes, we have 
opted to only use the dominant $(2,2)$ mode in this study. We also note that \TEOBResumE{} does not allow for variation of the argument of periapsis in initial conditions, and consequently the same was kept fixed in the parameter estimation analysis conducted with the model. This restriction can have a small effect on our results here~\cite{Clarke:2022fma, Islam_2021}, and we defer a detailed quantification of the same to future work.

\subsection{Parameter estimation}\label{sec:pe}

We use the python package \BILBY{}~\cite{bilby} to simulate the signals and
perform parameter estimation. \BILBY{} calculates the posterior distribution
$p(\bm{\theta}|y)$ for the set of source parameters $\bm{\theta}$ given the
data $y$, according to Bayes' theorem, $p(\bm{\theta}|y) \propto p(y|\bm{\theta})
p(\bm{\theta})$ where $p(y|\bm{\theta})$ is the likelihood and $p(\bm{\theta})$
is the prior. The likelihood function for
a set of $N$ detectors is the standard used for gravitational wave astronomy:
\begin{equation} \label{likelihood}
	p(y | \bm{\theta}) = \exp \left( -\frac{1}{2} \sum_{i=1}^N \langle \hat{y}_i(f)
	    -  \hat{s}_i(f, \bm{\theta}) | \hat{y}_i(f)
	    -  \hat{s}_i(f, \bm{\theta}) \rangle \right),
\end{equation}
where $\hat{y}_i(f)$ and $\hat{s}_i(f, \bm{\theta})$ are the
frequency-domain representations of the data and the model
waveform. The inner product
$\langle \cdot | \cdot \rangle$ is given by
\begin{equation} \label{eq:inner_product}
	\langle \hat{a}_i(f) | \hat{b}_i(f) \rangle = 4 \mathrm{Re} \int_{f_{min}}^{f_{max}}
	\frac{\hat{a}_i(f) \hat{b}_i(f) }{S_n^i (f)}
	\mathrm{d} f\,,
\end{equation}
with the weight factor $S_n^i(f)$ being the advanced LIGO power spectral density
(PSD). Superscript $i$ labels the detector, in our case either LIGO Hanford or
LIGO Livingston. We use the zero-detuning high-power noise
curve for LIGO detectors (labeled \texttt{ZERO\_DET\_high\_P} in~\cite{ligo_psd_data}).
The interpretation of this likelihood function is that a given
candidate waveform is likely to be present in the data, if subtracting it from
the data resembles colored Gaussian noise.

We sample from the posterior distribution using a nested sampling algorithm
implemented via the \dynesty{} package~\cite{dynesty} in \BILBY{}. We use
$1,000$ live points, and a stopping criterion of $\Delta \log{\mathcal{Z}}
< 0.1$ where $\mathcal{Z}$ is the estimated Bayesian evidence. To improve
efficiency, we use the option in \BILBY{} to analytically marginalize over
the coalescence time~\cite{time_marg}, coalescence phase~\cite{phase_marg}
and distance~\cite{thrane_talbot_2019} in the calculation of the likelihood
in Eq.~\eqref{likelihood}, and then reconstruct the posterior for these
parameters after the maximum likelihood calculation has completed.

\section{Results}

\subsection{Simulated signals}\label{sec:injection}

\begin{table}
\begin{ruledtabular}
	\begin{tabular*}{8.6cm}{c@{\extracolsep{\fill}} c}
		Parameter   & Simulated value                          \\ \hline
		$q$          & \{ 1, 0.5, 0.33 \}                      \\
		$e_{15}$     &\{ 0, 0.001, 0.01, 0.05, 0.1, 0.2, 0.3 \}\\
		$M_T$        & 20 \SolarMass{}                        \\
		$D_L$ & 400 Mpc                                        \\
		$\iota$         & 0.4                                  \\
		$\alpha$ & 0                                           \\
		$\delta$ & 0                                           \\
		$\psi$ & 0                                             \\
		$\phi_c$ & 0                                           \\
		$t_c$ & 7 s                                            \\ 
		Network SNR & 21-33 
	\end{tabular*}
\end{ruledtabular}
	\caption{Simulated signals values for the
	mass ratio $q$,
    eccentricity $e$,
    total mass $M_T$,
	luminosity distance $D_L$,
	inclination $\iota$,
	right ascension $\alpha$,
	declination $\delta$,
	polarization $\psi$,
	coalescence phase $\phi_c$,
	and coalescence time $t_c$.
	}
	
	\label{table:parameter_spec}
	
\end{table}

\begin{table}
\begin{ruledtabular}
	\begin{tabular*}{8.6cm}{l@{\extracolsep{\fill}} ccc}
		Parameter   & Prior &  Range              \\ \hline
		$q$          & Uniform & 0.25--1.              \\
		$M_c$          & Uniform & 5-- 10 \SolarMass{} \\
		$D_L$ & Uniform Comoving & 10 --1000 Mpc       \\
		$\iota$    &   Uniform sin  & 0--$\pi$         \\
		$\alpha$ & Uniform & 0--$2\pi$                 \\
		$\delta$ & Uniform cos & -$\pi / 2$--$\pi / 2$ \\
		$\psi$ & Uniform & 0--$\pi$                    \\
		$\phi_c$ & Uniform & 0--$2\pi$                 \\
		$t_c$ & Uniform & 0--8                         \\
	\end{tabular*}
\end{ruledtabular}
	\caption{Priors used in the parameter estimation
	for the
	mass ratio $q$,
	chirp mass $\mathcal{M}_c$,
	luminosity distance $D_L$,
	inclination $\iota$,
	right ascension $\alpha$,
	declination $\delta$,
	polarization $\psi$,
	coalescence phase $\phi_c$,
	and coalescence time $t_c$.}
	\label{table:priors}
\end{table}
\begin{figure}[b]
\includegraphics[width=8.6cm]{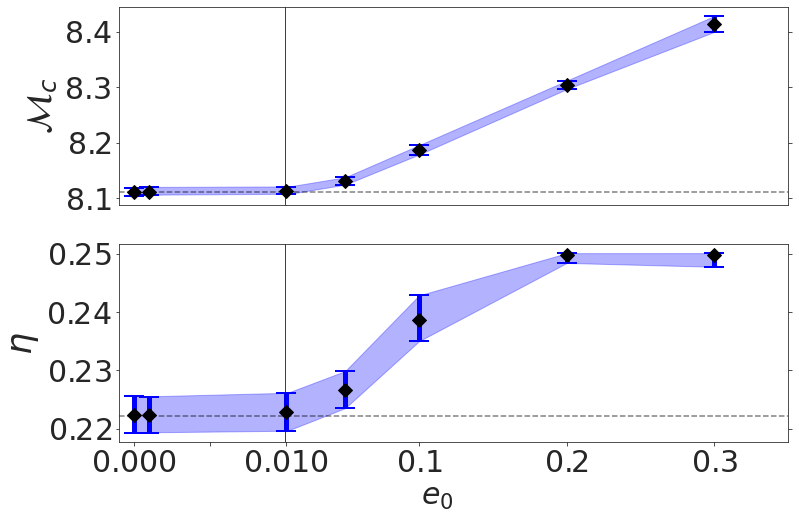}

\caption{Recovery of chirp mass $\mathcal{M}_c$ and symmetric mass ratio
$\eta$ at different eccentricities for a $q=0.5$ binary, when using
\ENIGMA{}-Inspiral as both template and simulated waveform model. 
The parameter estimation prior is constrained to zero orbital eccentricity.
Horizontal dashed lines indicate the value of the parameter for the 
signal. Error bars are 90\% confidence intervals,
with diamonds denoting the median recovered value. The shaded region
connecting the errorbars is to illustrate the trend. Note the different
scales between $[0., 0.01]$ and $[0.01, 0.3]$ on the $x$-axis. } 

\label{figure:enigma_enigma}
\end{figure}

\begin{figure*}[t]
	\includegraphics[width=\textwidth]{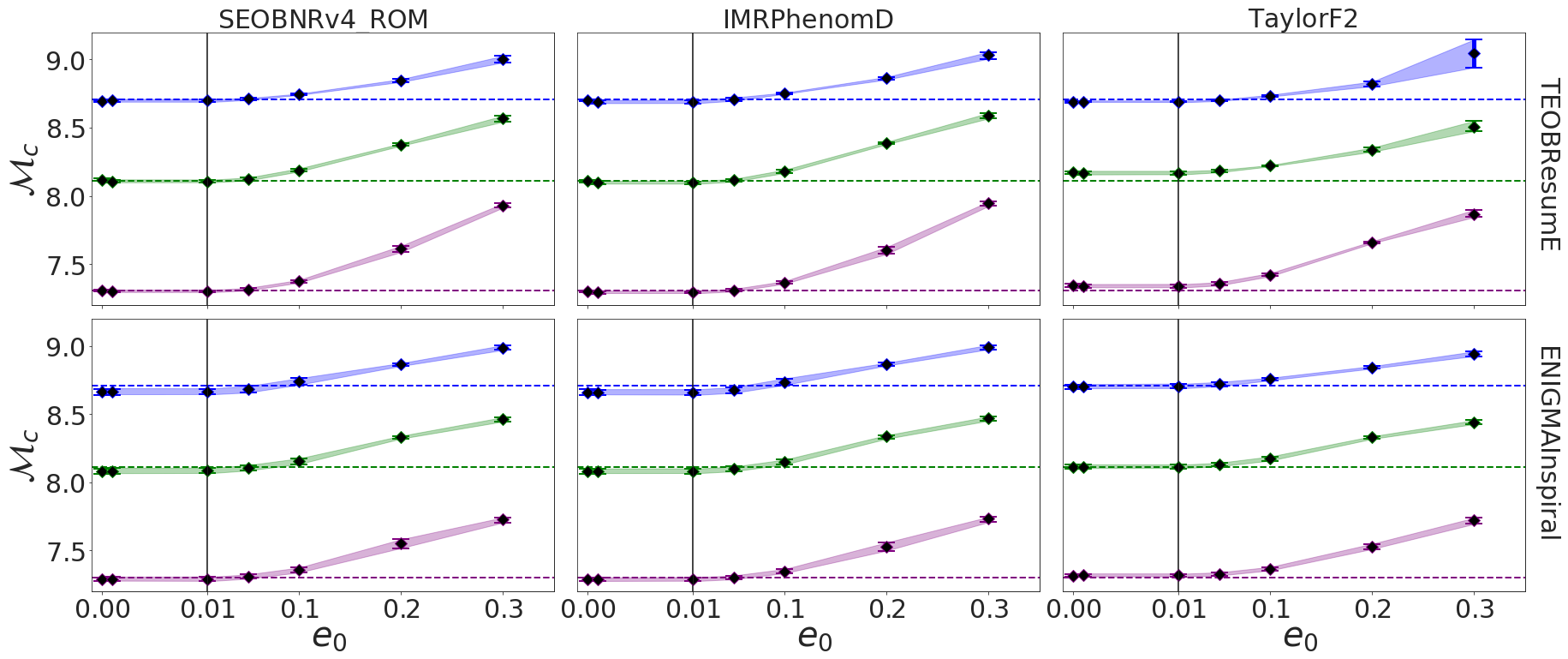}
	\caption{Estimated values of the chirp mass for signals at different initial eccentricities $e_0$ (defined at waveform frequency of $15$Hz). Dashed lines indicate the value of the simulated signal. The colors blue, green and purple correspond to the $q=1, 0.5, 0.33$ signals respectively. The bars represent the 90\% confidence intervals, with the diamond representing the median value. The shaded connects the ends of the error bars to illustrate the upward trend.
	The labels on top specify which model was used in the parameter estimation, and the labels on the right side of the figure specify which model was used to simulate the signal.  Note the different scales between $[0., 0.01]$ and $[0.01, 0.3]$ on the $x$-axes. 
	}

\label{figure:reduction_chirp}
\end{figure*}

\begin{figure*}[t]
	\centering
	\includegraphics[width=\textwidth]{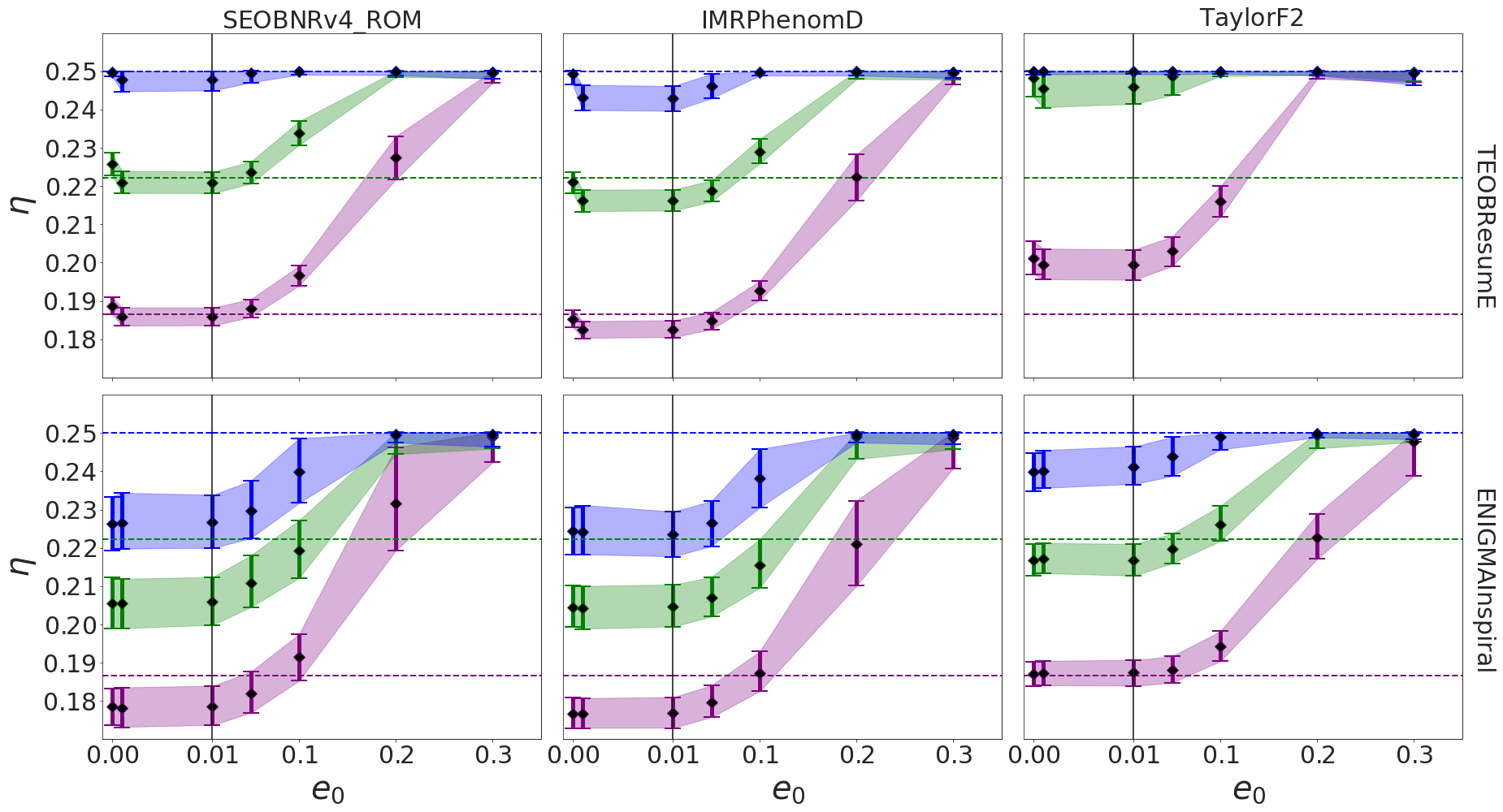}
	\caption{Estimated values of the symmetric mass ratio for signals at different eccentricities. Dashed lines indicate the value of the simulated signal. The colors blue, green and purple correspond to the $q=1, 0.5, 0.33$ signals respectively.
	The bars represent the 90\% confidence intervals, with diamonds representing the median value. The shaded region is to illustrate the upward trend. The labels on top specify which model was used in the parameter estimation, and the labels on the right side of the figure specify which model was used to simulate the signal.  Note the different scales between $[0., 0.01]$ and $[0.01, 0.3]$ on the $x$-axis. }
	
	\label{figure:reduction_q}
\end{figure*}

\begin{figure}[h]
    \centering
	\includegraphics[width=\columnwidth]{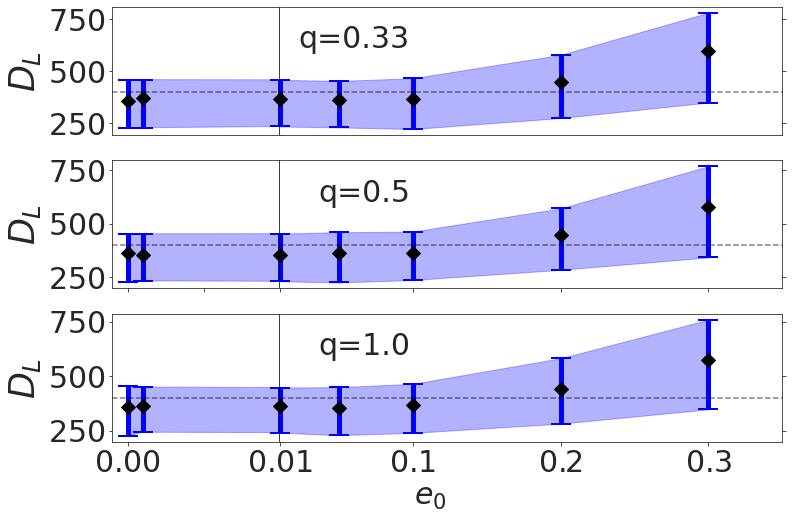}
	\caption{Estimated values of the luminosity distance for varying eccentricity using \TEOBResumE{} as the signal waveform and \Phenom{} in the likelihood model. 
	The dashed line is the signal's distance of 400 Mpc, with the bars represent the 90\% confidence intervals, and diamonds the median value.  Note the different scales between $[0., 0.01]$ and $[0.01, 0.3]$ on the $x$-axis. }

	\label{figure:reduction_distance}
\end{figure}

We have already observed binary black hole mergers with masses 
ranging from $14$~\SolarMass{} to $160$~\SolarMass{}~\cite{gwosc_catalog}.
In this study, we focus on binaries at the lower end of this spectrum, as a
longer inspiral signal will have a stronger eccentric signature.
Also, at lower masses the inspiral of the signal dominates the SNR as 
compared to the merger portion, which allows us to draw reliable 
conclusions using inspiral-only waveform models.

Since our goal is to understand the biases of existing GW parameter 
estimation pipelines that use quasi-circular waveforms as filter 
templates, we limit our waveform models in the likelihood evaluation
to \texttt{SEOBNRv4\_ROM}~\cite{seobnrv4_rom} and
\texttt{IMRPhenomD}~\cite{imrphenom_1, imrphenom_2}. We note that while
more recent EOB and Phenom models have become available more
recently~\cite{Pratten:2020ceb,Ossokine:2020kjp}, we do not employ them 
here for two reasons: (i) for the binary mass and spin parameter space
we consider, the \texttt{SEOBNRv4\_ROM} and \texttt{IMRPhenomD} 
models have good agreement with numerical relativity 
waveforms~\cite{seobnrv4_rom, imrphenom_1, imrphenom_2}, and (ii) the
newer EOB models were too computationally expensive for Bayesian parameter
estimation~\cite{Ossokine:2020kjp, lalsuite}.
Since the waveforms produced with our modified \ENIGMA{} model are 
inspiral-only, we also utilise the PN quasi-circular model 
\TaylorF{} to rule out any biases that may arise compared to 
inspiral-merger-ringdown templates. PE results with \TaylorF{} also
serve to nicely illustrate the kind of biases that result from ignoring
the merger-ringdown in template waveforms.
Using $3$ waveform models for the likelihood, $2$ waveform models
for the signals, and the direct product of the parameters described in
Table~\ref{table:parameter_spec} gives us 126 signals on which to perform
parameter estimation. As a control, we also study a further set of six
signals using \ENIGMA{}-Inspiral as the waveform approximant for the
signal {\it and} likelihood evaluation, with $q= 0.5$ and the eccentricity
taking the values in Table~\ref{table:parameter_spec}. The priors for all the parameter estimation analyses are shown in~\ref{table:priors}. We use a
non-spinning prior for our simulated signals since the majority of the
black holes detected thus far have had negligible spins.

Each of these signals is generated from a starting frequency 15 Hz using the given
waveform model, and then injected into zero noise. As described in Appendix C
of~\cite{Rodriguez_2014}, using zero noise eliminates any biases in parameter
estimation that would be caused by a particular noise realization. The data is
sampled at 2048 Hz, and we use the two-detector network of LIGO Hanford and
Livingston.
In the calculation of the likelihood, we use $f_{min} = 20$ Hz. 
We sample in chirp mass and mass ratio, rather than component 
masses. We use sampling priors shown in Table~\ref{table:priors}.
At a luminosity distance of $400$~Mpc, the signals are relatively loud,
and have a two-network SNR ranging from 21 to 33 at design sensitivity.
We deliberately choose for our signals to be loud in order to unambiguously quantify systematic correlations in the intrinsic parameters of non-spinning eccentric binaries. Low SNRs would lead the statistical uncertainties to dominate and obscure the systematic degeneracies. In comparison, the GW events we investigate in Sec.~\ref{sec:gw_events} are also located at a similar luminosity distance ($350-450$~Mpc), but have SNRs $\sim 15$ due to lower sensitivity of the detectors at the time of their observation. In a sense, our simulated signals represent the future versions of GW151226 and GW170608 if they were seen during the fourth observing run of LIGO-Virgo or beyond. It is worth noting that at SNRs $\sim 15$, we expect statistical uncertainties in parameter measurements that are $\sim \sqrt{2}$ times (i.e. 40\%) larger than those of the injections discussed here.

We start the discussion with Figure~\ref{figure:sample_plot} where we show the
posterior distributions recovered for four representative simulated binary mergers.
The signal waveform model is
\TEOBResumE{}, which we expect to be the best example since it is a
full IMR model, with zero and nonzero eccentricity, and the mass ratios
$q = 1$ and $ q=0.33$. We opt to plot the symmetric mass ratio
$\eta = \frac{q}{(1 + q)^2}$, instead of $q$ in this section as it is the parameter
that appears to leading order in a PN expansion.

In the $e=0$ cases, \Phenom{} well recovers all of
the component parameters. Since these signals are high SNR and have a long
inspiral, the chirp mass is extremely precisely and accurately estimated, with
the 90\% confidence interval within less than 0.1\% of the simulated value.

The $e=0.3$ cases, however, show a very precise but inaccurate inference of
the chirp mass, with the quasi-circular template drastically
\textit{overestimating} its value.
The median recovered value deviates from the simulated value by
$\approx 3\%$, whereas the 90\% confidence interval is about $0.2\%$ of the
median estimated value.  
$\eta$ is well determined to be $\frac{1}{4}$ in both the eccentric and non-eccentric cases. Moreover, in the asymmetric mass, non-eccentric case, the correct value of $\eta$ is inferred within our error-estimates. However, when the simulated signal is asymmetric and eccentric, the value of $\eta$ is confidently and incorrectly inferred to be that of a symmetric binary. 

Figure~\ref{figure:enigma_enigma} illustrates these trends for the case of using \ENIGMA{}-Inspiral for simulating the signal and performing parameter estimation. At low eccentricities, the mass parameters are correctly recovered, but at higher eccentricities, the recovered parameters are those of a heavier, symmetric binary.

\subsubsection{Chirp Mass Recovery}

Figure~\ref{figure:reduction_chirp} summarizes the chirp mass estimates
for the rest of our simulated signals, showing the six combinations of
signal and recovery waveform models. The top row shows the case where
the simulated signal is generated with the IMR model, \TEOBResumE{},
and the bottom row with the inspiral model \ENIGMA{}-Inspiral.

Let us consider the top row, where \TEOBResumE{} is used for the signal. No
matter the template model, we find a systematic bias in chirp mass recovery
that increases in proportion to orbital eccentricity. When
\TaylorF{} is used as the waveform model for templates, the error bars are wider

due to the missing merger-ringdown portion. Also in the $e_{15} = 0$ case, there is
a slight bias in the \TaylorF{} recovery due the missing merger. But since
we see very similar results in the case of \EOBROM{} and \Phenom{}, we 
conclude that this result is invariant under a change of waveform family.

We now consider the bottom row, using \ENIGMA{}-Inspiral for the signal.
In the left and middle panels, when using IMR models for inference, we 
see
a small bias in chirp mass recovery even at $e_{15} = 0$. This comes from
the merger-ringdown portion of the IMR templates fitting to the last few
cycles of the \ENIGMA{}-Inspiral signal, resulting in a lower chirp mass.
We see no such bias in the \TaylorF{} case at $e_{15} = 0$.

For each combination of signal and recovery model there is a clear trend:
as the eccentricity of the simulated signal increases, the circular templates
strongly overestimate the chirp mass beyond the values within the $90\%$
credible intervals. The effect becomes more pronounced at higher
eccentricities, and the maximum fractional deviation over the dataset is
$\sim 9\%$.

\subsubsection{Mass Ratio Recovery}

Figure~\ref{figure:reduction_q} shows the effects of increasing eccentricity on the estimation of the symmetric mass ratio, $\eta$. Overall, the trend is less systematic than in the case of the chirp mass.

Let us consider the top row with \TEOBResumE{} signals. At $e_{15} = 0$, the recovered $\eta$ agrees well with the injected value for the two IMR waveform templates. For \TaylorF{} there is a systematic shift of the posterior toward higher $\eta$ values due to the missing merger-ringdown cycles in templates. This illustrates that we need to use IMR waveform templates for measurably more precise parameter recovery from GW signals.
Another thing we notice is that there is a jump in the recovered $\eta$ posterior when the eccentricity changes from
$e_{15} = 0$ to $e_{15} = 10^{-3}$. This is due to a slight discontinuity in the
\TEOBResumE{} waveform as the eccentricity goes to $0$. For small
eccentricities, $e_{15} \leq 0.01$, there is good agreement between \TEOBResumE{}
and \EOBROM{}. There is a slight bias when $e_{15} \leq 0.01$ in the \Phenom{}
case, since the Phenom family of waveforms is constructed differently from the
EOB family. In the \TaylorF{} case, the recovered $\eta$ is overestimated as
compared to the simulated value except for $\eta = \frac{1}{4}$. This
systematic bias is due to the lack of the merger-ringdown portion in the
templates.

Now consider the bottom row of \ENIGMA{}-Inspiral signals. In both the
\EOBROM{} and \Phenom{} case, we see the recovered $\eta$ is underestimated
compared to that of the simulated signal. Again this is because the full
IMR templates fit their merger-ringdown portion to the last few cycles of
the inspiral-only signal. In these cases the bias in recovery between IMR
templates with an inspiral signal is greater than the bias caused by using different waveform families. The bias at $e_{15} = 0$ is much reduced in the case of \TaylorF{}, since both the signal and templates are inspiral-only.
However there is a slight bias in the recovery of the $\eta = \frac{1}{4}$ signal, so the two waveforms models do not entirely agree in this limit.

Despite these discrepancies, the trend we have already seen in
Figure~\ref{figure:enigma_enigma} clearly persists in all cases: as the
eccentricity of the signal increases, the recovered mass ratio becomes
more consistent with that of a symmetric binary system. This is quite
striking, particularly at $e_{15} = 0.3$, the recovered mass ratio is
consistent with an equal-mass binary, {\it regardless of the signal's
mass ratio for all combinations of waveform models.}

\subsubsection{Distance Estimates} 

From the representative corner plots of Figure~\ref{figure:sample_plot}, we 
can see that in eccentric case there is also a slight bias in the recovery 
of the luminosity distance. Figure~\ref{figure:reduction_distance} shows
the trend for the case of a \TEOBResumE{} signal with \Phenom{} template. 
This overestimation of luminosity distance is consistent with the templates
not capturing the entire signal SNR because of missing physical effects, i.e.,
orbital eccentricity. Having said that, even though there is a clear trend
upward in the distance recovery, we find it to broadly remain within the
statistical errors of the parameter estimation. We expect this bias to
become significant only at even higher SNR than we consider. In the worst
case, the median value is $\approx 50\%$ higher than that of the simulated
value. We also note that the behavior is largely identical for each of
the different simulations' mass ratio.

\subsubsection{Maximum a-posteriori Estimated Waveforms}

To help understand what features of the simulated and recovered best-fit
waveforms contribute to the biases in parameter estimation, it is useful to
look at the recovered maximum a posteriori (MAP) waveforms for some of our
signals.

%

\begin{figure*}[t]
	\centering
	\subfigure[$e = 0$ with \TEOBResumE{}]
	{\includegraphics[width = 85mm]{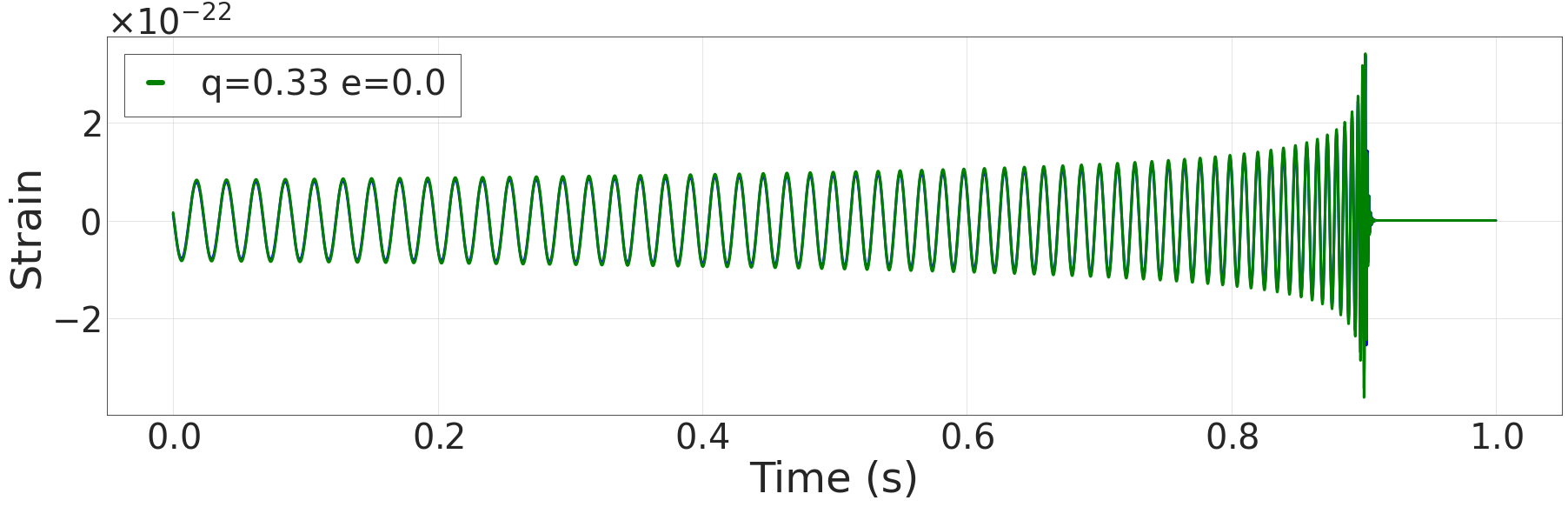}}
	\subfigure[$e = 0.3$ with \TEOBResumE{}]{
	\includegraphics[width = 85mm]{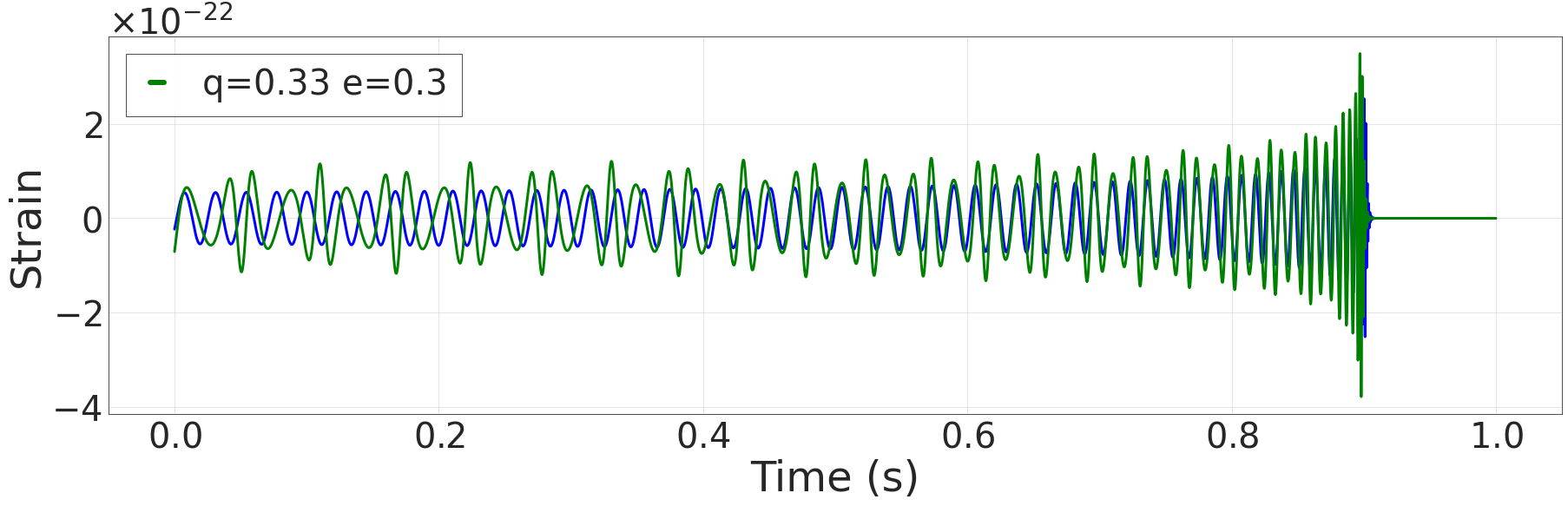}}

	\subfigure[$e = 0.$ with \ENIGMA{} Inspiral]{
	\includegraphics[width = 85mm]{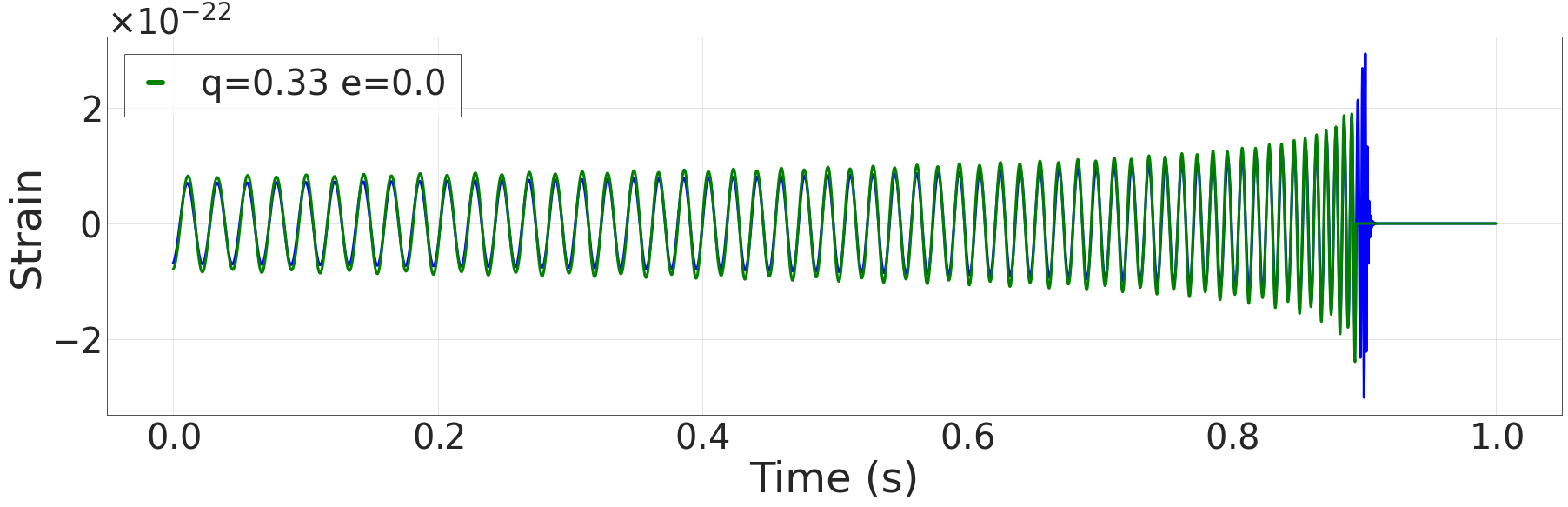}}
	\subfigure[$e = 0.3$ with \ENIGMA{} Inspiral]{
	\includegraphics[width = 85mm]{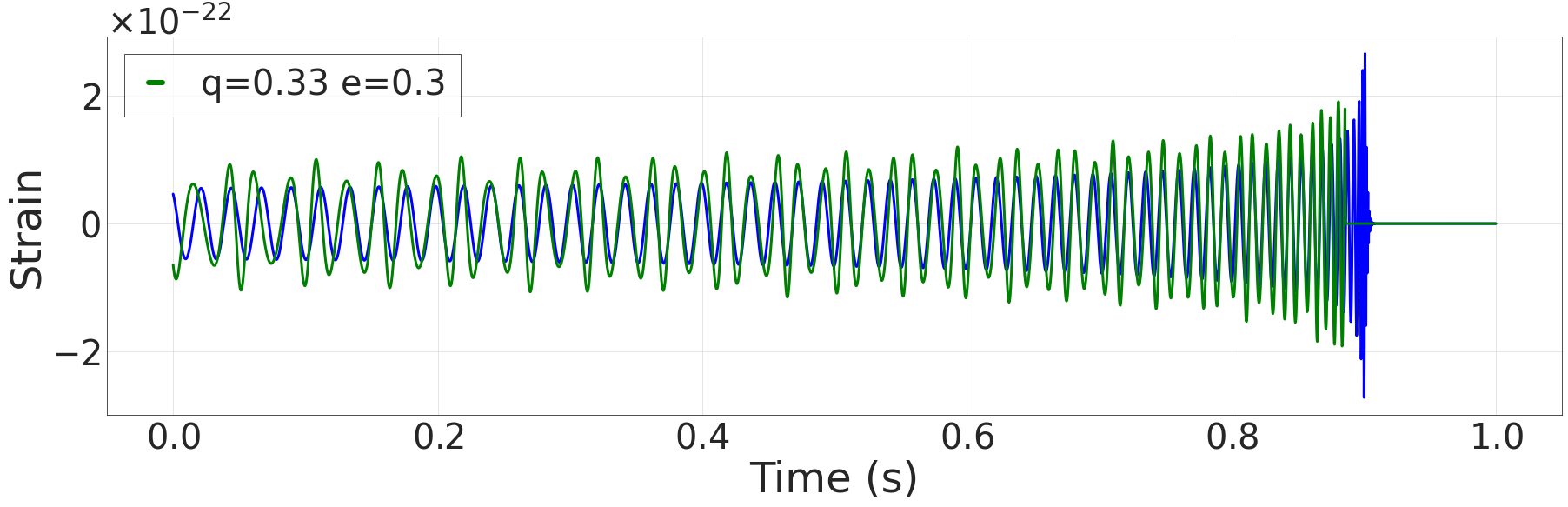}}
	\caption{Comparison of the maximum a posteriori \Phenom{} waveform (blue) to
	the signal waveform (green) in the case of zero and non-zero
	eccentricity, with mass ratio $q=0.33$. Only the last $1$s of the waveform is shown
	and the coalescence time and phase of the waveforms have been aligned.}

	\label{figure:MAP}
\end{figure*}

Figure~\ref{figure:MAP} shows the maximum likelihood estimated \Phenom{}
waveform from four representative runs with $q=0.33$, with \TEOBResumE{} and
\ENIGMA{}-Inspiral used for the signals. There is strong agreement in the
zero-eccentricity limit between the template and both signal waveform models,
as is to be expected. When the eccentricity is increased to $e=0.3$, the
quasi-circular model tries to best fit to the eccentric signal.
In the case of the full IMR signal with \TEOBResumE{}, the best
quasi-circular
template fits to the frequency of the {\it last few cycles} along with the
merger-ringdown. In the inspiral-only case, the best template tries to fit
to the {\it lower frequency cycles} of the last portion of the inspiral.

\subsection{ Gravitational-wave Events}\label{sec:gw_events}

\begin{figure*}
    \subfigure[GW151226, $|\chi_{1, 2}| = 0$]{
    \includegraphics[width=\columnwidth]{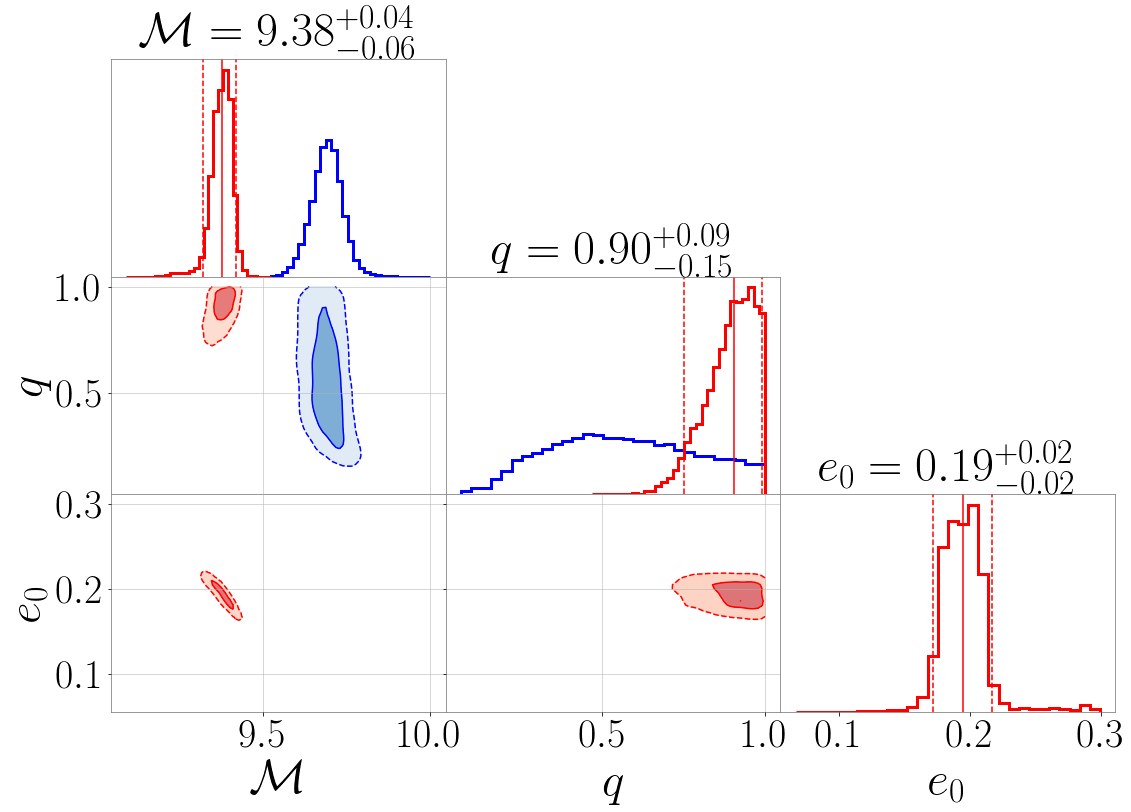}}
    \subfigure[GW170608, $|\chi_{1, 2}|= 0$]{
    \includegraphics[width=\columnwidth]{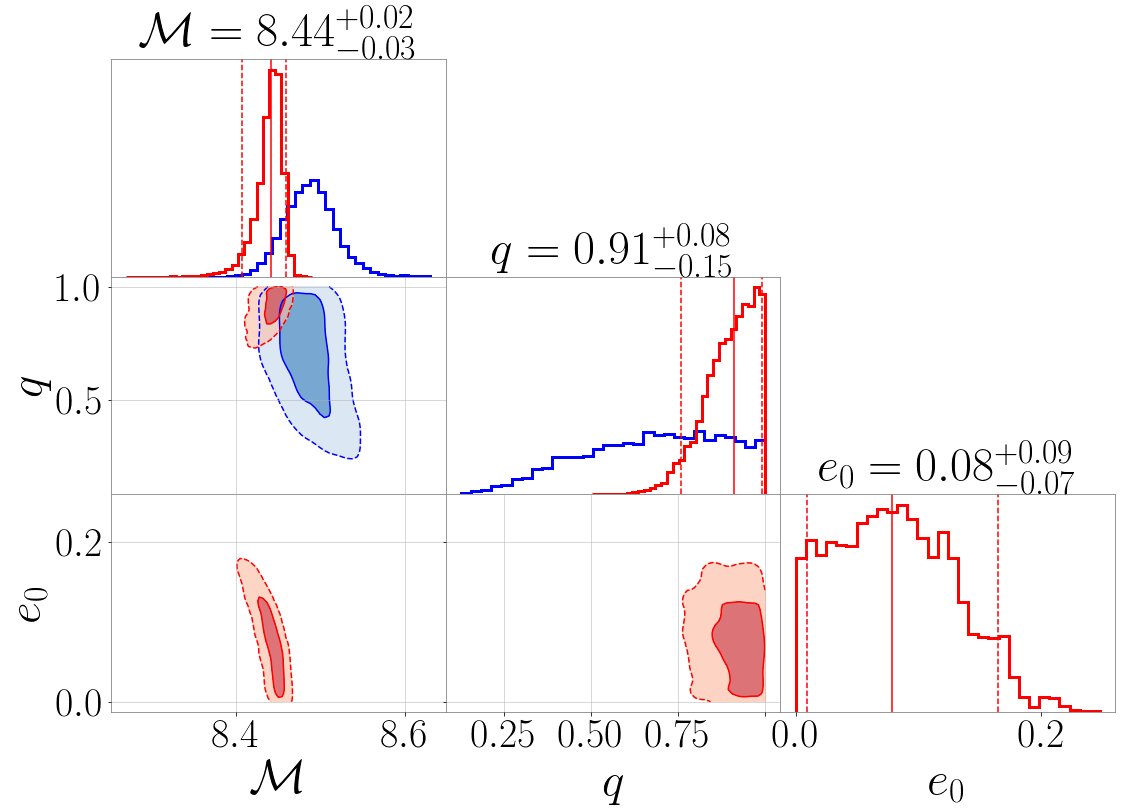}}
    \subfigure[GW151226, $|\chi_{1, 2}| < 0.3$]{
    \includegraphics[width=\columnwidth]{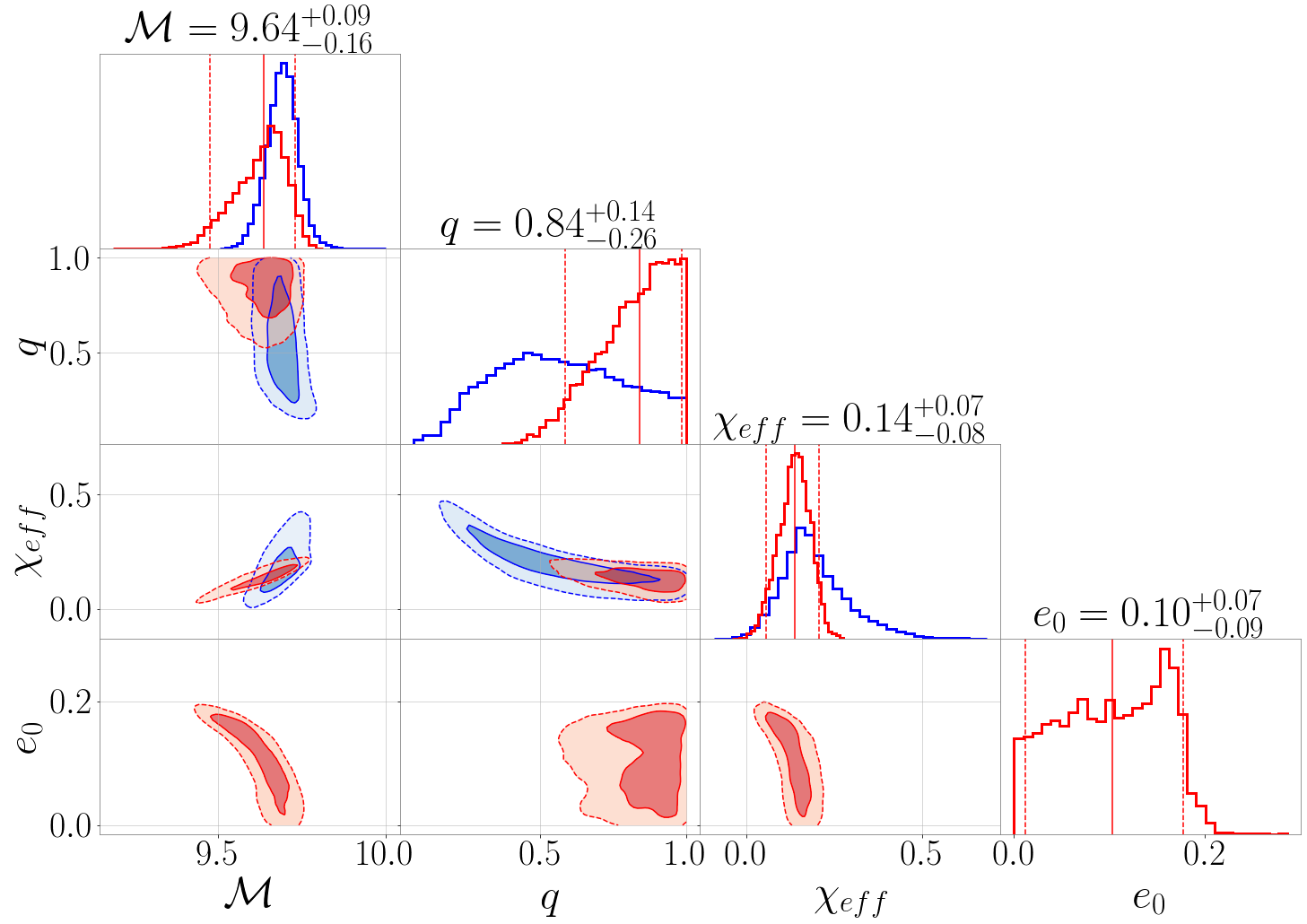}}
    \subfigure[GW170608, $|\chi_{1, 2}| < 0.3$]{
    \includegraphics[width=\columnwidth]{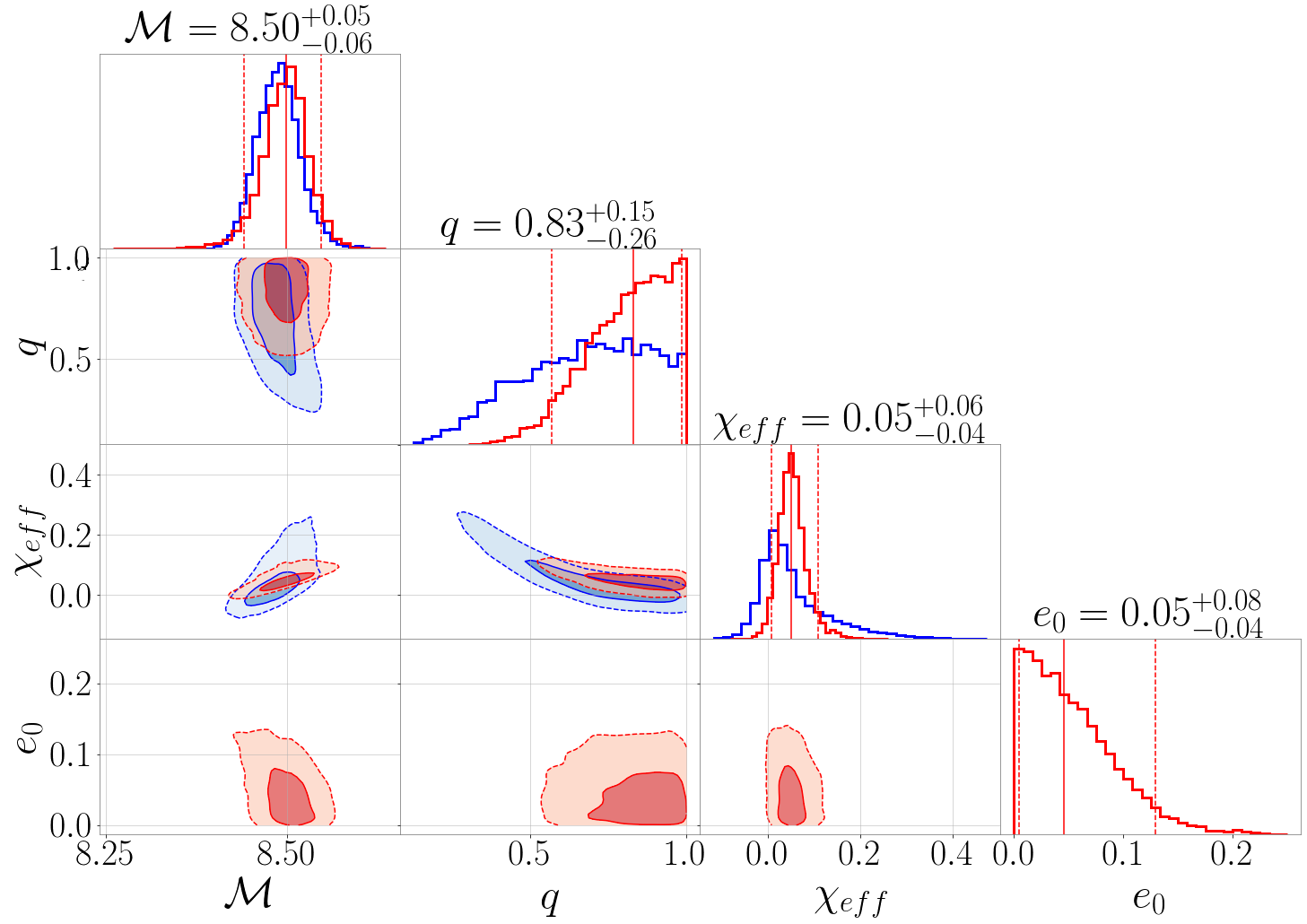}}
    \subfigure[GW151226, $|\chi_{1, 2}| < 0.99$]{
    \includegraphics[width=\columnwidth]{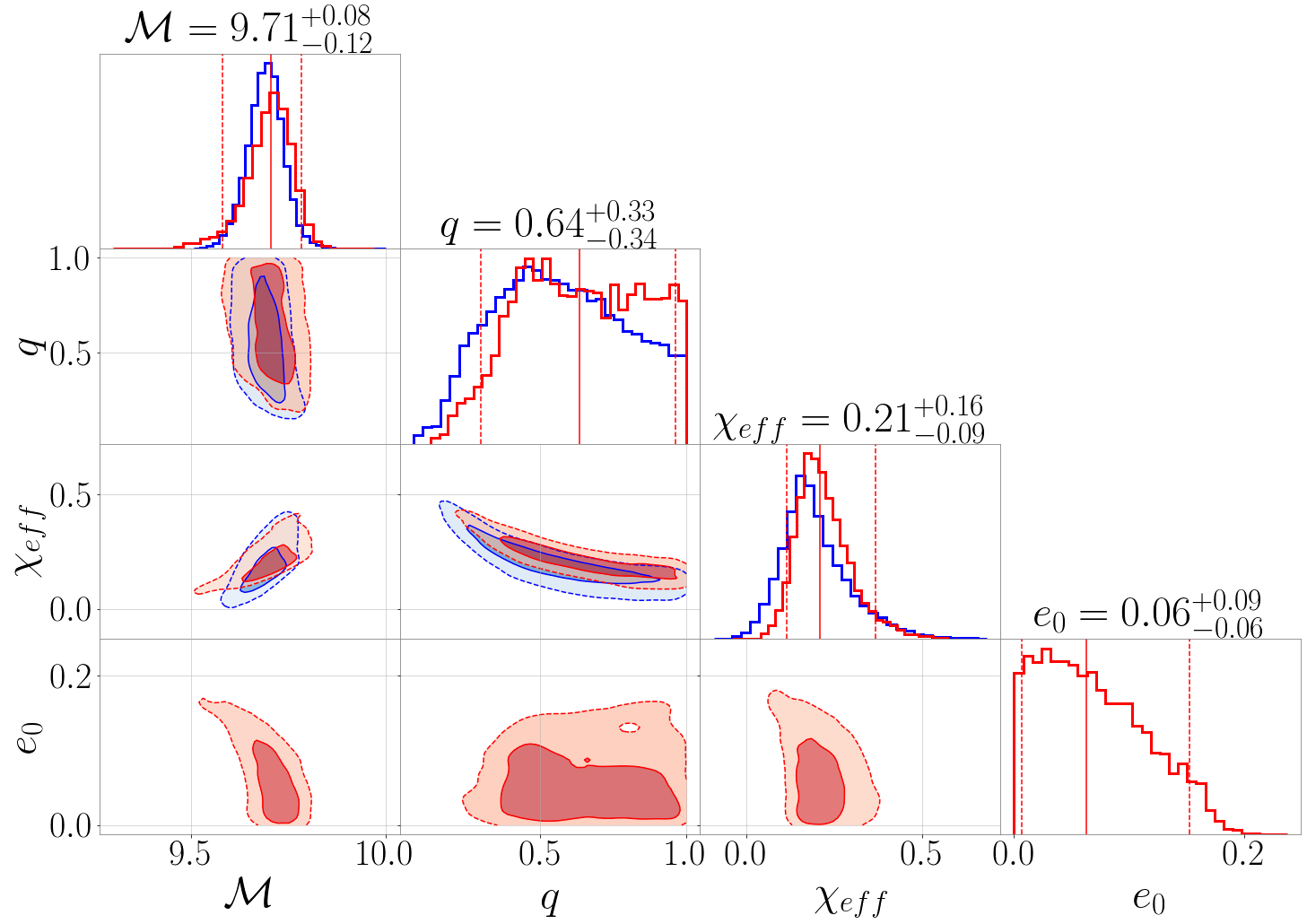}}
    \subfigure[GW170608, $|\chi_{1, 2}| < 0.99$]{
    \includegraphics[width=\columnwidth]{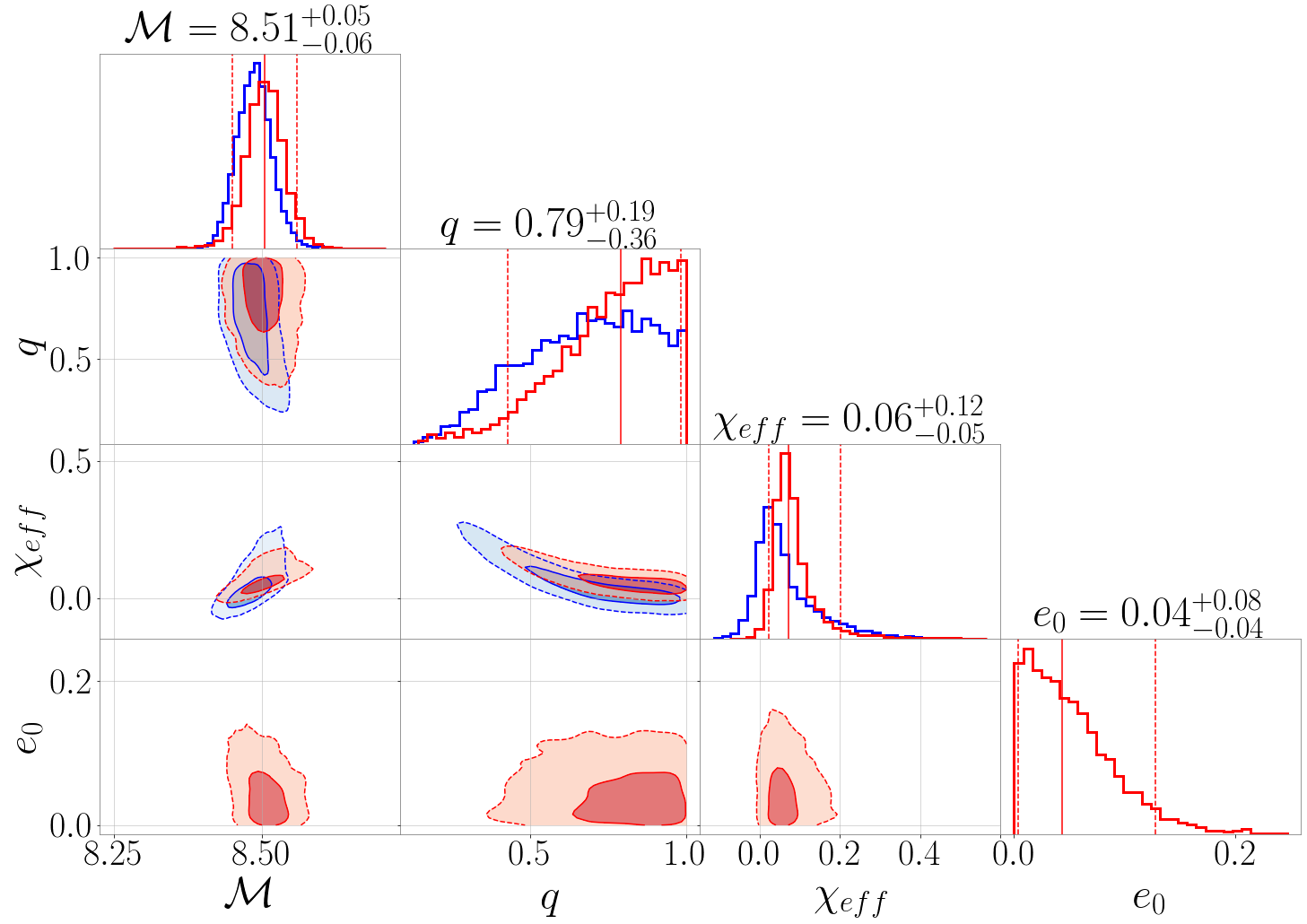}}
    \caption{ \textit{Red:} Posterior probability distribution using \TEOBResumE{} for GW151226 and GW170608. Panel captions indicate the maximum value of the $z-$component of BH spins, $\chi_{1,2}$, allowed in the parameter estimation prior. Median estimates with 90\% confidence intervals are shown for this data. \textit{Blue:} GWTC-2 data from~\protect{\cite{gwtc_data}}.  Contours show the regions of 50\% and 90\% confidence. See Table~\ref{tab:events} for a comparison with the exact medians and error estimates of the GWTC data. See section~\ref{sec:gw_events} for discussion.
    }
    \label{fig:events}
\end{figure*}


Recently, Ref.~\cite{Wu2020} has posited that events GW151226 and
G170608 in the first GW transients catalog~\cite{Abbott_2019} are
mergers of eccentric binary black holes. This contrasts the conclusion of Ref.~\cite{Romero_Shaw_2019}, that these two events are likely quasicircular binary black hole mergers. It is important to note that of these two studies, the first used an inspiral-only, non-spinning eccentric waveform model~\cite{EccentricFD}, while the latter used a complete inspiral-merger-ringdown model from the effective-one-body family: \texttt{SEOBNRE}~\cite{seobnre}. Because they use different waveform models, their eccentricity measurements are not directly comparable in a precise sense~\cite{Knee:2022hth}. But the claim of an event being eccentric at all could qualitatively withstand this difference, as its unlikely that any waveform model would assign zero eccentricity to a waveform that has modulations manifested by eccentric binaries. In addition, Ref.~\cite{Romero_Shaw_2019} does not sample from the posterior distribution directly as \texttt{SEOBNRE} is too expensive to evaluate, instead their eccentric posteriors were found by re-weighting posteriors from non-eccentric parameter estimation runs.

In this section we study these two events with a two-fold purpose.
We use the full IMR eccentric waveform model to
attempt to better understand the nature of these two events. And
we use them to further confirm the measurement degeneracy between
binary chirp mass and initial orbital eccentricity that we have
so far shed light on using simulated signals. Our application of \TEOBResumE{} 
is sufficiently fast to allow its direct use in parameter estimation runs with eccentricity and spins. 

We use the same Bayesian inferencing setup as described in
Sec.~\ref{sec:pe}. We use the open-source data available from the
Gravitational-Wave Open-Science Centre~\cite{gwosc_catalog}.
The estimated parameters with confidence intervals for this section are summarised in Table~\ref{tab:events}. 
\subsubsection{Non-spinning, Eccentric Inference}

First we perform parameter estimation using \TEOBResumE{}
with a non-spinning prior, but now we include eccentricity in the
analysis. Our prior for the eccentricity is uniform in the range
$0-0.3$, and the waveforms are generated from a lower frequency
of $10$ Hz, rather than the 15 Hz used in our simulated signals.
This is why we denote the measured initial eccentricity as $e_{10}$Hz.
This is to allow us to compare our eccentricity measurements with 
Refs~\cite{Romero_Shaw_2019} and~\cite{Wu_2020} who use this as their reference frequency. We again use $f_{min} = 20$ Hz in the evaluation of the likelihood integral in Eq.~\ref{likelihood}.

Panels (a) and (b) in Figure~\ref{fig:events} show the parameter
estimation results of these two events. In the case of GW151226 we
measure an eccentricity of $e_{10} \approx 0.2$, with $e_{10} = 0$
strongly excluded from the posterior by our non-spinning eccentric
waveform model (this measurement will change when we include the effect
of BH spins in the following section). Furthermore, in agreement with
our study of simulated eccentric signals, the recovered chirp mass
differs remarkably from that measured by the LVK in~\cite{gwosc_catalog},
by an amount greater than the statistical error of the measurement.
In the case of GW170608, the recovered eccentricity is very much
consistent with $e_{10} = 0.$ with an upper $90\%$ estimate of $e_{10} = 0.16$.
Its chirp mass is measured to be somewhat lower than the LVC estimate,
albeit consistent with $90\%$ credible intervals.
In both cases the symmetric mass ratio is determined to be closer to
$q=1$ with greater certainty than the LVC estimate~\cite{Abbott_2019}.

In both cases, the 2-dimensional marginalized posterior for the chirp
mass and eccentricity corroborate what our simulated signal study found
along with~\cite{Lenon_2020}: the measurement of the chirp mass and
eccentricity are correlated. These results show that GW151226 would be
interpreted as an eccentric system with a lower chirp mass than
previously thought~\cite{Abbott_2019}, if we disregard component spins.
The same effect is observed to a lesser degree in GW170608, which we
find to be consistent with having a non-eccentric origin. We also
note that allowing for eccentricity as a degree of freedom helps
make the measurement of both chirp mass and mass ratio more precise
for this event.

\subsubsection{The importance of including spin}

It is well known that the measurement of binary spins is correlated
with the measurement of the mass parameters when determining the
source parameters of gravitational-wave signals. As explained
by~\cite{Cutler_1994}, the error in the estimated chirp mass and
mass ratio increase when one includes the spins of compact
objects in their analysis, even if the system is non-spinning.
This makes it reasonable to expect our confident mass estimates
to be influenced by the exclusion of spin in our prior. Also,
as we have found previously, the chirp mass measurement is correlated
with orbital eccentricity, and so it is also possible that the
chirp mass measurement is correlated with the spin measurement.

We repeat our analysis of the two events GW151226 and GW170608,
with the same set of priors as in the previous section, but now
with an aligned-spin prior. We use the ``z-prior'' implemented in
Bilby and described in~\cite{lange2018rapid} for the aligned spins.
We perform several runs with different values of the maximum spin
magnitude for the prior to investigate how the resulting posterior
depends on the spin prior. We use $|\chi_{\mathrm{max}}| < 0.3, 0.7, 0.99$.
The value of 0.7 is chosen since the model \TEOBResumE{} is only
validated against NR waveforms with spin up to this value. 
In the results we will quote only the results of the effective spin,
which is the best recovered spin parameter~\cite{Abbott_2016}
and is defined as
\begin{equation} 
    \chi_{\rm eff} = \frac{m_1 \chi_1 + m_2 \chi_2}{M} 
    \label{eq:chi_eff}
\end{equation}
where $M$ is the total mass and $\chi_{1,2}$ are the components of
the spin aligned with the binary's orbital angular momentum.

Panels (c) and (d) in Figure~\ref{fig:events} show the results for the
two events with $|\chi_{1,2}| < 0.3$. Panel (c) shows the results from
GW151226. The posterior for eccentricity now has support for $e_{10} = 0$,
in contrast with the non-spinning prior, although the $e_{10}$ posterior
peaks at about $e_{10}=0.16$. From the $\mathcal{M}_c-e_{10}$ plane,
we can clearly see the negative correlation between the chirp mass and
eccentricity. The chirp mass estimate agrees better
with that of the LVC than in the non-spinning case, except the
posterior is slightly broader due to the higher eccentricity
contribution of the posterior. Most interesting for this event
is the $\chi_{\mathrm{eff}}-e_{10}$ plane: we see a negative
correlation between the spin and eccentricity measurements,
with $e_{10} = \chi_{\mathrm{eff}} = 0$ excluded. 

Panel (d) of Figure~\ref{fig:events} shows the results for
GW170608. We find that including spins has much improved the
agreement with the LVC chirp mass estimate. Also the posterior
for orbital eccentricity has become narrower, with the upper
$90\%$ credible limit decreasing from $e_{10}=0.18$ to $0.13$.
We can conclude that even when the event being analysed is
likely non-eccentric, including spin in the analysis improves
the confidence in the eccentricity constraints.


Panels (e) and (f) show the results from expanding the maximum spin in the prior to $0.99$. 
Panel (e) shows the event GW151226. We can immediately see that the constraint on the eccentricity is improved, and the recovery of the masses and effective spin are in better agreement with the GWTC-2 values. The correlation between the effective spin and eccentricity is still present in the 2D $\chi_{\rm eff}-e_{10}$ posterior but appears weaker than in the case of the restricted spin prior. This demonstrates further that the measurement of the eccentricity is correlated with that of the spin, and that to accurately measure the eccentricity one should sample from as much of the spin prior as possible, which requires waveform models that are reliable for moderate to large spin values. 

Panel (f) shows the event GW170608. In this case the posteriors are largely unchanged from the $|\chi_{1, 2}| < 0.3$ case, with the eccentricity constraint becoming $e_{10} < 0.12$. Since this event does not have appreciable spin, the more restrictive spin prior is well able to account for the full range of physics contributing to the posterior. 

Table~\ref{tab:events} summarises the results from each of our runs. We opt not to plot the posteriors for the case of $|\chi_{1,2}| < 0.7$ but the results are largely unchanged between this case and that of $|\chi_{1,2}| < 0.99$, despite the \TEOBResumE{} model not being validated to spins of that magnitude.


To evaluate the preference for the eccentric model over the non-eccentric one, we can calculate the Bayes factor: 
\begin{equation}
    \mathcal{B} = \frac{Z_{\mathrm{eccentric}}}{Z_{\mathrm{non-eccentric}}}. 
    \label{eq:bayes_factor}
\end{equation}
where $Z$ is the Bayesian evidence. Since the set of parameters for the non-eccentric model is nested within the full eccentric model, this ratio becomes the Savage-Dickey factor~\cite{dickey}:s
\begin{equation} 
    \mathcal{B} = \frac{\pi(e_{10} = 0)}{p(e_{10} = 0)}
    \label{eq:dickey}
\end{equation} 
which is simply the ratio of value of the prior weight $\pi(e_{10} = 0)$ to the posterior weight $p(e_{10} = 0)$. 

Table~\ref{table:bayes_factors} shows the estimated Bayes' factors for the two events as a function of increasing the maximum in the spin prior. In all cases we have $\mathcal{B} < 1$, indicating that the non-eccentric model is preferred. In the case of GW151226, the preference for the non-eccentric model increases as the maximum spin is increased in the prior. As expected from our previous results, increasing the spin has little effect on the results in the case of GW170608.

\begin{table}
\begin{ruledtabular}
	\begin{tabular*}{8.6cm}{l@{\extracolsep{\fill}} ccc}
		Prior   &  GW151226  & GW170608              \\ \hline
$| \chi_{1,2} | < 0.3 $ & 0.81 & 0.28 \\ 
$| \chi_{1,2} | < 0.7 $ & 0.45 & 0.28 \\ 
$| \chi_{1,2} | < 0.99 $ & 0.40 & 0.28 \\ 
	\end{tabular*}
\end{ruledtabular}
	\caption{Bayes factors for the eccentric versus non-eccentric model, with increasing value of the maximum spin in the prior. }
	\label{table:bayes_factors}
\end{table}

\subsection{Circular Simulated Signals for GW151226 and GW17060}
\label{sec:sim_events}

\begin{figure*}[t]
    \centering
    \subfigure[GW151226-like, $|\chi_{1, 2}| = 0$]{
    \includegraphics[width=8.6cm]{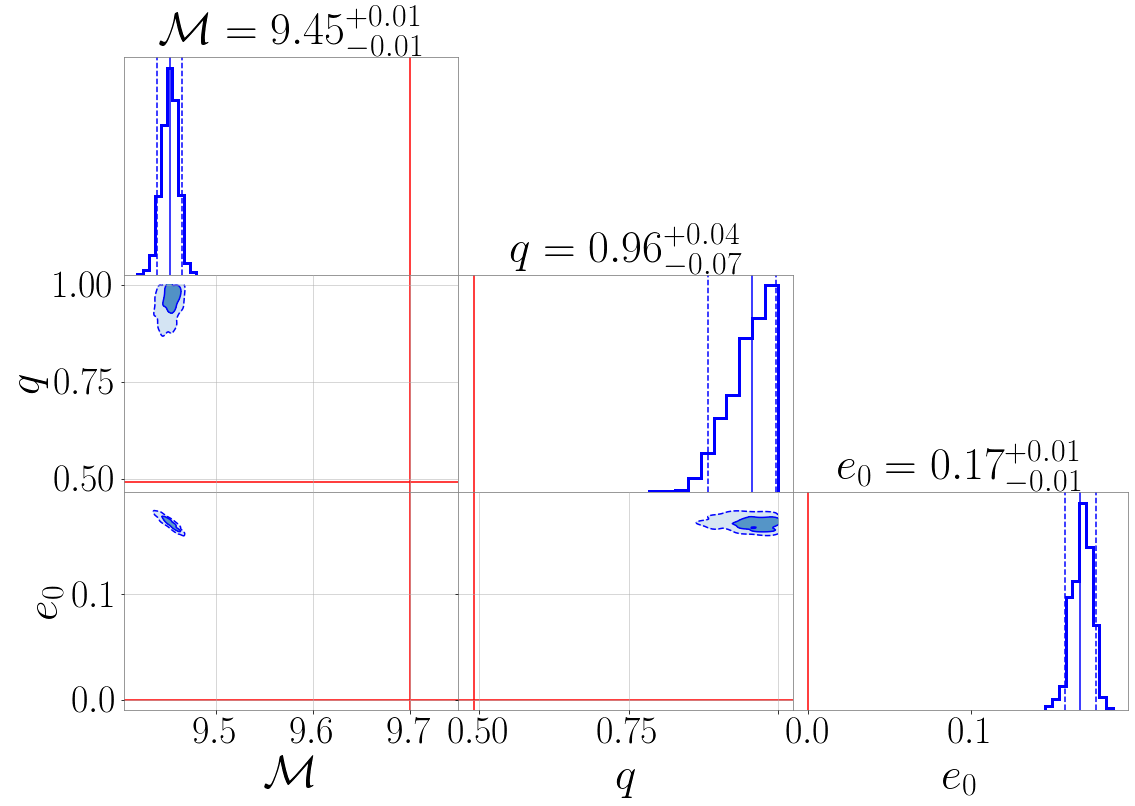}}\quad
    \subfigure[GW170608-like, $|\chi_{1, 2}| = 0$]{
    \includegraphics[width=8.6cm]{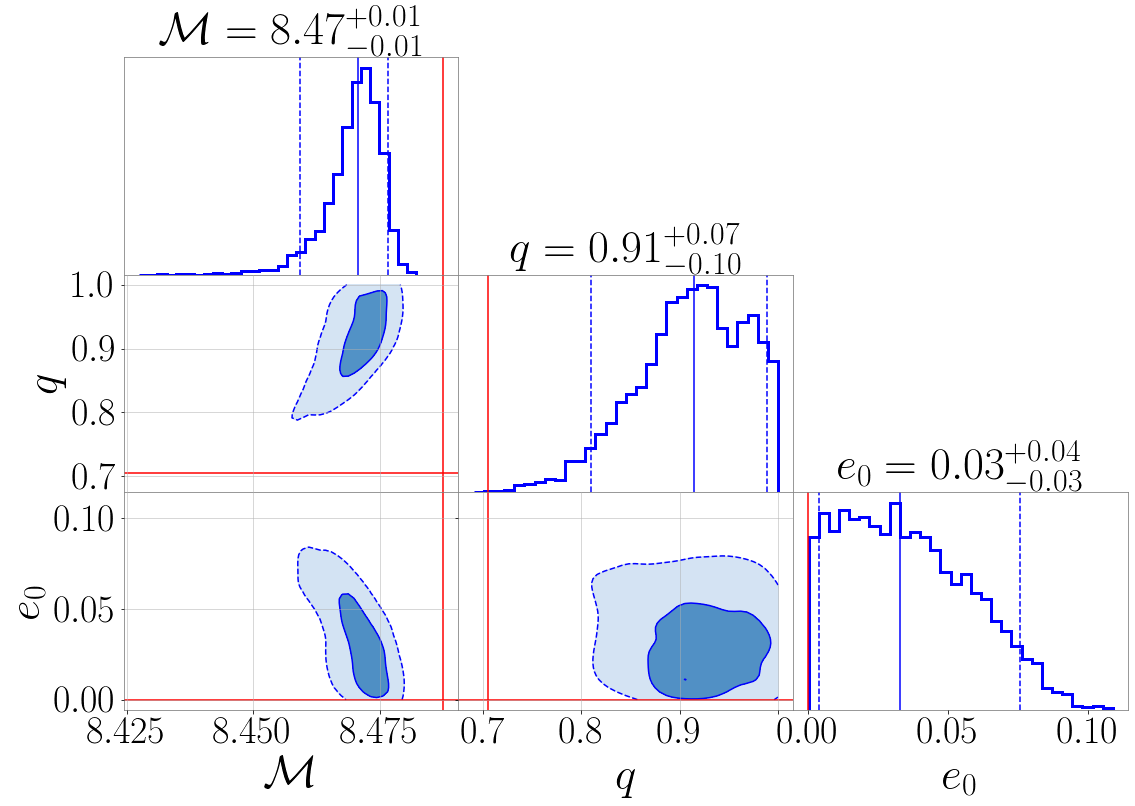}}
     \subfigure[GW151226-like, $|\chi_{1, 2}| < 0.3$]{
    \includegraphics[width=8.6cm]{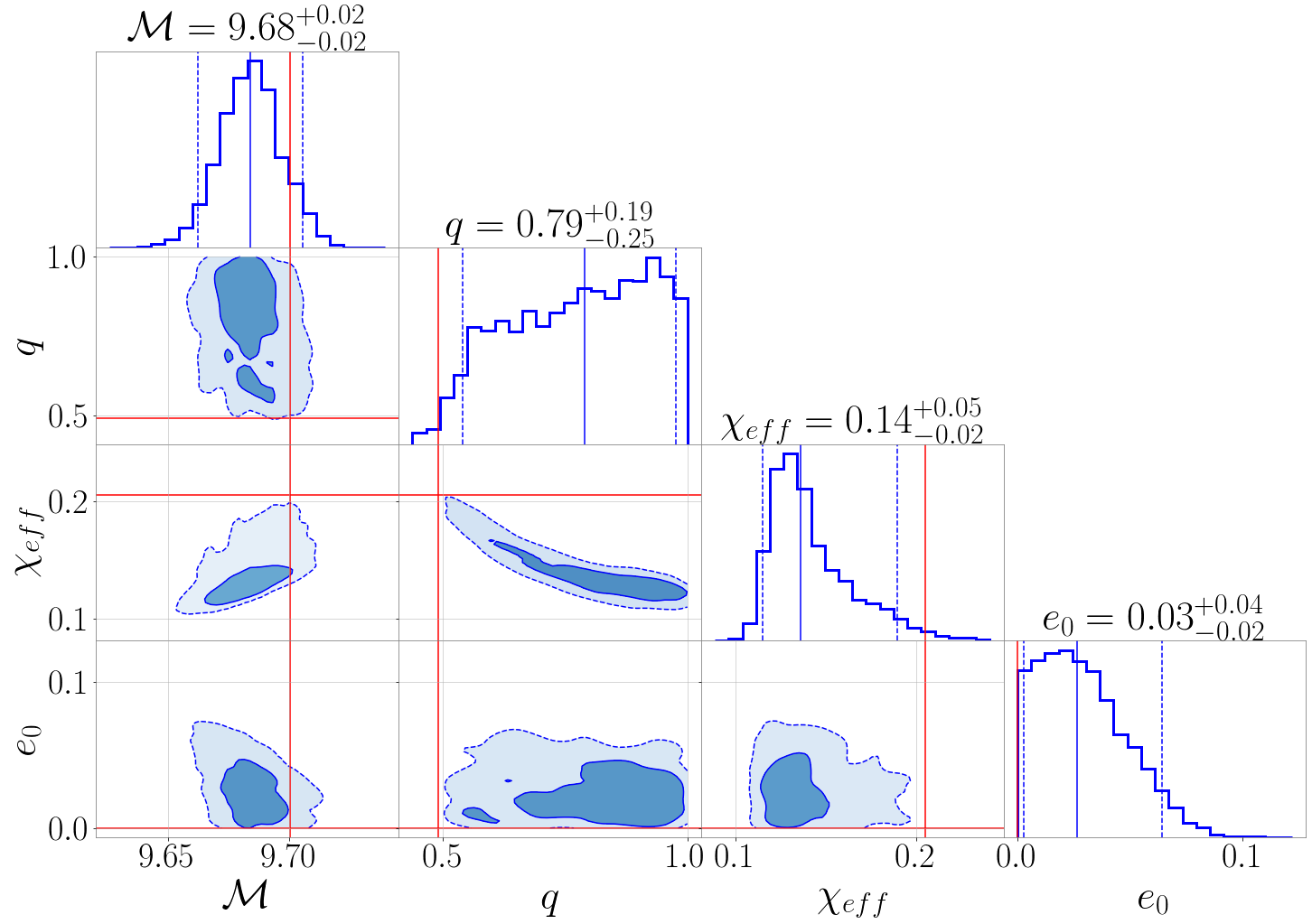}}\quad
    \subfigure[GW170608-like, $|\chi_{1, 2}| < 0.3$]{
    \includegraphics[width=8.6cm]{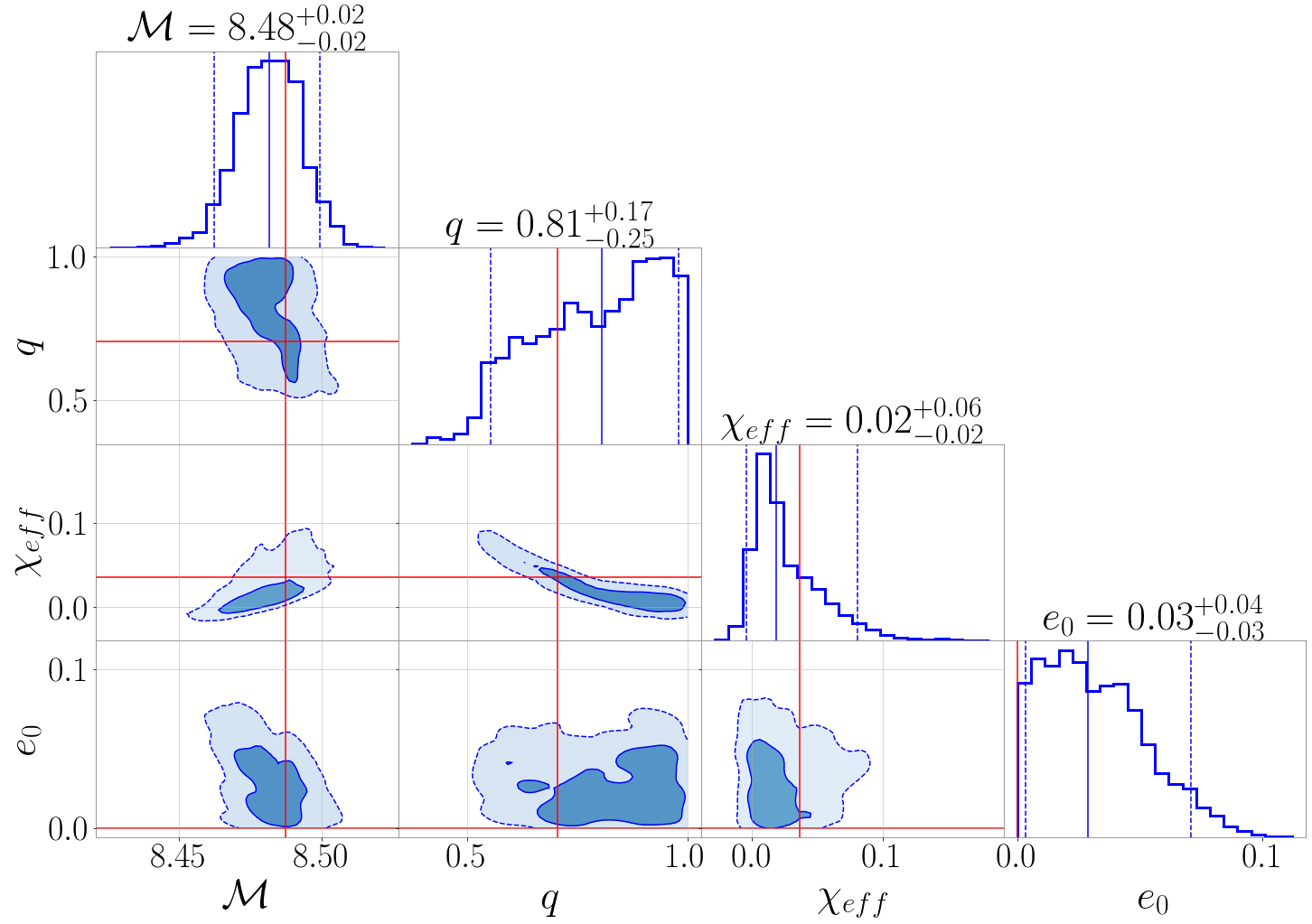}}
    \subfigure[GW151226-like, $|\chi_{1, 2}| < 0.99$]{
    \includegraphics[width=8.6cm]{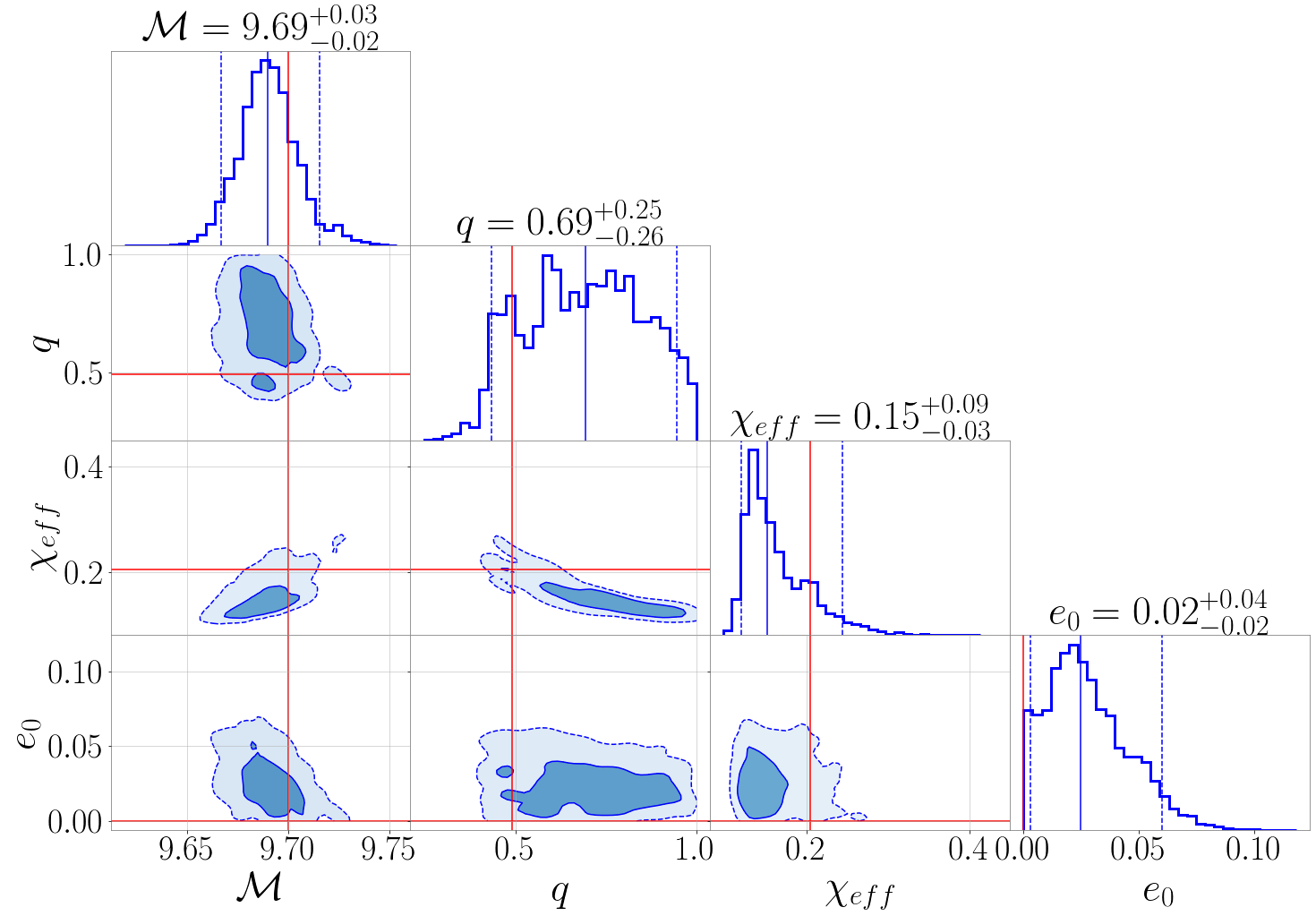}}\quad
    \subfigure[GW170608-like, $|\chi_{1, 2}| < 0.99$]{
    \includegraphics[width=8.6cm]{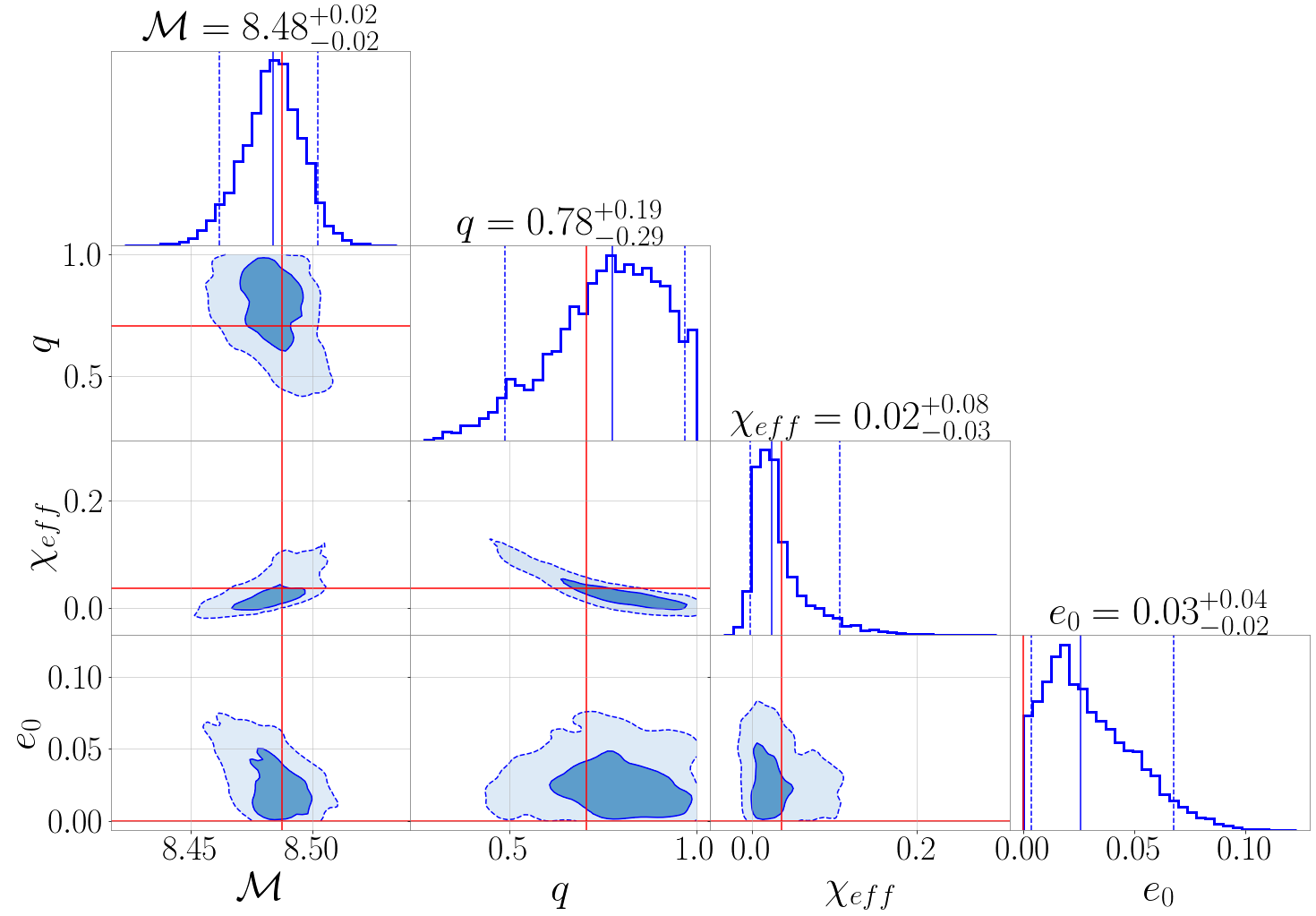}}

    \caption{ Posterior probability distribution using \TEOBResumE{} for GW151226-like and GW170608-like simulated signals, with an eccentric prior. Shown are the recovered distributions for source chirp mass, $\mathcal{M}$, mass ratio $q$, effective spin $\chi_{\mathrm eff}$ and eccentricity $e_{10}$. Red marks the simulated signal's values, which correspond to the MAP values given by the GWTC posteriors for the corresponding event. Shaded contours correspond to the 90\% and 50\% confidence regions. See section~\ref{sec:sim_events} for discussion.
  }
    \label{fig:circ}
\end{figure*}

We reaffirm our findings for GW151226 and GW170608 by follow-up
analyses with a set of simulated signal studies. The purpose of
this is to investigate how binary source parameters would be
recovered using \TEOBResumE{}, if the
signals were truly non-eccentric. We use the MAP source
parameters from the GWTC posteriors for GW170608 and GW151226,
and again inject the signals into zero-noise, and use the LIGO
Hanford and Livingston detectors noise curves.

We use the \TEOBResumE{} model with aligned spins to simulate the
signals, and then repeat our parameter estimation using the same
model, with the sampler setup from our previous runs. We wish for
the signal to have $e_{10} = 0$, but from Figure~\ref{figure:reduction_q}
we saw that \TEOBResumE{} has a discontinuity at $e_{10}=0$ so we
set the value to $10^{-10}$. We generate our signals from a lower frequency of 10 Hz, and 
use 20 Hz as the lower frequency cutoff in the evaluation of the likelihood integral. 
We use the same eccentricity prior
as before in all cases, and again perform two runs for each
event's parameters: with a non-spinning prior and with an aligned
spin prior.

Figure~\ref{fig:circ} shows the results for the two events' parameters using both a non-spinning and an aligned-spin prior with $\chi_{1,2} < 0.3$ and $<0.99$. Panels (a) and (b) show the non-spinning case. We see the same general trends as in Figure~\ref{fig:events} with the real event data: the chirp mass for both events is underestimated, the mass ratio is overestimated to be more consistent with that of a symmetric binary, and the presence of spin in the case of GW151226 leads to a spurious eccentricity measurement.

Panels (c) and (d) shows the results when the prior
is expanded to include aligned spins up to $0.3$. Since we are using the same
waveform model for the signal and the parameter estimation, we now
see a much better recovery of parameters as expected. Again the
non-zero eccentricity measurement for GW151226 disappears. 

In particular, the correlation between spin and eccentricity seen in panels (c) and (e) in Figure~\ref{fig:events} is not seen, and the
signal is well recovered as a spinning non-eccentric signal.

Panels (e) and (f) show the results when the maximum spin is increased to $0.99$. The results are qualitatively similar, but the recovery of the effective spin is improved.  


These results also further illustrate the fact the mass parameters,
spin and eccentricity are all correlated and it is important to
include each of these effects when performing parameter estimation.
We defer a more systematic investigation of this degeneracy to
future work.

\begin{table}[t]
\caption{\label{tab:events}%
Median estimates of source parameters from using \TEOBResumE{} for
parameter estimation. The left-hand column specifies the prior on the aligned spin, or whether the data is from GWTC-2. Upper and lower error estimates are the $90\%$
confidence intervals of the 1D marginalized posterior.
} 
\begin{ruledtabular}
\begin{tabular}{lllll}
 & \multicolumn{1}{c}{$\mathcal{M}_c$} & \multicolumn{1}{c}{$q$} & \multicolumn{1}{c}{$\chi_{\rm eff}$} & \multicolumn{1}{c}{$e_{10}$} \\ \cline{2-5} 
            &                        &                          &                          &                          \\
            & \multicolumn{4}{c}{GW151226}                                                                            \\ \cline{3-4}
            & \multicolumn{1}{c}{}   & \multicolumn{1}{c}{}     & \multicolumn{1}{c}{}     & \multicolumn{1}{c}{}     \\
$| \chi_{1,2} | = 0 $ & $9.38^{+0.04}_{-0.06}$ & $ 0.90^{+0.09}_{-0.15} $ & ---                      & $0.19^{+0.02}_{-0.02}$   \\
            &                        &                          &                          &                          \\
$| \chi_{1,2} | \leq 0.3 $     & $9.64^{+0.09}_{-0.16}$ & $0.84 ^{+0.14}_{-0.26} $ & $ 0.14^{+0.07}_{-0.08} $ & $ 0.10^{+0.07}_{-0.09} $ \\
            &                        &                          &                          &                          \\
$| \chi_{1,2} | \leq 0.7 $     & $9.70^{+0.08}_{-0.12}$ & $0.68 ^{+0.29}_{-0.37} $ & $ 0.20^{+0.15}_{-0.09} $ & $ 0.06^{+0.09}_{-0.06} $ \\
 &                        &                          &                          &                          \\
$| \chi_{1,2} | \leq 0.99 $     & $9.71^{+0.08}_{-0.12}$ & $0.64 ^{+0.33}_{-0.34} $ & $ 0.21^{+0.6}_{-0.09} $ & $ 0.06^{+0.09}_{-0.06} $ \\
            &                        &                          &                          &                          \\
GWTC-2:       & $9.69^{+0.08}_{-0.06}$ & $ 0.56^{+0.38}_{-0.23} $ & $ 0.18^{+0.20}_{-0.12} $ & ---                      \\
            &                        &                          &                          &                          \\
            & \multicolumn{4}{c}{GW170608}                                                                            \\ \cline{3-4}
            &                        &                          &                          &                          \\
$| \chi_{1,2} | = 0. $ & $8.44^{+0.02}_{-0.03}$ & $ 0.91^{+0.08}_{-0.15} $ & ---                      & $0.08^{+0.08}_{-0.07}$   \\
            &                        &                          &                          &                          \\
$| \chi_{1,2} | \leq 0.3 $    & $8.50^{+0.05}_{-0.06}$ & $0.83 ^{+0.15}_{-0.26} $ & $ 0.05^{+0.06}_{-0.04} $ & $ 0.05^{+0.08}_{-0.04} $ \\
            &                        &                          &                          &                          \\
$| \chi_{1,2} | \leq 0.7 $     & $8.50^{+0.05}_{-0.5}$ & $0.79 ^{+0.19}_{-0.34} $ & $ 0.06^{+0.11}_{-0.05} $ & $ 0.04^{+0.08}_{-0.04} $ \\
            &                        &                          &                          &                          \\
$| \chi_{1,2} | \leq 0.99 $     & $8.51^{+0.05}_{-0.6}$ & $0.79 ^{+0.19}_{-0.36} $ & $ 0.06^{+0.12}_{-0.05} $ & $ 0.04^{+0.08}_{-0.04} $ \\
            &                        &                          &    \\
GWTC-2:       & $8.49^{+0.05}_{-0.05}$ & $ 0.69^{+0.28}_{-0.36} $ & $ 0.03^{+0.18}_{-0.06} $ & ---           
\end{tabular}
\end{ruledtabular}
\end{table}

\section{Discussion} 

Lets consider a single case from our simulated signal recovery, with $M_{\mathrm{total}}^{\mathrm{signal}} = 20 \SolarMass{}$ and $q^{\mathrm{signal}} = 0.33$. This corresponds to $m_1^{\mathrm{signal}} \approx 15 \SolarMass{}  $ and $m_2^{\mathrm{signal}} \approx 5 \SolarMass{} $, but the recovered source masses would be $m_1 \approx m_2 \approx 9.2 \SolarMass{}$. Correct inference of the source masses is important for our understanding of the population of such objects, and in extreme cases such biases could be enough to incorrectly identify certain objects to inside or outside of the NS-BH mass gap of $\approx 2- 5 \SolarMass{}$ or the supernova pair-instability mass gap of $ \gtrsim 50 \SolarMass{}$. Thus, determining whether a signal is eccentric is important for the determination of the rest of its source parameters. As more detectors join the array of ground based detectors, the SNR of such eccentric signals will increase and systematic biases will become more pronounced when compared to the estimated error from the parameter estimation.

In section~\ref{sec:gw_events} and ~\ref{sec:sim_events} we could see the
$\mathcal{M}_c-e_{10}$ correlation manifest in the 2-D marginalized posterior
for the two gravitational wave events GW151226 and GW170608. Although
performing inference on these events with a non-spinning, eccentric prior
corroborated this correlation, neglecting to include spin in the analysis
leads to over-confident and biased estimates of the mass parameters.
Clearly a more systematic investigation of the relation effects
of eccentricity on the spin measurement and vice-versa are needed, which
we defer to a future study.

In future we will want to accurately measure the eccentricities of newly
detected signals, as well as those already present in GWTC2. It is clear
that we will not only need accurate and fast-to-evaluate waveform models
that can incorporate eccentricity, but also spins, and most likely precession.
By using, \TEOBResumE{} we have constrained the eccentricities of GW170608
and GW151226 to be $<0.12$ and $0.15$ 
at 90\% confidence
respectively. Our constraints are slightly tighter than the values of $<0.166$
and $<0.181$ found by~\cite{Wu_2020} which used a non-spinning, inspiral-only model. However, we find a much looser constraint than those set
by Ref.~\cite{Romero_Shaw_2019}, which found $e_{10} < 0.04$ for each of these events by re-weighting samples from a non-eccentric parameter estimation
run using the model \texttt{SEOBNRE}. From Fig~\ref{fig:circ},
we found that our method could only constrain a zero eccentricity to
about $e_{10} \approx 0.06$. In a similar light Ref~\cite{Lenon_2020}
also found that their measurement
of the eccentricity of GW190425 is a much looser constraint than that
of Ref.~\cite{Romero_Shaw_2019}, and explains that this could be due to the use
of a log-uniform prior on $e_{10}$, but also that the application of 
posterior re-weighting scheme might have failed to capture the
correlations \textit{between the chirp mass and eccentricity},
leading to overconfident
measurements as compared to their full MCMC calculation (although
one does find a hint of the correlation between eccentricity in the
supplementary figures~\cite{Romero_Shaw_2019_github_mchirp_gw151226,
Romero_Shaw_2019_github_mchirp_gw170608} of Ref.~\cite{Romero_Shaw_2019},
making the difference in priors a more likely explanation).
One way to compare posteriors which use different priors is to use rejection sampling to reweight the posterior from one prior to another. However, we found that the reweighting efficiency from a uniform to log-uniform prior is too low and results in only $\approx 40$ samples from the initial $\approx 20000$ samples. 
To fully investigate the choice of prior thus requires performing the parameter estimation with a log-uniform prior, which we defer to a future study. 

In this paper we chose to focus on the two lower mass black hole events from
the GWTC1 data. It would be interesting to continue this analysis by applying
\TEOBResumE{} to the parameter estimation of the rest of GW transient catalog, in particular
to see whether the eccentricity of events such as GW190521 can be established
or further constrained, and also whether the presence of eccentricity has
biased the existing LVC estimates of their source parameters. 

Finally, we list some of the limitations and caveats for the results
presented in this paper. {\it First}: our investigation of intrinsic 
parameter
space degeneracies focuses on non-spinning black hole binaries. This
was chosen in order to reduce computational time and simplify the
parameter space. We plan to extend this to include black hole spins
in the future. It is important to note though that in dynamical 
environments where eccentric binaries are expected to form, there is
not much reason for component BH spins to align with the orbit, and
therefore a thorough investigation including spins would need some
consideration of orbital precession as well, and no waveform model
exists yet that can model precession as well as eccentricity.
{\it Second}: our investigations do not include sub-dominant
waveform modes in templates. We also plan to extend this study to
include them, once suitable waveform models become available.
{\it Third}: the waveform model we use to study GW events, i.e. 
\TEOBResumE{}, does not allow for variation of mean anomaly and we
therefore kept it fixed in our parameter estimation analyses. This
can lead to small biases~\cite{Clarke:2022fma} that are not expected
to qualitatively change this paper's results. However, we also plan
to address this limitation in future work.

\section{Summary}

In section~\ref{sec:injection} we showed that there is correlation between
the measurement of the mass parameters of a binary compact object system
and its eccentricity. If one uses quasi-circular templates in their parameter
estimation, this correlation manifests as a bias in the measured chirp mass
and mass ratio. In the moderately eccentric case of $e=0.3$, a bias in the
chirp mass of up to $\approx 4 \%$ can occur, but the bias in the mass ratio
can be much more, since it tends to make the binary system appear more
symmetric.

In section~\ref{sec:gw_events} we sought to further establish this
correlation on real gravitational wave data. However, we found that
in order to correctly estimate the masses along with the eccentricity 
of such systems, we must employ a waveform model that can account for
eccentricity and spins simultaneously. 
Using the eccentric, aligned spin model \TEOBResumE{}, we measure
the eccentricity of GW170608 and GW151226 to be $<0.12$ and
$0.15$  at $90\%$ confidence respectively.

\begin{acknowledgments}
	We thank Alessandro Nagar for instructions on using the implementation
	of the \TEOBResumE{} model~\cite{EOBEBitbucket}, and Antoni Ramos Buades for helpful comments. The authors gratefully
	acknowledge the NSF Grants PHY-1912081 and OAC-193128 and a grant from
	the Sherman Fairchild Foundation at Cornell. P.K.’s research was also supported
	by the Department of Atomic Energy, Government of India, and by the
	Ashok and Gita Vaish Early Career Faculty Fellowship at the
	International Centre for Theoretical Sciences. The authors are grateful
	for computational resources provided by the	LIGO Laboratory and
	supported by National Science Foundation Grants PHY-0757058 and
	PHY-0823459.
\end{acknowledgments}

\appendix 
\section{ENIGMA Merger-ringdown}
\label{appendix:enigma}

\begin{figure} [b]
    \includegraphics[width=8.6cm]{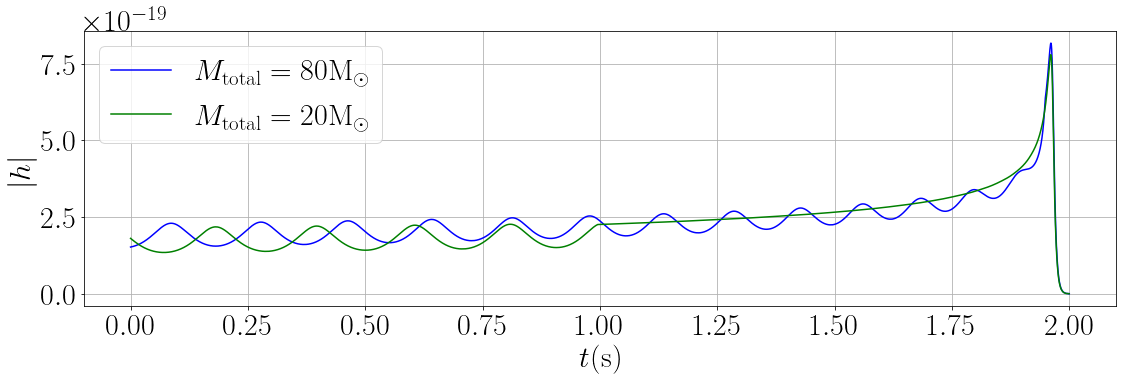}
    \caption{\textit{Green: }Amplitude of \ENIGMA{} generated waveform with $m_1=14 \SolarMass{}$, $m_2=6 \SolarMass{}$, $e_{10} = 0.3$. \textit{Blue:} The \ENIGMA{} generated waveform with the masses scaled by a factor of 4. The amplitude and samples times of the waveform are scaled by the same factor of 4 so they can be compared. }
    \label{fig:enigma_wrong}
\end{figure} 

The merger-ringdown portion of the \ENIGMA{} waveform is constructed
under the assumption that a moderately eccentric compact binary has
circularized by the time it has reached merger, which has been shown
to occur in numerical relativity simulations of eccentric
binaries~\cite{Hinder_2008}. Under this assumption, the parameter
space for the merger-ringdown portion of the waveform becomes
one-dimensional, depending only on the mass ratio $q$. A
one-dimensional Gaussian-process-regression surrogate model is constructed
to interpolate between a training set of quasi-circular numerical
relativity waveforms to any mass ratio up to $q=10$.

The final step in generating an \ENIGMA{} waveform is finding
the optimal attachment time between the PN inspiral waveform
and the surrogate merger-ringdown waveform. The optimal attachment
frequency is determined to be that which optimizes the overlap
between a circular \ENIGMA{} waveform and the corresponding
\texttt{SEOBNRv4} waveform. This overlap is calculated as in
equation \ref{eq:inner_product} with $f_{min} = 15$ Hz and
$f_{max} = 4096$ Hz. Due to the instantaneous monochromaticity
of these circular waveforms, the attachment frequency can be
converted to a corresponding attachment time.

In our studies, we found that \ENIGMA{} sometimes furnishes
qualitatively incorrect waveforms at smaller values of total mass,
and mass ratios $< 0.5$. This is because the optimal matching time is
found by optimizing the overlap of two waveforms, but with respect
to the LIGO PSD. At lower masses, which correspond to higher
gravitational wave frequencies, the LIGO PSD pushes the optimum
attachment frequency lower than it should. The attachment time is
then many cycles before the merger, and the assumption that the
eccentricity has radiated before attaching the merger-ringdown
waveform is no longer valid. This manifests as a cusp in the
amplitude of the waveform. Figure~\ref{fig:enigma_wrong}
illustrates the difference between \ENIGMA{}-generated waveforms
with $M_{\mathrm{total}} = 20 \SolarMass{}$ and
$M_{\mathrm{total}} = 80 \SolarMass{}$. The rescaled amplitudes
should be coincident, but in the low mass case the eccentricity
"switches off" in a non-smooth manner many cycles before the merger.
We find this same behavior regardless of the eccentricity value,
but have chosen $e_{10} = 0.3$ to accentuate the modulation of the
amplitude and make the difference more visible.

\bibliography{bibliography}

\begin{thebibliography}{110}%
\makeatletter
\providecommand \@ifxundefined [1]{%
 \@ifx{#1\undefined}
}%
\providecommand \@ifnum [1]{%
 \ifnum #1\expandafter \@firstoftwo
 \else \expandafter \@secondoftwo
 \fi
}%
\providecommand \@ifx [1]{%
 \ifx #1\expandafter \@firstoftwo
 \else \expandafter \@secondoftwo
 \fi
}%
\providecommand \natexlab [1]{#1}%
\providecommand \enquote  [1]{``#1''}%
\providecommand \bibnamefont  [1]{#1}%
\providecommand \bibfnamefont [1]{#1}%
\providecommand \citenamefont [1]{#1}%
\providecommand \href@noop [0]{\@secondoftwo}%
\providecommand \href [0]{\begingroup \@sanitize@url \@href}%
\providecommand \@href[1]{\@@startlink{#1}\@@href}%
\providecommand \@@href[1]{\endgroup#1\@@endlink}%
\providecommand \@sanitize@url [0]{\catcode `\\12\catcode `\$12\catcode
  `\&12\catcode `\#12\catcode `\^12\catcode `\_12\catcode `\%12\relax}%
\providecommand \@@startlink[1]{}%
\providecommand \@@endlink[0]{}%
\providecommand \url  [0]{\begingroup\@sanitize@url \@url }%
\providecommand \@url [1]{\endgroup\@href {#1}{\urlprefix }}%
\providecommand \urlprefix  [0]{URL }%
\providecommand \Eprint [0]{\href }%
\providecommand \doibase [0]{http://dx.doi.org/}%
\providecommand \selectlanguage [0]{\@gobble}%
\providecommand \bibinfo  [0]{\@secondoftwo}%
\providecommand \bibfield  [0]{\@secondoftwo}%
\providecommand \translation [1]{[#1]}%
\providecommand \BibitemOpen [0]{}%
\providecommand \bibitemStop [0]{}%
\providecommand \bibitemNoStop [0]{.\EOS\space}%
\providecommand \EOS [0]{\spacefactor3000\relax}%
\providecommand \BibitemShut  [1]{\csname bibitem#1\endcsname}%
\let\auto@bib@innerbib\@empty
\bibitem [{\citenamefont {Abbott}\ \emph
  {et~al.}(2019{\natexlab{a}})\citenamefont {Abbott}, \citenamefont {Abbott},
  \citenamefont {Abbott}, \citenamefont {Abraham}, \citenamefont {Acernese},
  \citenamefont {Ackley}, \citenamefont {Adams}, \citenamefont {Adhikari},
  \citenamefont {Adya}, \citenamefont {Affeldt},\ and\ \citenamefont
  {et~al.}}]{Abbott_2019}%
  \BibitemOpen
  \bibfield  {author} {\bibinfo {author} {\bibfnamefont {B.}~\bibnamefont
  {Abbott}}, \bibinfo {author} {\bibfnamefont {R.}~\bibnamefont {Abbott}},
  \bibinfo {author} {\bibfnamefont {T.}~\bibnamefont {Abbott}}, \bibinfo
  {author} {\bibfnamefont {S.}~\bibnamefont {Abraham}}, \bibinfo {author}
  {\bibfnamefont {F.}~\bibnamefont {Acernese}}, \bibinfo {author}
  {\bibfnamefont {K.}~\bibnamefont {Ackley}}, \bibinfo {author} {\bibfnamefont
  {C.}~\bibnamefont {Adams}}, \bibinfo {author} {\bibfnamefont
  {R.}~\bibnamefont {Adhikari}}, \bibinfo {author} {\bibfnamefont
  {V.}~\bibnamefont {Adya}}, \bibinfo {author} {\bibfnamefont {C.}~\bibnamefont
  {Affeldt}}, \ and\ \bibinfo {author} {\bibnamefont {et~al.}},\ }\href
  {\doibase 10.1103/physrevx.9.031040} {\bibfield  {journal} {\bibinfo
  {journal} {Phys. Rev. X}\ }\textbf {\bibinfo {volume} {9}} (\bibinfo {year}
  {2019}{\natexlab{a}}),\ 10.1103/physrevx.9.031040}\BibitemShut {NoStop}%
\bibitem [{\citenamefont {et. al}(2020)}]{abbott2020gwtc2}%
  \BibitemOpen
  \bibfield  {author} {\bibinfo {author} {\bibfnamefont {R.~A.}\ \bibnamefont
  {et. al}},\ }\href@noop {} {\enquote {\bibinfo {title} {Gwtc-2: Compact
  binary coalescences observed by ligo and virgo during the first half of the
  third observing run},}\ } (\bibinfo {year} {2020}),\ \Eprint
  {http://arxiv.org/abs/2010.14527} {arXiv:2010.14527 [gr-qc]} \BibitemShut
  {NoStop}%
\bibitem [{\citenamefont {Abbott}\ \emph {et~al.}(2021)\citenamefont {Abbott}
  \emph {et~al.}}]{LIGOScientific:2021djp}%
  \BibitemOpen
  \bibfield  {author} {\bibinfo {author} {\bibfnamefont {R.}~\bibnamefont
  {Abbott}} \emph {et~al.} (\bibinfo {collaboration} {LIGO Scientific, VIRGO,
  KAGRA}),\ }\href@noop {} {\  (\bibinfo {year} {2021})},\ \Eprint
  {http://arxiv.org/abs/2111.03606} {arXiv:2111.03606 [gr-qc]} \BibitemShut
  {NoStop}%
\bibitem [{kag(2019)}]{kagra}%
  \BibitemOpen
  \href {\doibase 10.1038/s41550-018-0658-y} {\bibfield  {journal} {\bibinfo
  {journal} {Nature Astronomy}\ }\textbf {\bibinfo {volume} {3}},\ \bibinfo
  {pages} {35–40} (\bibinfo {year} {2019})}\BibitemShut {NoStop}%
\bibitem [{\citenamefont {Unnikrishnan}(2013)}]{ligoIndia}%
  \BibitemOpen
  \bibfield  {author} {\bibinfo {author} {\bibfnamefont {C.~S.}\ \bibnamefont
  {Unnikrishnan}},\ }\href {\doibase 10.1142/s0218271813410101} {\bibfield
  {journal} {\bibinfo  {journal} {International Journal of Modern Physics D}\
  }\textbf {\bibinfo {volume} {22}},\ \bibinfo {pages} {1341010} (\bibinfo
  {year} {2013})}\BibitemShut {NoStop}%
\bibitem [{\citenamefont {Reitze}\ \emph {et~al.}(2019)\citenamefont {Reitze},
  \citenamefont {Adhikari}, \citenamefont {Ballmer}, \citenamefont {Barish},
  \citenamefont {Barsotti}, \citenamefont {Billingsley}, \citenamefont {Brown},
  \citenamefont {Chen}, \citenamefont {Coyne}, \citenamefont {Eisenstein},
  \citenamefont {Evans}, \citenamefont {Fritschel}, \citenamefont {Hall},
  \citenamefont {Lazzarini}, \citenamefont {Lovelace}, \citenamefont {Read},
  \citenamefont {Sathyaprakash}, \citenamefont {Shoemaker}, \citenamefont
  {Smith}, \citenamefont {Torrie}, \citenamefont {Vitale}, \citenamefont
  {Weiss}, \citenamefont {Wipf},\ and\ \citenamefont
  {Zucker}}]{cosmicExplorer}%
  \BibitemOpen
  \bibfield  {author} {\bibinfo {author} {\bibfnamefont {D.}~\bibnamefont
  {Reitze}}, \bibinfo {author} {\bibfnamefont {R.~X.}\ \bibnamefont
  {Adhikari}}, \bibinfo {author} {\bibfnamefont {S.}~\bibnamefont {Ballmer}},
  \bibinfo {author} {\bibfnamefont {B.}~\bibnamefont {Barish}}, \bibinfo
  {author} {\bibfnamefont {L.}~\bibnamefont {Barsotti}}, \bibinfo {author}
  {\bibfnamefont {G.}~\bibnamefont {Billingsley}}, \bibinfo {author}
  {\bibfnamefont {D.~A.}\ \bibnamefont {Brown}}, \bibinfo {author}
  {\bibfnamefont {Y.}~\bibnamefont {Chen}}, \bibinfo {author} {\bibfnamefont
  {D.}~\bibnamefont {Coyne}}, \bibinfo {author} {\bibfnamefont
  {R.}~\bibnamefont {Eisenstein}}, \bibinfo {author} {\bibfnamefont
  {M.}~\bibnamefont {Evans}}, \bibinfo {author} {\bibfnamefont
  {P.}~\bibnamefont {Fritschel}}, \bibinfo {author} {\bibfnamefont {E.~D.}\
  \bibnamefont {Hall}}, \bibinfo {author} {\bibfnamefont {A.}~\bibnamefont
  {Lazzarini}}, \bibinfo {author} {\bibfnamefont {G.}~\bibnamefont {Lovelace}},
  \bibinfo {author} {\bibfnamefont {J.}~\bibnamefont {Read}}, \bibinfo {author}
  {\bibfnamefont {B.~S.}\ \bibnamefont {Sathyaprakash}}, \bibinfo {author}
  {\bibfnamefont {D.}~\bibnamefont {Shoemaker}}, \bibinfo {author}
  {\bibfnamefont {J.}~\bibnamefont {Smith}}, \bibinfo {author} {\bibfnamefont
  {C.}~\bibnamefont {Torrie}}, \bibinfo {author} {\bibfnamefont
  {S.}~\bibnamefont {Vitale}}, \bibinfo {author} {\bibfnamefont
  {R.}~\bibnamefont {Weiss}}, \bibinfo {author} {\bibfnamefont
  {C.}~\bibnamefont {Wipf}}, \ and\ \bibinfo {author} {\bibfnamefont
  {M.}~\bibnamefont {Zucker}},\ }\href@noop {} {\enquote {\bibinfo {title}
  {Cosmic explorer: The u.s. contribution to gravitational-wave astronomy
  beyond ligo},}\ } (\bibinfo {year} {2019}),\ \Eprint
  {http://arxiv.org/abs/1907.04833} {arXiv:1907.04833 [astro-ph.IM]}
  \BibitemShut {NoStop}%
\bibitem [{\citenamefont {Maggiore}\ \emph {et~al.}(2020)\citenamefont
  {Maggiore}, \citenamefont {Broeck}, \citenamefont {Bartolo}, \citenamefont
  {Belgacem}, \citenamefont {Bertacca}, \citenamefont {Bizouard}, \citenamefont
  {Branchesi}, \citenamefont {Clesse}, \citenamefont {Foffa}, \citenamefont
  {García-Bellido},\ and\ \citenamefont {et~al.}}]{Maggiore_2020}%
  \BibitemOpen
  \bibfield  {author} {\bibinfo {author} {\bibfnamefont {M.}~\bibnamefont
  {Maggiore}}, \bibinfo {author} {\bibfnamefont {C.~V.~D.}\ \bibnamefont
  {Broeck}}, \bibinfo {author} {\bibfnamefont {N.}~\bibnamefont {Bartolo}},
  \bibinfo {author} {\bibfnamefont {E.}~\bibnamefont {Belgacem}}, \bibinfo
  {author} {\bibfnamefont {D.}~\bibnamefont {Bertacca}}, \bibinfo {author}
  {\bibfnamefont {M.~A.}\ \bibnamefont {Bizouard}}, \bibinfo {author}
  {\bibfnamefont {M.}~\bibnamefont {Branchesi}}, \bibinfo {author}
  {\bibfnamefont {S.}~\bibnamefont {Clesse}}, \bibinfo {author} {\bibfnamefont
  {S.}~\bibnamefont {Foffa}}, \bibinfo {author} {\bibfnamefont
  {J.}~\bibnamefont {García-Bellido}}, \ and\ \bibinfo {author} {\bibnamefont
  {et~al.}},\ }\href {\doibase 10.1088/1475-7516/2020/03/050} {\bibfield
  {journal} {\bibinfo  {journal} {Journal of Cosmology and Astroparticle
  Physics}\ }\textbf {\bibinfo {volume} {2020}},\ \bibinfo {pages} {050–050}
  (\bibinfo {year} {2020})}\BibitemShut {NoStop}%
\bibitem [{\citenamefont {Abbott}\ \emph
  {et~al.}(2016{\natexlab{a}})\citenamefont {Abbott} \emph
  {et~al.}}]{LIGOScientific:2016vpg}%
  \BibitemOpen
  \bibfield  {author} {\bibinfo {author} {\bibfnamefont {B.~P.}\ \bibnamefont
  {Abbott}} \emph {et~al.} (\bibinfo {collaboration} {LIGO Scientific,
  Virgo}),\ }\href {\doibase 10.3847/2041-8205/818/2/L22} {\bibfield  {journal}
  {\bibinfo  {journal} {Astrophys. J. Lett.}\ }\textbf {\bibinfo {volume}
  {818}},\ \bibinfo {pages} {L22} (\bibinfo {year} {2016}{\natexlab{a}})},\
  \Eprint {http://arxiv.org/abs/1602.03846} {arXiv:1602.03846 [astro-ph.HE]}
  \BibitemShut {NoStop}%
\bibitem [{\citenamefont {Marchant}\ \emph {et~al.}(2021)\citenamefont
  {Marchant}, \citenamefont {Pappas}, \citenamefont {Gallegos-Garcia},
  \citenamefont {Berry}, \citenamefont {Taam}, \citenamefont {Kalogera},\ and\
  \citenamefont {Podsiadlowski}}]{marchant2021role}%
  \BibitemOpen
  \bibfield  {author} {\bibinfo {author} {\bibfnamefont {P.}~\bibnamefont
  {Marchant}}, \bibinfo {author} {\bibfnamefont {K.~M.~W.}\ \bibnamefont
  {Pappas}}, \bibinfo {author} {\bibfnamefont {M.}~\bibnamefont
  {Gallegos-Garcia}}, \bibinfo {author} {\bibfnamefont {C.~P.~L.}\ \bibnamefont
  {Berry}}, \bibinfo {author} {\bibfnamefont {R.~E.}\ \bibnamefont {Taam}},
  \bibinfo {author} {\bibfnamefont {V.}~\bibnamefont {Kalogera}}, \ and\
  \bibinfo {author} {\bibfnamefont {P.}~\bibnamefont {Podsiadlowski}},\
  }\href@noop {} {\enquote {\bibinfo {title} {The role of mass transfer and
  common envelope evolution in the formation of merging binary black holes},}\
  } (\bibinfo {year} {2021}),\ \Eprint {http://arxiv.org/abs/2103.09243}
  {arXiv:2103.09243 [astro-ph.SR]} \BibitemShut {NoStop}%
\bibitem [{\citenamefont {Andrews}\ \emph {et~al.}(2020)\citenamefont
  {Andrews}, \citenamefont {Cronin}, \citenamefont {Kalogera}, \citenamefont
  {Berry},\ and\ \citenamefont {Zezas}}]{andrews2020targeted}%
  \BibitemOpen
  \bibfield  {author} {\bibinfo {author} {\bibfnamefont {J.~J.}\ \bibnamefont
  {Andrews}}, \bibinfo {author} {\bibfnamefont {J.}~\bibnamefont {Cronin}},
  \bibinfo {author} {\bibfnamefont {V.}~\bibnamefont {Kalogera}}, \bibinfo
  {author} {\bibfnamefont {C.}~\bibnamefont {Berry}}, \ and\ \bibinfo {author}
  {\bibfnamefont {A.}~\bibnamefont {Zezas}},\ }\href@noop {} {\enquote
  {\bibinfo {title} {Targeted modeling of gw150914's binary black hole source
  with dartboard},}\ } (\bibinfo {year} {2020}),\ \Eprint
  {http://arxiv.org/abs/2011.13918} {arXiv:2011.13918 [astro-ph.HE]}
  \BibitemShut {NoStop}%
\bibitem [{\citenamefont {Zevin}\ \emph
  {et~al.}(2021{\natexlab{a}})\citenamefont {Zevin}, \citenamefont {Bavera},
  \citenamefont {Berry}, \citenamefont {Kalogera}, \citenamefont {Fragos},
  \citenamefont {Marchant}, \citenamefont {Rodriguez}, \citenamefont
  {Antonini}, \citenamefont {Holz},\ and\ \citenamefont {Pankow}}]{Zevin_2021}%
  \BibitemOpen
  \bibfield  {author} {\bibinfo {author} {\bibfnamefont {M.}~\bibnamefont
  {Zevin}}, \bibinfo {author} {\bibfnamefont {S.~S.}\ \bibnamefont {Bavera}},
  \bibinfo {author} {\bibfnamefont {C.~P.~L.}\ \bibnamefont {Berry}}, \bibinfo
  {author} {\bibfnamefont {V.}~\bibnamefont {Kalogera}}, \bibinfo {author}
  {\bibfnamefont {T.}~\bibnamefont {Fragos}}, \bibinfo {author} {\bibfnamefont
  {P.}~\bibnamefont {Marchant}}, \bibinfo {author} {\bibfnamefont {C.~L.}\
  \bibnamefont {Rodriguez}}, \bibinfo {author} {\bibfnamefont {F.}~\bibnamefont
  {Antonini}}, \bibinfo {author} {\bibfnamefont {D.~E.}\ \bibnamefont {Holz}},
  \ and\ \bibinfo {author} {\bibfnamefont {C.}~\bibnamefont {Pankow}},\ }\href
  {\doibase 10.3847/1538-4357/abe40e} {\bibfield  {journal} {\bibinfo
  {journal} {The Astrophysical Journal}\ }\textbf {\bibinfo {volume} {910}},\
  \bibinfo {pages} {152} (\bibinfo {year} {2021}{\natexlab{a}})}\BibitemShut
  {NoStop}%
\bibitem [{\citenamefont {Ivanova}\ \emph {et~al.}(2013)\citenamefont
  {Ivanova}, \citenamefont {Justham}, \citenamefont {Chen}, \citenamefont
  {De~Marco}, \citenamefont {Fryer}, \citenamefont {Gaburov}, \citenamefont
  {Ge}, \citenamefont {Glebbeek}, \citenamefont {Han}, \citenamefont {Li},\
  and\ \citenamefont {et~al.}}]{Ivanova_2013}%
  \BibitemOpen
  \bibfield  {author} {\bibinfo {author} {\bibfnamefont {N.}~\bibnamefont
  {Ivanova}}, \bibinfo {author} {\bibfnamefont {S.}~\bibnamefont {Justham}},
  \bibinfo {author} {\bibfnamefont {X.}~\bibnamefont {Chen}}, \bibinfo {author}
  {\bibfnamefont {O.}~\bibnamefont {De~Marco}}, \bibinfo {author}
  {\bibfnamefont {C.~L.}\ \bibnamefont {Fryer}}, \bibinfo {author}
  {\bibfnamefont {E.}~\bibnamefont {Gaburov}}, \bibinfo {author} {\bibfnamefont
  {H.}~\bibnamefont {Ge}}, \bibinfo {author} {\bibfnamefont {E.}~\bibnamefont
  {Glebbeek}}, \bibinfo {author} {\bibfnamefont {Z.}~\bibnamefont {Han}},
  \bibinfo {author} {\bibfnamefont {X.-D.}\ \bibnamefont {Li}}, \ and\ \bibinfo
  {author} {\bibnamefont {et~al.}},\ }\href {\doibase
  10.1007/s00159-013-0059-2} {\bibfield  {journal} {\bibinfo  {journal} {The
  Astronomy and Astrophysics Review}\ }\textbf {\bibinfo {volume} {21}}
  (\bibinfo {year} {2013}),\ 10.1007/s00159-013-0059-2}\BibitemShut {NoStop}%
\bibitem [{\citenamefont {{Livio}}\ and\ \citenamefont
  {{Soker}}(1988)}]{livio}%
  \BibitemOpen
  \bibfield  {author} {\bibinfo {author} {\bibfnamefont {M.}~\bibnamefont
  {{Livio}}}\ and\ \bibinfo {author} {\bibfnamefont {N.}~\bibnamefont
  {{Soker}}},\ }\href {\doibase 10.1086/166419} {\bibfield  {journal} {\bibinfo
   {journal} {\apj}\ }\textbf {\bibinfo {volume} {329}},\ \bibinfo {pages}
  {764} (\bibinfo {year} {1988})}\BibitemShut {NoStop}%
\bibitem [{\citenamefont {{Kruckow, M. U.}}\ \emph {et~al.}(2016)\citenamefont
  {{Kruckow, M. U.}}, \citenamefont {{Tauris, T. M.}}, \citenamefont {{Langer,
  N.}}, \citenamefont {{Sz\'ecsi, D.}}, \citenamefont {{Marchant, P.}},\ and\
  \citenamefont {{Podsiadlowski, Ph.}}}]{kruckow}%
  \BibitemOpen
  \bibfield  {author} {\bibinfo {author} {\bibnamefont {{Kruckow, M. U.}}},
  \bibinfo {author} {\bibnamefont {{Tauris, T. M.}}}, \bibinfo {author}
  {\bibnamefont {{Langer, N.}}}, \bibinfo {author} {\bibnamefont {{Sz\'ecsi,
  D.}}}, \bibinfo {author} {\bibnamefont {{Marchant, P.}}}, \ and\ \bibinfo
  {author} {\bibnamefont {{Podsiadlowski, Ph.}}},\ }\href {\doibase
  10.1051/0004-6361/201629420} {\bibfield  {journal} {\bibinfo  {journal}
  {A\&A}\ }\textbf {\bibinfo {volume} {596}},\ \bibinfo {pages} {A58} (\bibinfo
  {year} {2016})}\BibitemShut {NoStop}%
\bibitem [{\citenamefont {Dominik}\ \emph {et~al.}(2012)\citenamefont
  {Dominik}, \citenamefont {Belczynski}, \citenamefont {Fryer}, \citenamefont
  {Holz}, \citenamefont {Berti}, \citenamefont {Bulik}, \citenamefont
  {Mandel},\ and\ \citenamefont {O'Shaughnessy}}]{Dominik:2012kk}%
  \BibitemOpen
  \bibfield  {author} {\bibinfo {author} {\bibfnamefont {M.}~\bibnamefont
  {Dominik}}, \bibinfo {author} {\bibfnamefont {K.}~\bibnamefont {Belczynski}},
  \bibinfo {author} {\bibfnamefont {C.}~\bibnamefont {Fryer}}, \bibinfo
  {author} {\bibfnamefont {D.}~\bibnamefont {Holz}}, \bibinfo {author}
  {\bibfnamefont {E.}~\bibnamefont {Berti}}, \bibinfo {author} {\bibfnamefont
  {T.}~\bibnamefont {Bulik}}, \bibinfo {author} {\bibfnamefont
  {I.}~\bibnamefont {Mandel}}, \ and\ \bibinfo {author} {\bibfnamefont
  {R.}~\bibnamefont {O'Shaughnessy}},\ }\href {\doibase
  10.1088/0004-637X/759/1/52} {\bibfield  {journal} {\bibinfo  {journal}
  {Astrophys. J.}\ }\textbf {\bibinfo {volume} {759}},\ \bibinfo {pages} {52}
  (\bibinfo {year} {2012})},\ \Eprint {http://arxiv.org/abs/1202.4901}
  {arXiv:1202.4901 [astro-ph.HE]} \BibitemShut {NoStop}%
\bibitem [{\citenamefont {de~Mink}\ \emph {et~al.}(2010)\citenamefont
  {de~Mink}, \citenamefont {Cantiello}, \citenamefont {Langer}, \citenamefont
  {Pols}, \citenamefont {Kologera},\ and\ \citenamefont {van~der
  Sluys}}]{de_Mink_2010}%
  \BibitemOpen
  \bibfield  {author} {\bibinfo {author} {\bibfnamefont {S.~E.}\ \bibnamefont
  {de~Mink}}, \bibinfo {author} {\bibfnamefont {M.}~\bibnamefont {Cantiello}},
  \bibinfo {author} {\bibfnamefont {N.}~\bibnamefont {Langer}}, \bibinfo
  {author} {\bibfnamefont {O.~R.}\ \bibnamefont {Pols}}, \bibinfo {author}
  {\bibfnamefont {V.}~\bibnamefont {Kologera}}, \ and\ \bibinfo {author}
  {\bibfnamefont {M.}~\bibnamefont {van~der Sluys}},\ }\href {\doibase
  10.1063/1.3536387} {\  (\bibinfo {year} {2010}),\
  10.1063/1.3536387}\BibitemShut {NoStop}%
\bibitem [{\citenamefont {de~Mink}\ and\ \citenamefont
  {Mandel}(2016)}]{de_Mink_2016}%
  \BibitemOpen
  \bibfield  {author} {\bibinfo {author} {\bibfnamefont {S.~E.}\ \bibnamefont
  {de~Mink}}\ and\ \bibinfo {author} {\bibfnamefont {I.}~\bibnamefont
  {Mandel}},\ }\href {\doibase 10.1093/mnras/stw1219} {\bibfield  {journal}
  {\bibinfo  {journal} {Monthly Notices of the Royal Astronomical Society}\
  }\textbf {\bibinfo {volume} {460}},\ \bibinfo {pages} {3545–3553} (\bibinfo
  {year} {2016})}\BibitemShut {NoStop}%
\bibitem [{\citenamefont {Zevin}\ \emph {et~al.}(2017)\citenamefont {Zevin},
  \citenamefont {Pankow}, \citenamefont {Rodriguez}, \citenamefont {Sampson},
  \citenamefont {Chase}, \citenamefont {Kalogera},\ and\ \citenamefont
  {Rasio}}]{Zevin_2017}%
  \BibitemOpen
  \bibfield  {author} {\bibinfo {author} {\bibfnamefont {M.}~\bibnamefont
  {Zevin}}, \bibinfo {author} {\bibfnamefont {C.}~\bibnamefont {Pankow}},
  \bibinfo {author} {\bibfnamefont {C.~L.}\ \bibnamefont {Rodriguez}}, \bibinfo
  {author} {\bibfnamefont {L.}~\bibnamefont {Sampson}}, \bibinfo {author}
  {\bibfnamefont {E.}~\bibnamefont {Chase}}, \bibinfo {author} {\bibfnamefont
  {V.}~\bibnamefont {Kalogera}}, \ and\ \bibinfo {author} {\bibfnamefont
  {F.~A.}\ \bibnamefont {Rasio}},\ }\href {\doibase 10.3847/1538-4357/aa8408}
  {\bibfield  {journal} {\bibinfo  {journal} {The Astrophysical Journal}\
  }\textbf {\bibinfo {volume} {846}},\ \bibinfo {pages} {82} (\bibinfo {year}
  {2017})}\BibitemShut {NoStop}%
\bibitem [{\citenamefont {Samsing}\ and\ \citenamefont
  {Hotokezaka}(2021)}]{Samsing:2020qqd}%
  \BibitemOpen
  \bibfield  {author} {\bibinfo {author} {\bibfnamefont {J.}~\bibnamefont
  {Samsing}}\ and\ \bibinfo {author} {\bibfnamefont {K.}~\bibnamefont
  {Hotokezaka}},\ }\href {\doibase 10.3847/1538-4357/ac2b27} {\bibfield
  {journal} {\bibinfo  {journal} {Astrophys. J.}\ }\textbf {\bibinfo {volume}
  {923}},\ \bibinfo {pages} {126} (\bibinfo {year} {2021})},\ \Eprint
  {http://arxiv.org/abs/2006.09744} {arXiv:2006.09744 [astro-ph.HE]}
  \BibitemShut {NoStop}%
\bibitem [{\citenamefont {Zwart}\ and\ \citenamefont
  {McMillan}(2000)}]{Portegies_Zwart_2000}%
  \BibitemOpen
  \bibfield  {author} {\bibinfo {author} {\bibfnamefont {S.~F.~P.}\
  \bibnamefont {Zwart}}\ and\ \bibinfo {author} {\bibfnamefont {S.~L.~W.}\
  \bibnamefont {McMillan}},\ }\href {\doibase 10.1086/312422} {\bibfield
  {journal} {\bibinfo  {journal} {The Astrophysical Journal}\ }\textbf
  {\bibinfo {volume} {528}},\ \bibinfo {pages} {L17} (\bibinfo {year}
  {2000})}\BibitemShut {NoStop}%
\bibitem [{\citenamefont {Rodriguez}\ \emph {et~al.}(2015)\citenamefont
  {Rodriguez}, \citenamefont {Morscher}, \citenamefont {Pattabiraman},
  \citenamefont {Chatterjee}, \citenamefont {Haster},\ and\ \citenamefont
  {Rasio}}]{Rodriguez_2015}%
  \BibitemOpen
  \bibfield  {author} {\bibinfo {author} {\bibfnamefont {C.~L.}\ \bibnamefont
  {Rodriguez}}, \bibinfo {author} {\bibfnamefont {M.}~\bibnamefont {Morscher}},
  \bibinfo {author} {\bibfnamefont {B.}~\bibnamefont {Pattabiraman}}, \bibinfo
  {author} {\bibfnamefont {S.}~\bibnamefont {Chatterjee}}, \bibinfo {author}
  {\bibfnamefont {C.-J.}\ \bibnamefont {Haster}}, \ and\ \bibinfo {author}
  {\bibfnamefont {F.~A.}\ \bibnamefont {Rasio}},\ }\href {\doibase
  10.1103/physrevlett.115.051101} {\bibfield  {journal} {\bibinfo  {journal}
  {Physical Review Letters}\ }\textbf {\bibinfo {volume} {115}} (\bibinfo
  {year} {2015}),\ 10.1103/physrevlett.115.051101}\BibitemShut {NoStop}%
\bibitem [{\citenamefont {Rodriguez}\ \emph
  {et~al.}(2016{\natexlab{a}})\citenamefont {Rodriguez}, \citenamefont {Zevin},
  \citenamefont {Pankow}, \citenamefont {Kalogera},\ and\ \citenamefont
  {Rasio}}]{Rodriguez_2016}%
  \BibitemOpen
  \bibfield  {author} {\bibinfo {author} {\bibfnamefont {C.~L.}\ \bibnamefont
  {Rodriguez}}, \bibinfo {author} {\bibfnamefont {M.}~\bibnamefont {Zevin}},
  \bibinfo {author} {\bibfnamefont {C.}~\bibnamefont {Pankow}}, \bibinfo
  {author} {\bibfnamefont {V.}~\bibnamefont {Kalogera}}, \ and\ \bibinfo
  {author} {\bibfnamefont {F.~A.}\ \bibnamefont {Rasio}},\ }\href {\doibase
  10.3847/2041-8205/832/1/l2} {\bibfield  {journal} {\bibinfo  {journal} {The
  Astrophysical Journal}\ }\textbf {\bibinfo {volume} {832}},\ \bibinfo {pages}
  {L2} (\bibinfo {year} {2016}{\natexlab{a}})}\BibitemShut {NoStop}%
\bibitem [{\citenamefont {Rodriguez}\ \emph
  {et~al.}(2016{\natexlab{b}})\citenamefont {Rodriguez}, \citenamefont
  {Haster}, \citenamefont {Chatterjee}, \citenamefont {Kalogera},\ and\
  \citenamefont {Rasio}}]{Rodriguez_2016_2}%
  \BibitemOpen
  \bibfield  {author} {\bibinfo {author} {\bibfnamefont {C.~L.}\ \bibnamefont
  {Rodriguez}}, \bibinfo {author} {\bibfnamefont {C.-J.}\ \bibnamefont
  {Haster}}, \bibinfo {author} {\bibfnamefont {S.}~\bibnamefont {Chatterjee}},
  \bibinfo {author} {\bibfnamefont {V.}~\bibnamefont {Kalogera}}, \ and\
  \bibinfo {author} {\bibfnamefont {F.~A.}\ \bibnamefont {Rasio}},\ }\href
  {\doibase 10.3847/2041-8205/824/1/l8} {\bibfield  {journal} {\bibinfo
  {journal} {The Astrophysical Journal}\ }\textbf {\bibinfo {volume} {824}},\
  \bibinfo {pages} {L8} (\bibinfo {year} {2016}{\natexlab{b}})}\BibitemShut
  {NoStop}%
\bibitem [{\citenamefont {O’Leary}\ \emph {et~al.}(2016)\citenamefont
  {O’Leary}, \citenamefont {Meiron},\ and\ \citenamefont
  {Kocsis}}]{O_Leary_2016}%
  \BibitemOpen
  \bibfield  {author} {\bibinfo {author} {\bibfnamefont {R.~M.}\ \bibnamefont
  {O’Leary}}, \bibinfo {author} {\bibfnamefont {Y.}~\bibnamefont {Meiron}}, \
  and\ \bibinfo {author} {\bibfnamefont {B.}~\bibnamefont {Kocsis}},\ }\href
  {\doibase 10.3847/2041-8205/824/1/l12} {\bibfield  {journal} {\bibinfo
  {journal} {The Astrophysical Journal}\ }\textbf {\bibinfo {volume} {824}},\
  \bibinfo {pages} {L12} (\bibinfo {year} {2016})}\BibitemShut {NoStop}%
\bibitem [{\citenamefont {Samsing}\ and\ \citenamefont
  {Ramirez-Ruiz}(2017)}]{Samsing2017}%
  \BibitemOpen
  \bibfield  {author} {\bibinfo {author} {\bibfnamefont {J.}~\bibnamefont
  {Samsing}}\ and\ \bibinfo {author} {\bibfnamefont {E.}~\bibnamefont
  {Ramirez-Ruiz}},\ }\href {\doibase 10.3847/2041-8213/aa6f0b} {\bibfield
  {journal} {\bibinfo  {journal} {The Astrophysical Journal}\ }\textbf
  {\bibinfo {volume} {840}},\ \bibinfo {pages} {L14} (\bibinfo {year}
  {2017})},\ \Eprint {http://arxiv.org/abs/1703.09703} {arXiv:1703.09703}
  \BibitemShut {NoStop}%
\bibitem [{\citenamefont {Yang}\ \emph {et~al.}(2019)\citenamefont {Yang},
  \citenamefont {Bartos}, \citenamefont {Haiman}, \citenamefont {Kocsis},
  \citenamefont {Márka}, \citenamefont {Stone},\ and\ \citenamefont
  {Márka}}]{Yang_2019}%
  \BibitemOpen
  \bibfield  {author} {\bibinfo {author} {\bibfnamefont {Y.}~\bibnamefont
  {Yang}}, \bibinfo {author} {\bibfnamefont {I.}~\bibnamefont {Bartos}},
  \bibinfo {author} {\bibfnamefont {Z.}~\bibnamefont {Haiman}}, \bibinfo
  {author} {\bibfnamefont {B.}~\bibnamefont {Kocsis}}, \bibinfo {author}
  {\bibfnamefont {Z.}~\bibnamefont {Márka}}, \bibinfo {author} {\bibfnamefont
  {N.~C.}\ \bibnamefont {Stone}}, \ and\ \bibinfo {author} {\bibfnamefont
  {S.}~\bibnamefont {Márka}},\ }\href {\doibase 10.3847/1538-4357/ab16e3}
  {\bibfield  {journal} {\bibinfo  {journal} {The Astrophysical Journal}\
  }\textbf {\bibinfo {volume} {876}},\ \bibinfo {pages} {122} (\bibinfo {year}
  {2019})}\BibitemShut {NoStop}%
\bibitem [{\citenamefont {McKernan}\ \emph {et~al.}(2020)\citenamefont
  {McKernan}, \citenamefont {Ford},\ and\ \citenamefont
  {O’Shaughnessy}}]{McKernan_2020}%
  \BibitemOpen
  \bibfield  {author} {\bibinfo {author} {\bibfnamefont {B.}~\bibnamefont
  {McKernan}}, \bibinfo {author} {\bibfnamefont {K.~E.~S.}\ \bibnamefont
  {Ford}}, \ and\ \bibinfo {author} {\bibfnamefont {R.}~\bibnamefont
  {O’Shaughnessy}},\ }\href {\doibase 10.1093/mnras/staa2681} {\bibfield
  {journal} {\bibinfo  {journal} {Monthly Notices of the Royal Astronomical
  Society}\ }\textbf {\bibinfo {volume} {498}},\ \bibinfo {pages} {4088–4094}
  (\bibinfo {year} {2020})}\BibitemShut {NoStop}%
\bibitem [{\citenamefont {Tagawa}\ \emph
  {et~al.}(2021{\natexlab{a}})\citenamefont {Tagawa}, \citenamefont {Kocsis},
  \citenamefont {Haiman}, \citenamefont {Bartos}, \citenamefont {Omukai},\ and\
  \citenamefont {Samsing}}]{Tagawa:2020jnc}%
  \BibitemOpen
  \bibfield  {author} {\bibinfo {author} {\bibfnamefont {H.}~\bibnamefont
  {Tagawa}}, \bibinfo {author} {\bibfnamefont {B.}~\bibnamefont {Kocsis}},
  \bibinfo {author} {\bibfnamefont {Z.}~\bibnamefont {Haiman}}, \bibinfo
  {author} {\bibfnamefont {I.}~\bibnamefont {Bartos}}, \bibinfo {author}
  {\bibfnamefont {K.}~\bibnamefont {Omukai}}, \ and\ \bibinfo {author}
  {\bibfnamefont {J.}~\bibnamefont {Samsing}},\ }\href {\doibase
  10.3847/2041-8213/abd4d3} {\bibfield  {journal} {\bibinfo  {journal}
  {Astrophys. J. Lett.}\ }\textbf {\bibinfo {volume} {907}},\ \bibinfo {pages}
  {L20} (\bibinfo {year} {2021}{\natexlab{a}})},\ \Eprint
  {http://arxiv.org/abs/2010.10526} {arXiv:2010.10526 [astro-ph.HE]}
  \BibitemShut {NoStop}%
\bibitem [{\citenamefont {Gond\'an}\ and\ \citenamefont
  {Kocsis}(2021)}]{Gondan:2020svr}%
  \BibitemOpen
  \bibfield  {author} {\bibinfo {author} {\bibfnamefont {L.}~\bibnamefont
  {Gond\'an}}\ and\ \bibinfo {author} {\bibfnamefont {B.}~\bibnamefont
  {Kocsis}},\ }\href {\doibase 10.1093/mnras/stab1722} {\bibfield  {journal}
  {\bibinfo  {journal} {Mon. Not. Roy. Astron. Soc.}\ }\textbf {\bibinfo
  {volume} {506}},\ \bibinfo {pages} {1665} (\bibinfo {year} {2021})},\ \Eprint
  {http://arxiv.org/abs/2011.02507} {arXiv:2011.02507 [astro-ph.HE]}
  \BibitemShut {NoStop}%
\bibitem [{\citenamefont {Vajpeyi}\ \emph {et~al.}(2022)\citenamefont
  {Vajpeyi}, \citenamefont {Thrane}, \citenamefont {Smith}, \citenamefont
  {McKernan},\ and\ \citenamefont {Ford}}]{Vajpeyi:2021qsw}%
  \BibitemOpen
  \bibfield  {author} {\bibinfo {author} {\bibfnamefont {A.}~\bibnamefont
  {Vajpeyi}}, \bibinfo {author} {\bibfnamefont {E.}~\bibnamefont {Thrane}},
  \bibinfo {author} {\bibfnamefont {R.}~\bibnamefont {Smith}}, \bibinfo
  {author} {\bibfnamefont {B.}~\bibnamefont {McKernan}}, \ and\ \bibinfo
  {author} {\bibfnamefont {K.~E.~S.}\ \bibnamefont {Ford}},\ }\href {\doibase
  10.3847/1538-4357/ac6180} {\bibfield  {journal} {\bibinfo  {journal}
  {Astrophys. J.}\ }\textbf {\bibinfo {volume} {931}},\ \bibinfo {pages} {82}
  (\bibinfo {year} {2022})},\ \Eprint {http://arxiv.org/abs/2111.03992}
  {arXiv:2111.03992 [gr-qc]} \BibitemShut {NoStop}%
\bibitem [{\citenamefont {Gröbner}\ \emph {et~al.}(2020)\citenamefont
  {Gröbner}, \citenamefont {Ishibashi}, \citenamefont {Tiwari}, \citenamefont
  {Haney},\ and\ \citenamefont {Jetzer}}]{Grobner_2020}%
  \BibitemOpen
  \bibfield  {author} {\bibinfo {author} {\bibfnamefont {M.}~\bibnamefont
  {Gröbner}}, \bibinfo {author} {\bibfnamefont {W.}~\bibnamefont {Ishibashi}},
  \bibinfo {author} {\bibfnamefont {S.}~\bibnamefont {Tiwari}}, \bibinfo
  {author} {\bibfnamefont {M.}~\bibnamefont {Haney}}, \ and\ \bibinfo {author}
  {\bibfnamefont {P.}~\bibnamefont {Jetzer}},\ }\href {\doibase
  10.1051/0004-6361/202037681} {\bibfield  {journal} {\bibinfo  {journal}
  {Astronomy \& Astrophysics}\ }\textbf {\bibinfo {volume} {638}},\ \bibinfo
  {pages} {A119} (\bibinfo {year} {2020})}\BibitemShut {NoStop}%
\bibitem [{\citenamefont {Gerosa}\ and\ \citenamefont
  {Berti}(2017)}]{Gerosa_2017}%
  \BibitemOpen
  \bibfield  {author} {\bibinfo {author} {\bibfnamefont {D.}~\bibnamefont
  {Gerosa}}\ and\ \bibinfo {author} {\bibfnamefont {E.}~\bibnamefont {Berti}},\
  }\href {\doibase 10.1103/PhysRevD.95.124046} {\bibfield  {journal} {\bibinfo
  {journal} {Phys. Rev. D}\ }\textbf {\bibinfo {volume} {95}},\ \bibinfo
  {pages} {124046} (\bibinfo {year} {2017})}\BibitemShut {NoStop}%
\bibitem [{\citenamefont {Rodriguez}\ \emph {et~al.}(2019)\citenamefont
  {Rodriguez}, \citenamefont {Zevin}, \citenamefont {Amaro-Seoane},
  \citenamefont {Chatterjee}, \citenamefont {Kremer}, \citenamefont {Rasio},\
  and\ \citenamefont {Ye}}]{Rodriguez_2019}%
  \BibitemOpen
  \bibfield  {author} {\bibinfo {author} {\bibfnamefont {C.~L.}\ \bibnamefont
  {Rodriguez}}, \bibinfo {author} {\bibfnamefont {M.}~\bibnamefont {Zevin}},
  \bibinfo {author} {\bibfnamefont {P.}~\bibnamefont {Amaro-Seoane}}, \bibinfo
  {author} {\bibfnamefont {S.}~\bibnamefont {Chatterjee}}, \bibinfo {author}
  {\bibfnamefont {K.}~\bibnamefont {Kremer}}, \bibinfo {author} {\bibfnamefont
  {F.~A.}\ \bibnamefont {Rasio}}, \ and\ \bibinfo {author} {\bibfnamefont
  {C.~S.}\ \bibnamefont {Ye}},\ }\href {\doibase 10.1103/physrevd.100.043027}
  {\bibfield  {journal} {\bibinfo  {journal} {Phys. Rev. D}\ }\textbf {\bibinfo
  {volume} {100}} (\bibinfo {year} {2019}),\
  10.1103/physrevd.100.043027}\BibitemShut {NoStop}%
\bibitem [{\citenamefont {Woosley}(2019)}]{Woosley_2019}%
  \BibitemOpen
  \bibfield  {author} {\bibinfo {author} {\bibfnamefont {S.~E.}\ \bibnamefont
  {Woosley}},\ }\href {\doibase 10.3847/1538-4357/ab1b41} {\bibfield  {journal}
  {\bibinfo  {journal} {The Astrophysical Journal}\ }\textbf {\bibinfo {volume}
  {878}},\ \bibinfo {pages} {49} (\bibinfo {year} {2019})}\BibitemShut
  {NoStop}%
\bibitem [{\citenamefont {Woosley}(2017)}]{Woosley_2017}%
  \BibitemOpen
  \bibfield  {author} {\bibinfo {author} {\bibfnamefont {S.~E.}\ \bibnamefont
  {Woosley}},\ }\href {\doibase 10.3847/1538-4357/836/2/244} {\bibfield
  {journal} {\bibinfo  {journal} {The Astrophysical Journal}\ }\textbf
  {\bibinfo {volume} {836}},\ \bibinfo {pages} {244} (\bibinfo {year}
  {2017})}\BibitemShut {NoStop}%
\bibitem [{\citenamefont {Talbot}\ and\ \citenamefont
  {Thrane}(2017)}]{talbot_thrane_2017}%
  \BibitemOpen
  \bibfield  {author} {\bibinfo {author} {\bibfnamefont {C.}~\bibnamefont
  {Talbot}}\ and\ \bibinfo {author} {\bibfnamefont {E.}~\bibnamefont
  {Thrane}},\ }\href {\doibase 10.1103/PhysRevD.96.023012} {\bibfield
  {journal} {\bibinfo  {journal} {Phys. Rev. D}\ }\textbf {\bibinfo {volume}
  {96}},\ \bibinfo {pages} {023012} (\bibinfo {year} {2017})}\BibitemShut
  {NoStop}%
\bibitem [{\citenamefont {Rodriguez}\ \emph
  {et~al.}(2018{\natexlab{a}})\citenamefont {Rodriguez}, \citenamefont
  {Amaro-Seoane}, \citenamefont {Chatterjee}, \citenamefont {Kremer},
  \citenamefont {Rasio}, \citenamefont {Samsing}, \citenamefont {Ye},\ and\
  \citenamefont {Zevin}}]{Rodriguez_2018}%
  \BibitemOpen
  \bibfield  {author} {\bibinfo {author} {\bibfnamefont {C.~L.}\ \bibnamefont
  {Rodriguez}}, \bibinfo {author} {\bibfnamefont {P.}~\bibnamefont
  {Amaro-Seoane}}, \bibinfo {author} {\bibfnamefont {S.}~\bibnamefont
  {Chatterjee}}, \bibinfo {author} {\bibfnamefont {K.}~\bibnamefont {Kremer}},
  \bibinfo {author} {\bibfnamefont {F.~A.}\ \bibnamefont {Rasio}}, \bibinfo
  {author} {\bibfnamefont {J.}~\bibnamefont {Samsing}}, \bibinfo {author}
  {\bibfnamefont {C.~S.}\ \bibnamefont {Ye}}, \ and\ \bibinfo {author}
  {\bibfnamefont {M.}~\bibnamefont {Zevin}},\ }\href {\doibase
  10.1103/PhysRevD.98.123005} {\bibfield  {journal} {\bibinfo  {journal} {Phys.
  Rev. D}\ }\textbf {\bibinfo {volume} {98}},\ \bibinfo {pages} {123005}
  (\bibinfo {year} {2018}{\natexlab{a}})}\BibitemShut {NoStop}%
\bibitem [{\citenamefont {Samsing}\ \emph {et~al.}(2018)\citenamefont
  {Samsing}, \citenamefont {D'Orazio}, \citenamefont {Askar},\ and\
  \citenamefont {Giersz}}]{samsing2018black}%
  \BibitemOpen
  \bibfield  {author} {\bibinfo {author} {\bibfnamefont {J.}~\bibnamefont
  {Samsing}}, \bibinfo {author} {\bibfnamefont {D.~J.}\ \bibnamefont
  {D'Orazio}}, \bibinfo {author} {\bibfnamefont {A.}~\bibnamefont {Askar}}, \
  and\ \bibinfo {author} {\bibfnamefont {M.}~\bibnamefont {Giersz}},\
  }\href@noop {} {\enquote {\bibinfo {title} {Black hole mergers from globular
  clusters observable by lisa and ligo: Results from post-newtonian
  binary-single scatterings},}\ } (\bibinfo {year} {2018}),\ \Eprint
  {http://arxiv.org/abs/1802.08654} {arXiv:1802.08654 [astro-ph.HE]}
  \BibitemShut {NoStop}%
\bibitem [{\citenamefont {Zevin}\ \emph {et~al.}(2019)\citenamefont {Zevin},
  \citenamefont {Samsing}, \citenamefont {Rodriguez}, \citenamefont {Haster},\
  and\ \citenamefont {Ramirez-Ruiz}}]{Zevin_2019}%
  \BibitemOpen
  \bibfield  {author} {\bibinfo {author} {\bibfnamefont {M.}~\bibnamefont
  {Zevin}}, \bibinfo {author} {\bibfnamefont {J.}~\bibnamefont {Samsing}},
  \bibinfo {author} {\bibfnamefont {C.}~\bibnamefont {Rodriguez}}, \bibinfo
  {author} {\bibfnamefont {C.-J.}\ \bibnamefont {Haster}}, \ and\ \bibinfo
  {author} {\bibfnamefont {E.}~\bibnamefont {Ramirez-Ruiz}},\ }\href {\doibase
  10.3847/1538-4357/aaf6ec} {\bibfield  {journal} {\bibinfo  {journal} {The
  Astrophysical Journal}\ }\textbf {\bibinfo {volume} {871}},\ \bibinfo {pages}
  {91} (\bibinfo {year} {2019})}\BibitemShut {NoStop}%
\bibitem [{\citenamefont {{Lidov}}(1962)}]{lidov}%
  \BibitemOpen
  \bibfield  {author} {\bibinfo {author} {\bibfnamefont {M.~L.}\ \bibnamefont
  {{Lidov}}},\ }\href {\doibase 10.1016/0032-0633(62)90129-0} {\bibfield
  {journal} {\bibinfo  {journal} {Planetary and Space Science}\ }\textbf
  {\bibinfo {volume} {9}},\ \bibinfo {pages} {719} (\bibinfo {year}
  {1962})}\BibitemShut {NoStop}%
\bibitem [{\citenamefont {{Kozai}}(1962)}]{kozai}%
  \BibitemOpen
  \bibfield  {author} {\bibinfo {author} {\bibfnamefont {Y.}~\bibnamefont
  {{Kozai}}},\ }\href {\doibase 10.1086/108790} {\bibfield  {journal} {\bibinfo
   {journal} {The Astronomical Journal}\ }\textbf {\bibinfo {volume} {67}},\
  \bibinfo {pages} {591} (\bibinfo {year} {1962})}\BibitemShut {NoStop}%
\bibitem [{\citenamefont {Samsing}(2018)}]{PhysRevD.97.103014}%
  \BibitemOpen
  \bibfield  {author} {\bibinfo {author} {\bibfnamefont {J.}~\bibnamefont
  {Samsing}},\ }\href {\doibase 10.1103/PhysRevD.97.103014} {\bibfield
  {journal} {\bibinfo  {journal} {Phys. Rev. D}\ }\textbf {\bibinfo {volume}
  {97}},\ \bibinfo {pages} {103014} (\bibinfo {year} {2018})}\BibitemShut
  {NoStop}%
\bibitem [{\citenamefont {Rodriguez}\ \emph
  {et~al.}(2018{\natexlab{b}})\citenamefont {Rodriguez}, \citenamefont
  {Amaro-Seoane}, \citenamefont {Chatterjee},\ and\ \citenamefont
  {Rasio}}]{PhysRevLett.120.151101}%
  \BibitemOpen
  \bibfield  {author} {\bibinfo {author} {\bibfnamefont {C.~L.}\ \bibnamefont
  {Rodriguez}}, \bibinfo {author} {\bibfnamefont {P.}~\bibnamefont
  {Amaro-Seoane}}, \bibinfo {author} {\bibfnamefont {S.}~\bibnamefont
  {Chatterjee}}, \ and\ \bibinfo {author} {\bibfnamefont {F.~A.}\ \bibnamefont
  {Rasio}},\ }\href {\doibase 10.1103/PhysRevLett.120.151101} {\bibfield
  {journal} {\bibinfo  {journal} {Phys. Rev. Lett.}\ }\textbf {\bibinfo
  {volume} {120}},\ \bibinfo {pages} {151101} (\bibinfo {year}
  {2018}{\natexlab{b}})}\BibitemShut {NoStop}%
\bibitem [{\citenamefont {Tagawa}\ \emph
  {et~al.}(2021{\natexlab{b}})\citenamefont {Tagawa}, \citenamefont {Kocsis},
  \citenamefont {Haiman}, \citenamefont {Bartos}, \citenamefont {Omukai},\ and\
  \citenamefont {Samsing}}]{Tagawa_2021}%
  \BibitemOpen
  \bibfield  {author} {\bibinfo {author} {\bibfnamefont {H.}~\bibnamefont
  {Tagawa}}, \bibinfo {author} {\bibfnamefont {B.}~\bibnamefont {Kocsis}},
  \bibinfo {author} {\bibfnamefont {Z.}~\bibnamefont {Haiman}}, \bibinfo
  {author} {\bibfnamefont {I.}~\bibnamefont {Bartos}}, \bibinfo {author}
  {\bibfnamefont {K.}~\bibnamefont {Omukai}}, \ and\ \bibinfo {author}
  {\bibfnamefont {J.}~\bibnamefont {Samsing}},\ }\href {\doibase
  10.3847/2041-8213/abd4d3} {\bibfield  {journal} {\bibinfo  {journal} {The
  Astrophysical Journal}\ }\textbf {\bibinfo {volume} {907}},\ \bibinfo {pages}
  {L20} (\bibinfo {year} {2021}{\natexlab{b}})}\BibitemShut {NoStop}%
\bibitem [{\citenamefont {Samsing}\ \emph {et~al.}(2020)\citenamefont
  {Samsing}, \citenamefont {Bartos}, \citenamefont {D'Orazio}, \citenamefont
  {Haiman}, \citenamefont {Kocsis}, \citenamefont {Leigh}, \citenamefont {Liu},
  \citenamefont {Pessah},\ and\ \citenamefont {Tagawa}}]{samsing2020active}%
  \BibitemOpen
  \bibfield  {author} {\bibinfo {author} {\bibfnamefont {J.}~\bibnamefont
  {Samsing}}, \bibinfo {author} {\bibfnamefont {I.}~\bibnamefont {Bartos}},
  \bibinfo {author} {\bibfnamefont {D.~J.}\ \bibnamefont {D'Orazio}}, \bibinfo
  {author} {\bibfnamefont {Z.}~\bibnamefont {Haiman}}, \bibinfo {author}
  {\bibfnamefont {B.}~\bibnamefont {Kocsis}}, \bibinfo {author} {\bibfnamefont
  {N.~W.~C.}\ \bibnamefont {Leigh}}, \bibinfo {author} {\bibfnamefont
  {B.}~\bibnamefont {Liu}}, \bibinfo {author} {\bibfnamefont {M.~E.}\
  \bibnamefont {Pessah}}, \ and\ \bibinfo {author} {\bibfnamefont
  {H.}~\bibnamefont {Tagawa}},\ }\href@noop {} {\enquote {\bibinfo {title}
  {Active galactic nuclei as factories for eccentric black hole mergers},}\ }
  (\bibinfo {year} {2020}),\ \Eprint {http://arxiv.org/abs/2010.09765}
  {arXiv:2010.09765 [astro-ph.HE]} \BibitemShut {NoStop}%
\bibitem [{\citenamefont {Zevin}\ \emph
  {et~al.}(2021{\natexlab{b}})\citenamefont {Zevin}, \citenamefont
  {Romero-Shaw}, \citenamefont {Kremer}, \citenamefont {Thrane},\ and\
  \citenamefont {Lasky}}]{Zevin:2021rtf}%
  \BibitemOpen
  \bibfield  {author} {\bibinfo {author} {\bibfnamefont {M.}~\bibnamefont
  {Zevin}}, \bibinfo {author} {\bibfnamefont {I.~M.}\ \bibnamefont
  {Romero-Shaw}}, \bibinfo {author} {\bibfnamefont {K.}~\bibnamefont {Kremer}},
  \bibinfo {author} {\bibfnamefont {E.}~\bibnamefont {Thrane}}, \ and\ \bibinfo
  {author} {\bibfnamefont {P.~D.}\ \bibnamefont {Lasky}},\ }\href {\doibase
  10.3847/2041-8213/ac32dc} {\bibfield  {journal} {\bibinfo  {journal}
  {Astrophys. J. Lett.}\ }\textbf {\bibinfo {volume} {921}},\ \bibinfo {pages}
  {L43} (\bibinfo {year} {2021}{\natexlab{b}})},\ \Eprint
  {http://arxiv.org/abs/2106.09042} {arXiv:2106.09042 [astro-ph.HE]}
  \BibitemShut {NoStop}%
\bibitem [{\citenamefont {Tiwari}\ \emph
  {et~al.}(2016{\natexlab{a}})\citenamefont {Tiwari} \emph
  {et~al.}}]{Tiwari:2015gal}%
  \BibitemOpen
  \bibfield  {author} {\bibinfo {author} {\bibfnamefont {V.}~\bibnamefont
  {Tiwari}} \emph {et~al.},\ }\href {\doibase 10.1103/PhysRevD.93.043007}
  {\bibfield  {journal} {\bibinfo  {journal} {Phys. Rev. D}\ }\textbf {\bibinfo
  {volume} {93}},\ \bibinfo {pages} {043007} (\bibinfo {year}
  {2016}{\natexlab{a}})},\ \Eprint {http://arxiv.org/abs/1511.09240}
  {arXiv:1511.09240 [gr-qc]} \BibitemShut {NoStop}%
\bibitem [{\citenamefont {Abbott}\ \emph
  {et~al.}(2019{\natexlab{b}})\citenamefont {Abbott} \emph
  {et~al.}}]{LIGOScientific:2019dag}%
  \BibitemOpen
  \bibfield  {author} {\bibinfo {author} {\bibfnamefont {B.~P.}\ \bibnamefont
  {Abbott}} \emph {et~al.} (\bibinfo {collaboration} {LIGO Scientific,
  Virgo}),\ }\href {\doibase 10.3847/1538-4357/ab3c2d} {\bibfield  {journal}
  {\bibinfo  {journal} {Astrophys. J.}\ }\textbf {\bibinfo {volume} {883}},\
  \bibinfo {pages} {149} (\bibinfo {year} {2019}{\natexlab{b}})},\ \Eprint
  {http://arxiv.org/abs/1907.09384} {arXiv:1907.09384 [astro-ph.HE]}
  \BibitemShut {NoStop}%
\bibitem [{\citenamefont {Cheeseboro}\ and\ \citenamefont
  {Baker}(2021)}]{Cheeseboro:2021rey}%
  \BibitemOpen
  \bibfield  {author} {\bibinfo {author} {\bibfnamefont {B.~D.}\ \bibnamefont
  {Cheeseboro}}\ and\ \bibinfo {author} {\bibfnamefont {P.~T.}\ \bibnamefont
  {Baker}},\ }\href {\doibase 10.1103/PhysRevD.104.104016} {\bibfield
  {journal} {\bibinfo  {journal} {Phys. Rev. D}\ }\textbf {\bibinfo {volume}
  {104}},\ \bibinfo {pages} {104016} (\bibinfo {year} {2021})},\ \Eprint
  {http://arxiv.org/abs/2108.01050} {arXiv:2108.01050 [gr-qc]} \BibitemShut
  {NoStop}%
\bibitem [{\citenamefont {Wang}\ and\ \citenamefont
  {Nitz}(2021)}]{Wang:2021qsu}%
  \BibitemOpen
  \bibfield  {author} {\bibinfo {author} {\bibfnamefont {Y.-F.}\ \bibnamefont
  {Wang}}\ and\ \bibinfo {author} {\bibfnamefont {A.~H.}\ \bibnamefont
  {Nitz}},\ }\href {\doibase 10.3847/1538-4357/abe939} {\bibfield  {journal}
  {\bibinfo  {journal} {Astrophys. J.}\ }\textbf {\bibinfo {volume} {912}},\
  \bibinfo {pages} {53} (\bibinfo {year} {2021})},\ \Eprint
  {http://arxiv.org/abs/2101.12269} {arXiv:2101.12269 [astro-ph.HE]}
  \BibitemShut {NoStop}%
\bibitem [{\citenamefont {Ravichandran}\ \emph {et~al.}(2023)\citenamefont
  {Ravichandran}, \citenamefont {Vijaykumar}, \citenamefont {Kapadia},\ and\
  \citenamefont {Kumar}}]{Ravichandran:2023qma}%
  \BibitemOpen
  \bibfield  {author} {\bibinfo {author} {\bibfnamefont {A.}~\bibnamefont
  {Ravichandran}}, \bibinfo {author} {\bibfnamefont {A.}~\bibnamefont
  {Vijaykumar}}, \bibinfo {author} {\bibfnamefont {S.~J.}\ \bibnamefont
  {Kapadia}}, \ and\ \bibinfo {author} {\bibfnamefont {P.}~\bibnamefont
  {Kumar}},\ }\href@noop {} {\  (\bibinfo {year} {2023})},\ \Eprint
  {http://arxiv.org/abs/2302.00666} {arXiv:2302.00666 [gr-qc]} \BibitemShut
  {NoStop}%
\bibitem [{\citenamefont {Abbott}\ \emph {et~al.}(2020)\citenamefont {Abbott}
  \emph {et~al.}}]{LIGOScientific:2020ufj}%
  \BibitemOpen
  \bibfield  {author} {\bibinfo {author} {\bibfnamefont {R.}~\bibnamefont
  {Abbott}} \emph {et~al.} (\bibinfo {collaboration} {LIGO Scientific,
  Virgo}),\ }\href {\doibase 10.3847/2041-8213/aba493} {\bibfield  {journal}
  {\bibinfo  {journal} {Astrophys. J. Lett.}\ }\textbf {\bibinfo {volume}
  {900}},\ \bibinfo {pages} {L13} (\bibinfo {year} {2020})},\ \Eprint
  {http://arxiv.org/abs/2009.01190} {arXiv:2009.01190 [astro-ph.HE]}
  \BibitemShut {NoStop}%
\bibitem [{\citenamefont {Nitz}\ \emph {et~al.}(2020)\citenamefont {Nitz},
  \citenamefont {Lenon},\ and\ \citenamefont {Brown}}]{Nitz_2020}%
  \BibitemOpen
  \bibfield  {author} {\bibinfo {author} {\bibfnamefont {A.~H.}\ \bibnamefont
  {Nitz}}, \bibinfo {author} {\bibfnamefont {A.}~\bibnamefont {Lenon}}, \ and\
  \bibinfo {author} {\bibfnamefont {D.~A.}\ \bibnamefont {Brown}},\ }\href
  {\doibase 10.3847/1538-4357/ab6611} {\bibfield  {journal} {\bibinfo
  {journal} {The Astrophysical Journal}\ }\textbf {\bibinfo {volume} {890}},\
  \bibinfo {pages} {1} (\bibinfo {year} {2020})}\BibitemShut {NoStop}%
\bibitem [{\citenamefont {Abac}\ \emph {et~al.}(2023)\citenamefont {Abac} \emph
  {et~al.}}]{LIGOScientific:2023lpe}%
  \BibitemOpen
  \bibfield  {author} {\bibinfo {author} {\bibfnamefont {A.~G.}\ \bibnamefont
  {Abac}} \emph {et~al.} (\bibinfo {collaboration} {LIGO Scientific, VIRGO,
  KAGRA}),\ }\href@noop {} {\  (\bibinfo {year} {2023})},\ \Eprint
  {http://arxiv.org/abs/2308.03822} {arXiv:2308.03822 [astro-ph.HE]}
  \BibitemShut {NoStop}%
\bibitem [{\citenamefont {Romero-Shaw}\ \emph {et~al.}(2019)\citenamefont
  {Romero-Shaw}, \citenamefont {Lasky},\ and\ \citenamefont
  {Thrane}}]{Romero_Shaw_2019}%
  \BibitemOpen
  \bibfield  {author} {\bibinfo {author} {\bibfnamefont {I.~M.}\ \bibnamefont
  {Romero-Shaw}}, \bibinfo {author} {\bibfnamefont {P.~D.}\ \bibnamefont
  {Lasky}}, \ and\ \bibinfo {author} {\bibfnamefont {E.}~\bibnamefont
  {Thrane}},\ }\href {\doibase 10.1093/mnras/stz2996} {\bibfield  {journal}
  {\bibinfo  {journal} {Monthly Notices of the Royal Astronomical Society}\
  }\textbf {\bibinfo {volume} {490}},\ \bibinfo {pages} {5210–5216} (\bibinfo
  {year} {2019})}\BibitemShut {NoStop}%
\bibitem [{\citenamefont {Wu}\ \emph {et~al.}(2020{\natexlab{a}})\citenamefont
  {Wu}, \citenamefont {Cao},\ and\ \citenamefont {Zhu}}]{Wu_2020}%
  \BibitemOpen
  \bibfield  {author} {\bibinfo {author} {\bibfnamefont {S.}~\bibnamefont
  {Wu}}, \bibinfo {author} {\bibfnamefont {Z.}~\bibnamefont {Cao}}, \ and\
  \bibinfo {author} {\bibfnamefont {Z.-H.}\ \bibnamefont {Zhu}},\ }\href
  {\doibase 10.1093/mnras/staa1176} {\bibfield  {journal} {\bibinfo  {journal}
  {Monthly Notices of the Royal Astronomical Society}\ }\textbf {\bibinfo
  {volume} {495}},\ \bibinfo {pages} {466–478} (\bibinfo {year}
  {2020}{\natexlab{a}})}\BibitemShut {NoStop}%
\bibitem [{\citenamefont {Romero-Shaw}\ \emph {et~al.}(2020)\citenamefont
  {Romero-Shaw}, \citenamefont {Lasky}, \citenamefont {Thrane},\ and\
  \citenamefont {Bustillo}}]{Romero-Shaw2020}%
  \BibitemOpen
  \bibfield  {author} {\bibinfo {author} {\bibfnamefont {I.~M.}\ \bibnamefont
  {Romero-Shaw}}, \bibinfo {author} {\bibfnamefont {P.~D.}\ \bibnamefont
  {Lasky}}, \bibinfo {author} {\bibfnamefont {E.}~\bibnamefont {Thrane}}, \
  and\ \bibinfo {author} {\bibfnamefont {J.~C.}\ \bibnamefont {Bustillo}},\
  }\href {http://arxiv.org/abs/2009.04771} {\  (\bibinfo {year} {2020})},\
  \Eprint {http://arxiv.org/abs/2009.04771} {arXiv:2009.04771} \BibitemShut
  {NoStop}%
\bibitem [{\citenamefont {Romero-Shaw}\ \emph {et~al.}(2021)\citenamefont
  {Romero-Shaw}, \citenamefont {Lasky},\ and\ \citenamefont
  {Thrane}}]{Romero-Shaw:2021ual}%
  \BibitemOpen
  \bibfield  {author} {\bibinfo {author} {\bibfnamefont {I.~M.}\ \bibnamefont
  {Romero-Shaw}}, \bibinfo {author} {\bibfnamefont {P.~D.}\ \bibnamefont
  {Lasky}}, \ and\ \bibinfo {author} {\bibfnamefont {E.}~\bibnamefont
  {Thrane}},\ }\href {\doibase 10.3847/2041-8213/ac3138} {\bibfield  {journal}
  {\bibinfo  {journal} {Astrophys. J. Lett.}\ }\textbf {\bibinfo {volume}
  {921}},\ \bibinfo {pages} {L31} (\bibinfo {year} {2021})},\ \Eprint
  {http://arxiv.org/abs/2108.01284} {arXiv:2108.01284 [astro-ph.HE]}
  \BibitemShut {NoStop}%
\bibitem [{\citenamefont {Iglesias}\ \emph {et~al.}(2022)\citenamefont
  {Iglesias} \emph {et~al.}}]{Iglesias:2022xfc}%
  \BibitemOpen
  \bibfield  {author} {\bibinfo {author} {\bibfnamefont {H.~L.}\ \bibnamefont
  {Iglesias}} \emph {et~al.},\ }\href@noop {} {\  (\bibinfo {year} {2022})},\
  \Eprint {http://arxiv.org/abs/2208.01766} {arXiv:2208.01766 [gr-qc]}
  \BibitemShut {NoStop}%
\bibitem [{\citenamefont {Romero-Shaw}\ \emph {et~al.}(2022)\citenamefont
  {Romero-Shaw}, \citenamefont {Lasky},\ and\ \citenamefont
  {Thrane}}]{Romero-Shaw:2022xko}%
  \BibitemOpen
  \bibfield  {author} {\bibinfo {author} {\bibfnamefont {I.~M.}\ \bibnamefont
  {Romero-Shaw}}, \bibinfo {author} {\bibfnamefont {P.~D.}\ \bibnamefont
  {Lasky}}, \ and\ \bibinfo {author} {\bibfnamefont {E.}~\bibnamefont
  {Thrane}},\ }\href {\doibase 10.3847/1538-4357/ac9798} {\bibfield  {journal}
  {\bibinfo  {journal} {Astrophys. J.}\ }\textbf {\bibinfo {volume} {940}},\
  \bibinfo {pages} {171} (\bibinfo {year} {2022})},\ \Eprint
  {http://arxiv.org/abs/2206.14695} {arXiv:2206.14695 [astro-ph.HE]}
  \BibitemShut {NoStop}%
\bibitem [{\citenamefont {Lenon}\ \emph {et~al.}(2020)\citenamefont {Lenon},
  \citenamefont {Nitz},\ and\ \citenamefont {Brown}}]{Lenon_2020}%
  \BibitemOpen
  \bibfield  {author} {\bibinfo {author} {\bibfnamefont {A.~K.}\ \bibnamefont
  {Lenon}}, \bibinfo {author} {\bibfnamefont {A.~H.}\ \bibnamefont {Nitz}}, \
  and\ \bibinfo {author} {\bibfnamefont {D.~A.}\ \bibnamefont {Brown}},\ }\href
  {\doibase 10.1093/mnras/staa2120} {\bibfield  {journal} {\bibinfo  {journal}
  {Monthly Notices of the Royal Astronomical Society}\ }\textbf {\bibinfo
  {volume} {497}},\ \bibinfo {pages} {1966–1971} (\bibinfo {year}
  {2020})}\BibitemShut {NoStop}%
\bibitem [{\citenamefont {Kimball}\ \emph {et~al.}(2020)\citenamefont
  {Kimball}, \citenamefont {Talbot}, \citenamefont {Berry}, \citenamefont
  {Zevin}, \citenamefont {Thrane}, \citenamefont {Kalogera}, \citenamefont
  {Buscicchio}, \citenamefont {Carney}, \citenamefont {Dent}, \citenamefont
  {Middleton}, \citenamefont {Payne}, \citenamefont {Veitch},\ and\
  \citenamefont {Williams}}]{kimball2020evidence}%
  \BibitemOpen
  \bibfield  {author} {\bibinfo {author} {\bibfnamefont {C.}~\bibnamefont
  {Kimball}}, \bibinfo {author} {\bibfnamefont {C.}~\bibnamefont {Talbot}},
  \bibinfo {author} {\bibfnamefont {C.~P.~L.}\ \bibnamefont {Berry}}, \bibinfo
  {author} {\bibfnamefont {M.}~\bibnamefont {Zevin}}, \bibinfo {author}
  {\bibfnamefont {E.}~\bibnamefont {Thrane}}, \bibinfo {author} {\bibfnamefont
  {V.}~\bibnamefont {Kalogera}}, \bibinfo {author} {\bibfnamefont
  {R.}~\bibnamefont {Buscicchio}}, \bibinfo {author} {\bibfnamefont
  {M.}~\bibnamefont {Carney}}, \bibinfo {author} {\bibfnamefont
  {T.}~\bibnamefont {Dent}}, \bibinfo {author} {\bibfnamefont {H.}~\bibnamefont
  {Middleton}}, \bibinfo {author} {\bibfnamefont {E.}~\bibnamefont {Payne}},
  \bibinfo {author} {\bibfnamefont {J.}~\bibnamefont {Veitch}}, \ and\ \bibinfo
  {author} {\bibfnamefont {D.}~\bibnamefont {Williams}},\ }\href@noop {}
  {\enquote {\bibinfo {title} {Evidence for hierarchical black hole mergers in
  the second ligo--virgo gravitational-wave catalog},}\ } (\bibinfo {year}
  {2020}),\ \Eprint {http://arxiv.org/abs/2011.05332} {arXiv:2011.05332
  [astro-ph.HE]} \BibitemShut {NoStop}%
\bibitem [{\citenamefont {Gamba}\ \emph {et~al.}(2021)\citenamefont {Gamba},
  \citenamefont {Breschi}, \citenamefont {Carullo}, \citenamefont {Rettegno},
  \citenamefont {Albanesi}, \citenamefont {Bernuzzi},\ and\ \citenamefont
  {Nagar}}]{Gamba:2021gap}%
  \BibitemOpen
  \bibfield  {author} {\bibinfo {author} {\bibfnamefont {R.}~\bibnamefont
  {Gamba}}, \bibinfo {author} {\bibfnamefont {M.}~\bibnamefont {Breschi}},
  \bibinfo {author} {\bibfnamefont {G.}~\bibnamefont {Carullo}}, \bibinfo
  {author} {\bibfnamefont {P.}~\bibnamefont {Rettegno}}, \bibinfo {author}
  {\bibfnamefont {S.}~\bibnamefont {Albanesi}}, \bibinfo {author}
  {\bibfnamefont {S.}~\bibnamefont {Bernuzzi}}, \ and\ \bibinfo {author}
  {\bibfnamefont {A.}~\bibnamefont {Nagar}},\ }\href@noop {} {\  (\bibinfo
  {year} {2021})},\ \Eprint {http://arxiv.org/abs/2106.05575} {arXiv:2106.05575
  [gr-qc]} \BibitemShut {NoStop}%
\bibitem [{\citenamefont {Gayathri}\ \emph {et~al.}(2020)\citenamefont
  {Gayathri}, \citenamefont {Healy}, \citenamefont {Lange}, \citenamefont
  {O'Brien}, \citenamefont {Szczepanczyk}, \citenamefont {Bartos},
  \citenamefont {Campanelli}, \citenamefont {Klimenko}, \citenamefont
  {Lousto},\ and\ \citenamefont {O'Shaughnessy}}]{gayathri2020gw190521}%
  \BibitemOpen
  \bibfield  {author} {\bibinfo {author} {\bibfnamefont {V.}~\bibnamefont
  {Gayathri}}, \bibinfo {author} {\bibfnamefont {J.}~\bibnamefont {Healy}},
  \bibinfo {author} {\bibfnamefont {J.}~\bibnamefont {Lange}}, \bibinfo
  {author} {\bibfnamefont {B.}~\bibnamefont {O'Brien}}, \bibinfo {author}
  {\bibfnamefont {M.}~\bibnamefont {Szczepanczyk}}, \bibinfo {author}
  {\bibfnamefont {I.}~\bibnamefont {Bartos}}, \bibinfo {author} {\bibfnamefont
  {M.}~\bibnamefont {Campanelli}}, \bibinfo {author} {\bibfnamefont
  {S.}~\bibnamefont {Klimenko}}, \bibinfo {author} {\bibfnamefont
  {C.}~\bibnamefont {Lousto}}, \ and\ \bibinfo {author} {\bibfnamefont
  {R.}~\bibnamefont {O'Shaughnessy}},\ }\href@noop {} {\enquote {\bibinfo
  {title} {Gw190521 as a highly eccentric black hole merger},}\ } (\bibinfo
  {year} {2020}),\ \Eprint {http://arxiv.org/abs/2009.05461} {arXiv:2009.05461
  [astro-ph.HE]} \BibitemShut {NoStop}%
\bibitem [{\citenamefont {Sato}\ \emph {et~al.}(2017)\citenamefont {Sato},
  \citenamefont {Kawamura}, \citenamefont {Ando}, \citenamefont {Nakamura},
  \citenamefont {Tsubono}, \citenamefont {Araya}, \citenamefont {Funaki},
  \citenamefont {Ioka}, \citenamefont {Kanda}, \citenamefont {Moriwaki},
  \citenamefont {Musha}, \citenamefont {Nakazawa}, \citenamefont {Numata},
  \citenamefont {ichiro Sakai}, \citenamefont {Seto}, \citenamefont
  {Takashima}, \citenamefont {Tanaka}, \citenamefont {Agatsuma}, \citenamefont
  {suke Aoyanagi}, \citenamefont {Arai}, \citenamefont {Asada}, \citenamefont
  {Aso}, \citenamefont {Chiba}, \citenamefont {Ebisuzaki}, \citenamefont
  {Ejiri}, \citenamefont {Enoki}, \citenamefont {Eriguchi}, \citenamefont
  {Fujimoto}, \citenamefont {Fujita}, \citenamefont {Fukushima}, \citenamefont
  {Futamase}, \citenamefont {Ganzu}, \citenamefont {Harada}, \citenamefont
  {Hashimoto}, \citenamefont {Hayama}, \citenamefont {Hikida}, \citenamefont
  {Himemoto}, \citenamefont {Hirabayashi}, \citenamefont {Hiramatsu},
  \citenamefont {Hong}, \citenamefont {Horisawa}, \citenamefont {Hosokawa},
  \citenamefont {Ichiki}, \citenamefont {Ikegami}, \citenamefont {Inoue},
  \citenamefont {Ishidoshiro}, \citenamefont {Ishihara}, \citenamefont
  {Ishikawa}, \citenamefont {Ishizaki}, \citenamefont {Ito}, \citenamefont
  {Itoh}, \citenamefont {Kawashima}, \citenamefont {Kawazoe}, \citenamefont
  {Kishimoto}, \citenamefont {Kiuchi}, \citenamefont {Kobayashi}, \citenamefont
  {Kohri}, \citenamefont {Koizumi}, \citenamefont {Kojima}, \citenamefont
  {Kokeyama}, \citenamefont {Kokuyama}, \citenamefont {Kotake}, \citenamefont
  {Kozai}, \citenamefont {Kudoh}, \citenamefont {Kunimori}, \citenamefont
  {Kuninaka}, \citenamefont {Kuroda}, \citenamefont {ichi Maeda}, \citenamefont
  {Matsuhara}, \citenamefont {Mino}, \citenamefont {Miyakawa}, \citenamefont
  {Miyoki}, \citenamefont {Morimoto}, \citenamefont {Morioka}, \citenamefont
  {Morisawa}, \citenamefont {Mukohyama}, \citenamefont {Nagano}, \citenamefont
  {Naito}, \citenamefont {Nakamura}, \citenamefont {Nakano}, \citenamefont
  {Nakao}, \citenamefont {Nakasuka}, \citenamefont {Nakayama}, \citenamefont
  {Nishida}, \citenamefont {Nishiyama}, \citenamefont {Nishizawa},
  \citenamefont {Niwa}, \citenamefont {Noumi}, \citenamefont {Obuchi},
  \citenamefont {Ohashi}, \citenamefont {Ohishi}, \citenamefont {Ohkawa},
  \citenamefont {Okada}, \citenamefont {Onozato}, \citenamefont {Oohara},
  \citenamefont {Sago}, \citenamefont {Saijo}, \citenamefont {Sakagami},
  \citenamefont {Sakata}, \citenamefont {Sasaki}, \citenamefont {Sato},
  \citenamefont {Shibata}, \citenamefont {Shinkai}, \citenamefont {Somiya},
  \citenamefont {Sotani}, \citenamefont {Sugiyama}, \citenamefont {Suwa},
  \citenamefont {Suzuki}, \citenamefont {Tagoshi}, \citenamefont {Takahashi},
  \citenamefont {Takahashi}, \citenamefont {Takahashi}, \citenamefont
  {Takahashi}, \citenamefont {Takahashi}, \citenamefont {Takahashi},
  \citenamefont {Takahashi}, \citenamefont {Akiteru}, \citenamefont {Takano},
  \citenamefont {Taniguchi}, \citenamefont {Taruya}, \citenamefont {Tashiro},
  \citenamefont {Torii}, \citenamefont {Toyoshima}, \citenamefont {Tsujikawa},
  \citenamefont {Tsunesada}, \citenamefont {Ueda}, \citenamefont {ichi Ueda},
  \citenamefont {Utashima}, \citenamefont {Wakabayashi}, \citenamefont
  {Yamakawa}, \citenamefont {Yamamoto}, \citenamefont {Yamazaki}, \citenamefont
  {Yokoyama}, \citenamefont {Yoo}, \citenamefont {Yoshida},\ and\ \citenamefont
  {Yoshino}}]{Sato_2017}%
  \BibitemOpen
  \bibfield  {author} {\bibinfo {author} {\bibfnamefont {S.}~\bibnamefont
  {Sato}}, \bibinfo {author} {\bibfnamefont {S.}~\bibnamefont {Kawamura}},
  \bibinfo {author} {\bibfnamefont {M.}~\bibnamefont {Ando}}, \bibinfo {author}
  {\bibfnamefont {T.}~\bibnamefont {Nakamura}}, \bibinfo {author}
  {\bibfnamefont {K.}~\bibnamefont {Tsubono}}, \bibinfo {author} {\bibfnamefont
  {A.}~\bibnamefont {Araya}}, \bibinfo {author} {\bibfnamefont
  {I.}~\bibnamefont {Funaki}}, \bibinfo {author} {\bibfnamefont
  {K.}~\bibnamefont {Ioka}}, \bibinfo {author} {\bibfnamefont {N.}~\bibnamefont
  {Kanda}}, \bibinfo {author} {\bibfnamefont {S.}~\bibnamefont {Moriwaki}},
  \bibinfo {author} {\bibfnamefont {M.}~\bibnamefont {Musha}}, \bibinfo
  {author} {\bibfnamefont {K.}~\bibnamefont {Nakazawa}}, \bibinfo {author}
  {\bibfnamefont {K.}~\bibnamefont {Numata}}, \bibinfo {author} {\bibfnamefont
  {S.}~\bibnamefont {ichiro Sakai}}, \bibinfo {author} {\bibfnamefont
  {N.}~\bibnamefont {Seto}}, \bibinfo {author} {\bibfnamefont {T.}~\bibnamefont
  {Takashima}}, \bibinfo {author} {\bibfnamefont {T.}~\bibnamefont {Tanaka}},
  \bibinfo {author} {\bibfnamefont {K.}~\bibnamefont {Agatsuma}}, \bibinfo
  {author} {\bibfnamefont {K.}~\bibnamefont {suke Aoyanagi}}, \bibinfo {author}
  {\bibfnamefont {K.}~\bibnamefont {Arai}}, \bibinfo {author} {\bibfnamefont
  {H.}~\bibnamefont {Asada}}, \bibinfo {author} {\bibfnamefont
  {Y.}~\bibnamefont {Aso}}, \bibinfo {author} {\bibfnamefont {T.}~\bibnamefont
  {Chiba}}, \bibinfo {author} {\bibfnamefont {T.}~\bibnamefont {Ebisuzaki}},
  \bibinfo {author} {\bibfnamefont {Y.}~\bibnamefont {Ejiri}}, \bibinfo
  {author} {\bibfnamefont {M.}~\bibnamefont {Enoki}}, \bibinfo {author}
  {\bibfnamefont {Y.}~\bibnamefont {Eriguchi}}, \bibinfo {author}
  {\bibfnamefont {M.-K.}\ \bibnamefont {Fujimoto}}, \bibinfo {author}
  {\bibfnamefont {R.}~\bibnamefont {Fujita}}, \bibinfo {author} {\bibfnamefont
  {M.}~\bibnamefont {Fukushima}}, \bibinfo {author} {\bibfnamefont
  {T.}~\bibnamefont {Futamase}}, \bibinfo {author} {\bibfnamefont
  {K.}~\bibnamefont {Ganzu}}, \bibinfo {author} {\bibfnamefont
  {T.}~\bibnamefont {Harada}}, \bibinfo {author} {\bibfnamefont
  {T.}~\bibnamefont {Hashimoto}}, \bibinfo {author} {\bibfnamefont
  {K.}~\bibnamefont {Hayama}}, \bibinfo {author} {\bibfnamefont
  {W.}~\bibnamefont {Hikida}}, \bibinfo {author} {\bibfnamefont
  {Y.}~\bibnamefont {Himemoto}}, \bibinfo {author} {\bibfnamefont
  {H.}~\bibnamefont {Hirabayashi}}, \bibinfo {author} {\bibfnamefont
  {T.}~\bibnamefont {Hiramatsu}}, \bibinfo {author} {\bibfnamefont {F.-L.}\
  \bibnamefont {Hong}}, \bibinfo {author} {\bibfnamefont {H.}~\bibnamefont
  {Horisawa}}, \bibinfo {author} {\bibfnamefont {M.}~\bibnamefont {Hosokawa}},
  \bibinfo {author} {\bibfnamefont {K.}~\bibnamefont {Ichiki}}, \bibinfo
  {author} {\bibfnamefont {T.}~\bibnamefont {Ikegami}}, \bibinfo {author}
  {\bibfnamefont {K.~T.}\ \bibnamefont {Inoue}}, \bibinfo {author}
  {\bibfnamefont {K.}~\bibnamefont {Ishidoshiro}}, \bibinfo {author}
  {\bibfnamefont {H.}~\bibnamefont {Ishihara}}, \bibinfo {author}
  {\bibfnamefont {T.}~\bibnamefont {Ishikawa}}, \bibinfo {author}
  {\bibfnamefont {H.}~\bibnamefont {Ishizaki}}, \bibinfo {author}
  {\bibfnamefont {H.}~\bibnamefont {Ito}}, \bibinfo {author} {\bibfnamefont
  {Y.}~\bibnamefont {Itoh}}, \bibinfo {author} {\bibfnamefont {N.}~\bibnamefont
  {Kawashima}}, \bibinfo {author} {\bibfnamefont {F.}~\bibnamefont {Kawazoe}},
  \bibinfo {author} {\bibfnamefont {N.}~\bibnamefont {Kishimoto}}, \bibinfo
  {author} {\bibfnamefont {K.}~\bibnamefont {Kiuchi}}, \bibinfo {author}
  {\bibfnamefont {S.}~\bibnamefont {Kobayashi}}, \bibinfo {author}
  {\bibfnamefont {K.}~\bibnamefont {Kohri}}, \bibinfo {author} {\bibfnamefont
  {H.}~\bibnamefont {Koizumi}}, \bibinfo {author} {\bibfnamefont
  {Y.}~\bibnamefont {Kojima}}, \bibinfo {author} {\bibfnamefont
  {K.}~\bibnamefont {Kokeyama}}, \bibinfo {author} {\bibfnamefont
  {W.}~\bibnamefont {Kokuyama}}, \bibinfo {author} {\bibfnamefont
  {K.}~\bibnamefont {Kotake}}, \bibinfo {author} {\bibfnamefont
  {Y.}~\bibnamefont {Kozai}}, \bibinfo {author} {\bibfnamefont
  {H.}~\bibnamefont {Kudoh}}, \bibinfo {author} {\bibfnamefont
  {H.}~\bibnamefont {Kunimori}}, \bibinfo {author} {\bibfnamefont
  {H.}~\bibnamefont {Kuninaka}}, \bibinfo {author} {\bibfnamefont
  {K.}~\bibnamefont {Kuroda}}, \bibinfo {author} {\bibfnamefont
  {K.}~\bibnamefont {ichi Maeda}}, \bibinfo {author} {\bibfnamefont
  {H.}~\bibnamefont {Matsuhara}}, \bibinfo {author} {\bibfnamefont
  {Y.}~\bibnamefont {Mino}}, \bibinfo {author} {\bibfnamefont {O.}~\bibnamefont
  {Miyakawa}}, \bibinfo {author} {\bibfnamefont {S.}~\bibnamefont {Miyoki}},
  \bibinfo {author} {\bibfnamefont {M.~Y.}\ \bibnamefont {Morimoto}}, \bibinfo
  {author} {\bibfnamefont {T.}~\bibnamefont {Morioka}}, \bibinfo {author}
  {\bibfnamefont {T.}~\bibnamefont {Morisawa}}, \bibinfo {author}
  {\bibfnamefont {S.}~\bibnamefont {Mukohyama}}, \bibinfo {author}
  {\bibfnamefont {S.}~\bibnamefont {Nagano}}, \bibinfo {author} {\bibfnamefont
  {I.}~\bibnamefont {Naito}}, \bibinfo {author} {\bibfnamefont
  {K.}~\bibnamefont {Nakamura}}, \bibinfo {author} {\bibfnamefont
  {H.}~\bibnamefont {Nakano}}, \bibinfo {author} {\bibfnamefont
  {K.}~\bibnamefont {Nakao}}, \bibinfo {author} {\bibfnamefont
  {S.}~\bibnamefont {Nakasuka}}, \bibinfo {author} {\bibfnamefont
  {Y.}~\bibnamefont {Nakayama}}, \bibinfo {author} {\bibfnamefont
  {E.}~\bibnamefont {Nishida}}, \bibinfo {author} {\bibfnamefont
  {K.}~\bibnamefont {Nishiyama}}, \bibinfo {author} {\bibfnamefont
  {A.}~\bibnamefont {Nishizawa}}, \bibinfo {author} {\bibfnamefont
  {Y.}~\bibnamefont {Niwa}}, \bibinfo {author} {\bibfnamefont {T.}~\bibnamefont
  {Noumi}}, \bibinfo {author} {\bibfnamefont {Y.}~\bibnamefont {Obuchi}},
  \bibinfo {author} {\bibfnamefont {M.}~\bibnamefont {Ohashi}}, \bibinfo
  {author} {\bibfnamefont {N.}~\bibnamefont {Ohishi}}, \bibinfo {author}
  {\bibfnamefont {M.}~\bibnamefont {Ohkawa}}, \bibinfo {author} {\bibfnamefont
  {N.}~\bibnamefont {Okada}}, \bibinfo {author} {\bibfnamefont
  {K.}~\bibnamefont {Onozato}}, \bibinfo {author} {\bibfnamefont
  {K.}~\bibnamefont {Oohara}}, \bibinfo {author} {\bibfnamefont
  {N.}~\bibnamefont {Sago}}, \bibinfo {author} {\bibfnamefont {M.}~\bibnamefont
  {Saijo}}, \bibinfo {author} {\bibfnamefont {M.}~\bibnamefont {Sakagami}},
  \bibinfo {author} {\bibfnamefont {S.}~\bibnamefont {Sakata}}, \bibinfo
  {author} {\bibfnamefont {M.}~\bibnamefont {Sasaki}}, \bibinfo {author}
  {\bibfnamefont {T.}~\bibnamefont {Sato}}, \bibinfo {author} {\bibfnamefont
  {M.}~\bibnamefont {Shibata}}, \bibinfo {author} {\bibfnamefont
  {H.}~\bibnamefont {Shinkai}}, \bibinfo {author} {\bibfnamefont
  {K.}~\bibnamefont {Somiya}}, \bibinfo {author} {\bibfnamefont
  {H.}~\bibnamefont {Sotani}}, \bibinfo {author} {\bibfnamefont
  {N.}~\bibnamefont {Sugiyama}}, \bibinfo {author} {\bibfnamefont
  {Y.}~\bibnamefont {Suwa}}, \bibinfo {author} {\bibfnamefont {R.}~\bibnamefont
  {Suzuki}}, \bibinfo {author} {\bibfnamefont {H.}~\bibnamefont {Tagoshi}},
  \bibinfo {author} {\bibfnamefont {F.}~\bibnamefont {Takahashi}}, \bibinfo
  {author} {\bibfnamefont {K.}~\bibnamefont {Takahashi}}, \bibinfo {author}
  {\bibfnamefont {K.}~\bibnamefont {Takahashi}}, \bibinfo {author}
  {\bibfnamefont {R.}~\bibnamefont {Takahashi}}, \bibinfo {author}
  {\bibfnamefont {R.}~\bibnamefont {Takahashi}}, \bibinfo {author}
  {\bibfnamefont {T.}~\bibnamefont {Takahashi}}, \bibinfo {author}
  {\bibfnamefont {H.}~\bibnamefont {Takahashi}}, \bibinfo {author}
  {\bibfnamefont {T.}~\bibnamefont {Akiteru}}, \bibinfo {author} {\bibfnamefont
  {T.}~\bibnamefont {Takano}}, \bibinfo {author} {\bibfnamefont
  {K.}~\bibnamefont {Taniguchi}}, \bibinfo {author} {\bibfnamefont
  {A.}~\bibnamefont {Taruya}}, \bibinfo {author} {\bibfnamefont
  {H.}~\bibnamefont {Tashiro}}, \bibinfo {author} {\bibfnamefont
  {Y.}~\bibnamefont {Torii}}, \bibinfo {author} {\bibfnamefont
  {M.}~\bibnamefont {Toyoshima}}, \bibinfo {author} {\bibfnamefont
  {S.}~\bibnamefont {Tsujikawa}}, \bibinfo {author} {\bibfnamefont
  {Y.}~\bibnamefont {Tsunesada}}, \bibinfo {author} {\bibfnamefont
  {A.}~\bibnamefont {Ueda}}, \bibinfo {author} {\bibfnamefont {K.}~\bibnamefont
  {ichi Ueda}}, \bibinfo {author} {\bibfnamefont {M.}~\bibnamefont {Utashima}},
  \bibinfo {author} {\bibfnamefont {Y.}~\bibnamefont {Wakabayashi}}, \bibinfo
  {author} {\bibfnamefont {H.}~\bibnamefont {Yamakawa}}, \bibinfo {author}
  {\bibfnamefont {K.}~\bibnamefont {Yamamoto}}, \bibinfo {author}
  {\bibfnamefont {T.}~\bibnamefont {Yamazaki}}, \bibinfo {author}
  {\bibfnamefont {J.}~\bibnamefont {Yokoyama}}, \bibinfo {author}
  {\bibfnamefont {C.-M.}\ \bibnamefont {Yoo}}, \bibinfo {author} {\bibfnamefont
  {S.}~\bibnamefont {Yoshida}}, \ and\ \bibinfo {author} {\bibfnamefont
  {T.}~\bibnamefont {Yoshino}},\ }\href {\doibase
  10.1088/1742-6596/840/1/012010} {\bibfield  {journal} {\bibinfo  {journal}
  {Journal of Physics: Conference Series}\ }\textbf {\bibinfo {volume} {840}},\
  \bibinfo {pages} {012010} (\bibinfo {year} {2017})}\BibitemShut {NoStop}%
\bibitem [{\citenamefont {Kawamura}\ \emph {et~al.}(2020)\citenamefont
  {Kawamura} \emph {et~al.}}]{Kawamura:2020pcg}%
  \BibitemOpen
  \bibfield  {author} {\bibinfo {author} {\bibfnamefont {S.}~\bibnamefont
  {Kawamura}} \emph {et~al.},\ }\href@noop {} {\  (\bibinfo {year} {2020})},\
  \Eprint {http://arxiv.org/abs/2006.13545} {arXiv:2006.13545 [gr-qc]}
  \BibitemShut {NoStop}%
\bibitem [{\citenamefont {Harms}\ \emph {et~al.}(2021)\citenamefont {Harms}
  \emph {et~al.}}]{LGWA:2020mma}%
  \BibitemOpen
  \bibfield  {author} {\bibinfo {author} {\bibfnamefont {J.}~\bibnamefont
  {Harms}} \emph {et~al.} (\bibinfo {collaboration} {LGWA}),\ }\href {\doibase
  10.3847/1538-4357/abe5a7} {\bibfield  {journal} {\bibinfo  {journal}
  {Astrophys. J.}\ }\textbf {\bibinfo {volume} {910}},\ \bibinfo {pages} {1}
  (\bibinfo {year} {2021})},\ \Eprint {http://arxiv.org/abs/2010.13726}
  {arXiv:2010.13726 [gr-qc]} \BibitemShut {NoStop}%
\bibitem [{\citenamefont {Chen}\ \emph {et~al.}(2021)\citenamefont {Chen},
  \citenamefont {Huerta}, \citenamefont {Adamo}, \citenamefont {Haas},
  \citenamefont {O'Shea}, \citenamefont {Kumar},\ and\ \citenamefont
  {Moore}}]{chen2020observation}%
  \BibitemOpen
  \bibfield  {author} {\bibinfo {author} {\bibfnamefont {Z.}~\bibnamefont
  {Chen}}, \bibinfo {author} {\bibfnamefont {E.~A.}\ \bibnamefont {Huerta}},
  \bibinfo {author} {\bibfnamefont {J.}~\bibnamefont {Adamo}}, \bibinfo
  {author} {\bibfnamefont {R.}~\bibnamefont {Haas}}, \bibinfo {author}
  {\bibfnamefont {E.}~\bibnamefont {O'Shea}}, \bibinfo {author} {\bibfnamefont
  {P.}~\bibnamefont {Kumar}}, \ and\ \bibinfo {author} {\bibfnamefont
  {C.}~\bibnamefont {Moore}},\ }\href {\doibase 10.1103/PhysRevD.103.084018}
  {\bibfield  {journal} {\bibinfo  {journal} {Phys. Rev. D}\ }\textbf {\bibinfo
  {volume} {103}},\ \bibinfo {pages} {084018} (\bibinfo {year} {2021})},\
  \Eprint {http://arxiv.org/abs/2008.03313} {arXiv:2008.03313 [gr-qc]}
  \BibitemShut {NoStop}%
\bibitem [{\citenamefont {Huerta}\ and\ \citenamefont
  {Brown}(2013)}]{Huerta_2013}%
  \BibitemOpen
  \bibfield  {author} {\bibinfo {author} {\bibfnamefont {E.~A.}\ \bibnamefont
  {Huerta}}\ and\ \bibinfo {author} {\bibfnamefont {D.~A.}\ \bibnamefont
  {Brown}},\ }\href {\doibase 10.1103/physrevd.87.127501} {\bibfield  {journal}
  {\bibinfo  {journal} {Phys. Rev. D}\ }\textbf {\bibinfo {volume} {87}}
  (\bibinfo {year} {2013}),\ 10.1103/physrevd.87.127501}\BibitemShut {NoStop}%
\bibitem [{\citenamefont {Tiwari}\ \emph
  {et~al.}(2016{\natexlab{b}})\citenamefont {Tiwari}, \citenamefont {Klimenko},
  \citenamefont {Christensen}, \citenamefont {Huerta}, \citenamefont
  {Mohapatra}, \citenamefont {Gopakumar}, \citenamefont {Haney}, \citenamefont
  {Ajith}, \citenamefont {McWilliams}, \citenamefont {Vedovato}, \citenamefont
  {Drago}, \citenamefont {Salemi}, \citenamefont {Prodi}, \citenamefont
  {Lazzaro}, \citenamefont {Tiwari}, \citenamefont {Mitselmakher},\ and\
  \citenamefont {{Da Silva}}}]{Tiwari2016}%
  \BibitemOpen
  \bibfield  {author} {\bibinfo {author} {\bibfnamefont {V.}~\bibnamefont
  {Tiwari}}, \bibinfo {author} {\bibfnamefont {S.}~\bibnamefont {Klimenko}},
  \bibinfo {author} {\bibfnamefont {N.}~\bibnamefont {Christensen}}, \bibinfo
  {author} {\bibfnamefont {E.~A.}\ \bibnamefont {Huerta}}, \bibinfo {author}
  {\bibfnamefont {S.~R.}\ \bibnamefont {Mohapatra}}, \bibinfo {author}
  {\bibfnamefont {A.}~\bibnamefont {Gopakumar}}, \bibinfo {author}
  {\bibfnamefont {M.}~\bibnamefont {Haney}}, \bibinfo {author} {\bibfnamefont
  {P.}~\bibnamefont {Ajith}}, \bibinfo {author} {\bibfnamefont {S.~T.}\
  \bibnamefont {McWilliams}}, \bibinfo {author} {\bibfnamefont
  {G.}~\bibnamefont {Vedovato}}, \bibinfo {author} {\bibfnamefont
  {M.}~\bibnamefont {Drago}}, \bibinfo {author} {\bibfnamefont
  {F.}~\bibnamefont {Salemi}}, \bibinfo {author} {\bibfnamefont {G.~A.}\
  \bibnamefont {Prodi}}, \bibinfo {author} {\bibfnamefont {C.}~\bibnamefont
  {Lazzaro}}, \bibinfo {author} {\bibfnamefont {S.}~\bibnamefont {Tiwari}},
  \bibinfo {author} {\bibfnamefont {G.}~\bibnamefont {Mitselmakher}}, \ and\
  \bibinfo {author} {\bibfnamefont {F.}~\bibnamefont {{Da Silva}}},\ }\href
  {\doibase 10.1103/PhysRevD.93.043007} {\bibfield  {journal} {\bibinfo
  {journal} {Phys. Rev. D}\ }\textbf {\bibinfo {volume} {93}} (\bibinfo {year}
  {2016}{\natexlab{b}}),\ 10.1103/PhysRevD.93.043007}\BibitemShut {NoStop}%
\bibitem [{\citenamefont {Huerta}\ \emph {et~al.}(2017)\citenamefont {Huerta},
  \citenamefont {Moore}, \citenamefont {Kumar}, \citenamefont {George},
  \citenamefont {Chua}, \citenamefont {Haas}, \citenamefont {Wessel},
  \citenamefont {Johnson}, \citenamefont {Glennon}, \citenamefont {Rebei},
  \citenamefont {Holgado}, \citenamefont {Gair},\ and\ \citenamefont
  {Pfeiffer}}]{ENIGMA}%
  \BibitemOpen
  \bibfield  {author} {\bibinfo {author} {\bibfnamefont {E.~A.}\ \bibnamefont
  {Huerta}}, \bibinfo {author} {\bibfnamefont {C.~J.}\ \bibnamefont {Moore}},
  \bibinfo {author} {\bibfnamefont {P.}~\bibnamefont {Kumar}}, \bibinfo
  {author} {\bibfnamefont {D.}~\bibnamefont {George}}, \bibinfo {author}
  {\bibfnamefont {A.~J.~K.}\ \bibnamefont {Chua}}, \bibinfo {author}
  {\bibfnamefont {R.}~\bibnamefont {Haas}}, \bibinfo {author} {\bibfnamefont
  {E.}~\bibnamefont {Wessel}}, \bibinfo {author} {\bibfnamefont
  {D.}~\bibnamefont {Johnson}}, \bibinfo {author} {\bibfnamefont
  {D.}~\bibnamefont {Glennon}}, \bibinfo {author} {\bibfnamefont
  {A.}~\bibnamefont {Rebei}}, \bibinfo {author} {\bibfnamefont {A.~M.}\
  \bibnamefont {Holgado}}, \bibinfo {author} {\bibfnamefont {J.~R.}\
  \bibnamefont {Gair}}, \ and\ \bibinfo {author} {\bibfnamefont {H.~P.}\
  \bibnamefont {Pfeiffer}},\ }\href {\doibase 10.1103/PhysRevD.97.024031}
  {\bibfield  {journal} {\bibinfo  {journal} {Phys. Rev. D}\ }\textbf {\bibinfo
  {volume} {97}} (\bibinfo {year} {2017}),\ 10.1103/PhysRevD.97.024031},\
  \Eprint {http://arxiv.org/abs/1711.06276} {1711.06276} \BibitemShut {NoStop}%
\bibitem [{gwo()}]{gwosc_catalog}%
  \BibitemOpen
  \href@noop {} {\enquote {\bibinfo {title} {The gravitational wave open
  science center catalog},}\ }\bibinfo {howpublished}
  {\url{https://www.gw-openscience.org/eventapi/html/allevents/}}\BibitemShut
  {NoStop}%
\bibitem [{\citenamefont {Peters}\ and\ \citenamefont
  {Mathews}(1963)}]{peters}%
  \BibitemOpen
  \bibfield  {author} {\bibinfo {author} {\bibfnamefont {P.~C.}\ \bibnamefont
  {Peters}}\ and\ \bibinfo {author} {\bibfnamefont {J.}~\bibnamefont
  {Mathews}},\ }\href {\doibase 10.1103/PhysRev.131.435} {\bibfield  {journal}
  {\bibinfo  {journal} {Phys. Rev.}\ }\textbf {\bibinfo {volume} {131}},\
  \bibinfo {pages} {435} (\bibinfo {year} {1963})}\BibitemShut {NoStop}%
\bibitem [{\citenamefont {Martel}\ and\ \citenamefont
  {Poisson}(1999)}]{Martel_1999}%
  \BibitemOpen
  \bibfield  {author} {\bibinfo {author} {\bibfnamefont {K.}~\bibnamefont
  {Martel}}\ and\ \bibinfo {author} {\bibfnamefont {E.}~\bibnamefont
  {Poisson}},\ }\href {\doibase 10.1103/physrevd.60.124008} {\bibfield
  {journal} {\bibinfo  {journal} {Phys. Rev. D}\ }\textbf {\bibinfo {volume}
  {60}} (\bibinfo {year} {1999}),\ 10.1103/physrevd.60.124008}\BibitemShut
  {NoStop}%
\bibitem [{\citenamefont {Abbott}\ \emph {et~al.}(2017)\citenamefont {Abbott}
  \emph {et~al.}}]{LIGOScientific:2016ebw}%
  \BibitemOpen
  \bibfield  {author} {\bibinfo {author} {\bibfnamefont {B.~P.}\ \bibnamefont
  {Abbott}} \emph {et~al.} (\bibinfo {collaboration} {LIGO Scientific,
  Virgo}),\ }\href {\doibase 10.1088/1361-6382/aa6854} {\bibfield  {journal}
  {\bibinfo  {journal} {Class. Quant. Grav.}\ }\textbf {\bibinfo {volume}
  {34}},\ \bibinfo {pages} {104002} (\bibinfo {year} {2017})},\ \Eprint
  {http://arxiv.org/abs/1611.07531} {arXiv:1611.07531 [gr-qc]} \BibitemShut
  {NoStop}%
\bibitem [{\citenamefont {Mora}\ and\ \citenamefont
  {Will}(2002)}]{Mora:2002gf}%
  \BibitemOpen
  \bibfield  {author} {\bibinfo {author} {\bibfnamefont {T.}~\bibnamefont
  {Mora}}\ and\ \bibinfo {author} {\bibfnamefont {C.~M.}\ \bibnamefont
  {Will}},\ }\href {\doibase 10.1103/PhysRevD.66.101501} {\bibfield  {journal}
  {\bibinfo  {journal} {Phys. Rev. D}\ }\textbf {\bibinfo {volume} {66}},\
  \bibinfo {pages} {101501} (\bibinfo {year} {2002})},\ \Eprint
  {http://arxiv.org/abs/gr-qc/0208089} {arXiv:gr-qc/0208089} \BibitemShut
  {NoStop}%
\bibitem [{\citenamefont {Knee}\ \emph {et~al.}(2022)\citenamefont {Knee},
  \citenamefont {Romero-Shaw}, \citenamefont {Lasky}, \citenamefont {McIver},\
  and\ \citenamefont {Thrane}}]{Knee:2022hth}%
  \BibitemOpen
  \bibfield  {author} {\bibinfo {author} {\bibfnamefont {A.~M.}\ \bibnamefont
  {Knee}}, \bibinfo {author} {\bibfnamefont {I.~M.}\ \bibnamefont
  {Romero-Shaw}}, \bibinfo {author} {\bibfnamefont {P.~D.}\ \bibnamefont
  {Lasky}}, \bibinfo {author} {\bibfnamefont {J.}~\bibnamefont {McIver}}, \
  and\ \bibinfo {author} {\bibfnamefont {E.}~\bibnamefont {Thrane}},\ }\href
  {\doibase 10.3847/1538-4357/ac8b02} {\bibfield  {journal} {\bibinfo
  {journal} {Astrophys. J.}\ }\textbf {\bibinfo {volume} {936}},\ \bibinfo
  {pages} {172} (\bibinfo {year} {2022})},\ \Eprint
  {http://arxiv.org/abs/2207.14346} {arXiv:2207.14346 [gr-qc]} \BibitemShut
  {NoStop}%
\bibitem [{\citenamefont {Shaikh}\ \emph {et~al.}(2023)\citenamefont {Shaikh},
  \citenamefont {Varma}, \citenamefont {Pfeiffer}, \citenamefont
  {Ramos-Buades},\ and\ \citenamefont {van~de Meent}}]{Shaikh:2023ypz}%
  \BibitemOpen
  \bibfield  {author} {\bibinfo {author} {\bibfnamefont {M.~A.}\ \bibnamefont
  {Shaikh}}, \bibinfo {author} {\bibfnamefont {V.}~\bibnamefont {Varma}},
  \bibinfo {author} {\bibfnamefont {H.~P.}\ \bibnamefont {Pfeiffer}}, \bibinfo
  {author} {\bibfnamefont {A.}~\bibnamefont {Ramos-Buades}}, \ and\ \bibinfo
  {author} {\bibfnamefont {M.}~\bibnamefont {van~de Meent}},\ }\href@noop {} {\
   (\bibinfo {year} {2023})},\ \Eprint {http://arxiv.org/abs/2302.11257}
  {arXiv:2302.11257 [gr-qc]} \BibitemShut {NoStop}%
\bibitem [{\citenamefont {Huerta}\ \emph {et~al.}(2014)\citenamefont {Huerta},
  \citenamefont {Kumar}, \citenamefont {McWilliams}, \citenamefont
  {O'Shaughnessy},\ and\ \citenamefont {Yunes}}]{EccentricFD}%
  \BibitemOpen
  \bibfield  {author} {\bibinfo {author} {\bibfnamefont {E.~A.}\ \bibnamefont
  {Huerta}}, \bibinfo {author} {\bibfnamefont {P.}~\bibnamefont {Kumar}},
  \bibinfo {author} {\bibfnamefont {S.~T.}\ \bibnamefont {McWilliams}},
  \bibinfo {author} {\bibfnamefont {R.}~\bibnamefont {O'Shaughnessy}}, \ and\
  \bibinfo {author} {\bibfnamefont {N.}~\bibnamefont {Yunes}},\ }\href
  {\doibase 10.1103/PhysRevD.90.084016} {\bibfield  {journal} {\bibinfo
  {journal} {Phys. Rev. D}\ }\textbf {\bibinfo {volume} {90}},\ \bibinfo
  {pages} {084016} (\bibinfo {year} {2014})}\BibitemShut {NoStop}%
\bibitem [{\citenamefont {Tanay}\ \emph {et~al.}(2016)\citenamefont {Tanay},
  \citenamefont {Haney},\ and\ \citenamefont {Gopakumar}}]{EccentricTD}%
  \BibitemOpen
  \bibfield  {author} {\bibinfo {author} {\bibfnamefont {S.}~\bibnamefont
  {Tanay}}, \bibinfo {author} {\bibfnamefont {M.}~\bibnamefont {Haney}}, \ and\
  \bibinfo {author} {\bibfnamefont {A.}~\bibnamefont {Gopakumar}},\ }\href
  {\doibase 10.1103/PhysRevD.93.064031} {\bibfield  {journal} {\bibinfo
  {journal} {Phys. Rev. D}\ }\textbf {\bibinfo {volume} {93}},\ \bibinfo
  {pages} {064031} (\bibinfo {year} {2016})}\BibitemShut {NoStop}%
\bibitem [{\citenamefont {{LIGO Scientific Collaboration}}(2018)}]{lalsuite}%
  \BibitemOpen
  \bibfield  {author} {\bibinfo {author} {\bibnamefont {{LIGO Scientific
  Collaboration}}},\ }\href {\doibase 10.7935/GT1W-FZ16} {\enquote {\bibinfo
  {title} {{LIGO} {A}lgorithm {L}ibrary - {LALS}uite},}\ }\bibinfo
  {howpublished} {free software (GPL)} (\bibinfo {year} {2018})\BibitemShut
  {NoStop}%
\bibitem [{\citenamefont {Cao}\ and\ \citenamefont {Han}(2017)}]{seobnre}%
  \BibitemOpen
  \bibfield  {author} {\bibinfo {author} {\bibfnamefont {Z.}~\bibnamefont
  {Cao}}\ and\ \bibinfo {author} {\bibfnamefont {W.-B.}\ \bibnamefont {Han}},\
  }\href {\doibase 10.1103/PhysRevD.96.044028} {\bibfield  {journal} {\bibinfo
  {journal} {Phys. Rev. D}\ }\textbf {\bibinfo {volume} {96}},\ \bibinfo
  {pages} {044028} (\bibinfo {year} {2017})}\BibitemShut {NoStop}%
\bibitem [{\citenamefont {Chiaramello}\ and\ \citenamefont
  {Nagar}(2020{\natexlab{a}})}]{Chiaramello2020}%
  \BibitemOpen
  \bibfield  {author} {\bibinfo {author} {\bibfnamefont {D.}~\bibnamefont
  {Chiaramello}}\ and\ \bibinfo {author} {\bibfnamefont {A.}~\bibnamefont
  {Nagar}},\ }\href {\doibase 10.1103/PhysRevD.101.101501} {\bibfield
  {journal} {\bibinfo  {journal} {Phys. Rev. D}\ }\textbf {\bibinfo {volume}
  {101}} (\bibinfo {year} {2020}{\natexlab{a}}),\
  10.1103/PhysRevD.101.101501},\ \Eprint {http://arxiv.org/abs/2001.11736}
  {arXiv:2001.11736} \BibitemShut {NoStop}%
\bibitem [{\citenamefont {Islam}\ \emph {et~al.}(2021)\citenamefont {Islam},
  \citenamefont {Varma}, \citenamefont {Lodman}, \citenamefont {Field},
  \citenamefont {Khanna}, \citenamefont {Scheel}, \citenamefont {Pfeiffer},
  \citenamefont {Gerosa},\ and\ \citenamefont {Kidder}}]{Islam_2021}%
  \BibitemOpen
  \bibfield  {author} {\bibinfo {author} {\bibfnamefont {T.}~\bibnamefont
  {Islam}}, \bibinfo {author} {\bibfnamefont {V.}~\bibnamefont {Varma}},
  \bibinfo {author} {\bibfnamefont {J.}~\bibnamefont {Lodman}}, \bibinfo
  {author} {\bibfnamefont {S.~E.}\ \bibnamefont {Field}}, \bibinfo {author}
  {\bibfnamefont {G.}~\bibnamefont {Khanna}}, \bibinfo {author} {\bibfnamefont
  {M.~A.}\ \bibnamefont {Scheel}}, \bibinfo {author} {\bibfnamefont {H.~P.}\
  \bibnamefont {Pfeiffer}}, \bibinfo {author} {\bibfnamefont {D.}~\bibnamefont
  {Gerosa}}, \ and\ \bibinfo {author} {\bibfnamefont {L.~E.}\ \bibnamefont
  {Kidder}},\ }\href {\doibase 10.1103/physrevd.103.064022} {\bibfield
  {journal} {\bibinfo  {journal} {Phys. Rev. D}\ }\textbf {\bibinfo {volume}
  {103}} (\bibinfo {year} {2021}),\ 10.1103/physrevd.103.064022}\BibitemShut
  {NoStop}%
\bibitem [{\citenamefont {Nagar}\ \emph {et~al.}(2020)\citenamefont {Nagar},
  \citenamefont {Pratten}, \citenamefont {Riemenschneider},\ and\ \citenamefont
  {Gamba}}]{TEOBResum}%
  \BibitemOpen
  \bibfield  {author} {\bibinfo {author} {\bibfnamefont {A.}~\bibnamefont
  {Nagar}}, \bibinfo {author} {\bibfnamefont {G.}~\bibnamefont {Pratten}},
  \bibinfo {author} {\bibfnamefont {G.}~\bibnamefont {Riemenschneider}}, \ and\
  \bibinfo {author} {\bibfnamefont {R.}~\bibnamefont {Gamba}},\ }\href
  {\doibase 10.1103/PhysRevD.101.024041} {\bibfield  {journal} {\bibinfo
  {journal} {Phys. Rev. D}\ }\textbf {\bibinfo {volume} {101}},\ \bibinfo
  {pages} {024041} (\bibinfo {year} {2020})}\BibitemShut {NoStop}%
\bibitem [{\citenamefont {Chiaramello}\ and\ \citenamefont
  {Nagar}(2020{\natexlab{b}})}]{TEOBResumE}%
  \BibitemOpen
  \bibfield  {author} {\bibinfo {author} {\bibfnamefont {D.}~\bibnamefont
  {Chiaramello}}\ and\ \bibinfo {author} {\bibfnamefont {A.}~\bibnamefont
  {Nagar}},\ }\href {\doibase 10.1103/physrevd.101.101501} {\bibfield
  {journal} {\bibinfo  {journal} {Phys. Rev. D}\ }\textbf {\bibinfo {volume}
  {101}} (\bibinfo {year} {2020}{\natexlab{b}}),\
  10.1103/physrevd.101.101501}\BibitemShut {NoStop}%
\bibitem [{EOB(2021{\natexlab{a}})}]{EOBBitbucket}%
  \BibitemOpen
  \href@noop {} {\enquote {\bibinfo {title} {\texttt{TEOBResumS}:
  Effective-one-body model with spin and tidal interactions},}\ }\bibinfo
  {howpublished} {\url{https://bitbucket.org/eob_ihes/teobresums/src/master/}}
  (\bibinfo {year} {2021}{\natexlab{a}})\BibitemShut {NoStop}%
\bibitem [{\citenamefont {Clarke}\ \emph {et~al.}(2022)\citenamefont {Clarke},
  \citenamefont {Romero-Shaw}, \citenamefont {Lasky},\ and\ \citenamefont
  {Thrane}}]{Clarke:2022fma}%
  \BibitemOpen
  \bibfield  {author} {\bibinfo {author} {\bibfnamefont {T.~A.}\ \bibnamefont
  {Clarke}}, \bibinfo {author} {\bibfnamefont {I.~M.}\ \bibnamefont
  {Romero-Shaw}}, \bibinfo {author} {\bibfnamefont {P.~D.}\ \bibnamefont
  {Lasky}}, \ and\ \bibinfo {author} {\bibfnamefont {E.}~\bibnamefont
  {Thrane}},\ }\href {\doibase 10.1093/mnras/stac2965} {\bibfield  {journal}
  {\bibinfo  {journal} {Mon. Not. Roy. Astron. Soc.}\ }\textbf {\bibinfo
  {volume} {517}},\ \bibinfo {pages} {3778} (\bibinfo {year} {2022})},\ \Eprint
  {http://arxiv.org/abs/2206.14006} {arXiv:2206.14006 [gr-qc]} \BibitemShut
  {NoStop}%
\bibitem [{\citenamefont {{Ashton}}\ \emph {et~al.}(2019)\citenamefont
  {{Ashton}}, \citenamefont {{H{\"u}bner}}, \citenamefont {{Lasky}},
  \citenamefont {{Talbot}}, \citenamefont {{Ackley}}, \citenamefont
  {{Biscoveanu}}, \citenamefont {{Chu}}, \citenamefont {{Divakarla}},
  \citenamefont {{Easter}}, \citenamefont {{Goncharov}}, \citenamefont
  {{Hernandez Vivanco}}, \citenamefont {{Harms}}, \citenamefont {{Lower}},
  \citenamefont {{Meadors}}, \citenamefont {{Melchor}}, \citenamefont
  {{Payne}}, \citenamefont {{Pitkin}}, \citenamefont {{Powell}}, \citenamefont
  {{Sarin}}, \citenamefont {{Smith}},\ and\ \citenamefont {{Thrane}}}]{bilby}%
  \BibitemOpen
  \bibfield  {author} {\bibinfo {author} {\bibfnamefont {G.}~\bibnamefont
  {{Ashton}}}, \bibinfo {author} {\bibfnamefont {M.}~\bibnamefont
  {{H{\"u}bner}}}, \bibinfo {author} {\bibfnamefont {P.~D.}\ \bibnamefont
  {{Lasky}}}, \bibinfo {author} {\bibfnamefont {C.}~\bibnamefont {{Talbot}}},
  \bibinfo {author} {\bibfnamefont {K.}~\bibnamefont {{Ackley}}}, \bibinfo
  {author} {\bibfnamefont {S.}~\bibnamefont {{Biscoveanu}}}, \bibinfo {author}
  {\bibfnamefont {Q.}~\bibnamefont {{Chu}}}, \bibinfo {author} {\bibfnamefont
  {A.}~\bibnamefont {{Divakarla}}}, \bibinfo {author} {\bibfnamefont {P.~J.}\
  \bibnamefont {{Easter}}}, \bibinfo {author} {\bibfnamefont {B.}~\bibnamefont
  {{Goncharov}}}, \bibinfo {author} {\bibfnamefont {F.}~\bibnamefont
  {{Hernandez Vivanco}}}, \bibinfo {author} {\bibfnamefont {J.}~\bibnamefont
  {{Harms}}}, \bibinfo {author} {\bibfnamefont {M.~E.}\ \bibnamefont
  {{Lower}}}, \bibinfo {author} {\bibfnamefont {G.~D.}\ \bibnamefont
  {{Meadors}}}, \bibinfo {author} {\bibfnamefont {D.}~\bibnamefont
  {{Melchor}}}, \bibinfo {author} {\bibfnamefont {E.}~\bibnamefont {{Payne}}},
  \bibinfo {author} {\bibfnamefont {M.~D.}\ \bibnamefont {{Pitkin}}}, \bibinfo
  {author} {\bibfnamefont {J.}~\bibnamefont {{Powell}}}, \bibinfo {author}
  {\bibfnamefont {N.}~\bibnamefont {{Sarin}}}, \bibinfo {author} {\bibfnamefont
  {R.~J.~E.}\ \bibnamefont {{Smith}}}, \ and\ \bibinfo {author} {\bibfnamefont
  {E.}~\bibnamefont {{Thrane}}},\ }\href {\doibase 10.3847/1538-4365/ab06fc}
  {\bibfield  {journal} {\bibinfo  {journal} {\apjs}\ }\textbf {\bibinfo
  {volume} {241}},\ \bibinfo {eid} {27} (\bibinfo {year} {2019})},\ \Eprint
  {http://arxiv.org/abs/1811.02042} {arXiv:1811.02042 [astro-ph.IM]}
  \BibitemShut {NoStop}%
\bibitem [{lig()}]{ligo_psd_data}%
  \BibitemOpen
  \href@noop {} {\enquote {\bibinfo {title} {Advanced ligo anticipated
  sensitivity curves},}\ }\bibinfo {howpublished}
  {\url{https://dcc.ligo.org/LIGO-T0900288/public}}\BibitemShut {NoStop}%
\bibitem [{\citenamefont {Speagle}(2020)}]{dynesty}%
  \BibitemOpen
  \bibfield  {author} {\bibinfo {author} {\bibfnamefont {J.~S.}\ \bibnamefont
  {Speagle}},\ }\href {\doibase 10.1093/mnras/staa278} {\bibfield  {journal}
  {\bibinfo  {journal} {Monthly Notices of the Royal Astronomical Society}\
  }\textbf {\bibinfo {volume} {493}},\ \bibinfo {pages} {3132–3158} (\bibinfo
  {year} {2020})}\BibitemShut {NoStop}%
\bibitem [{\citenamefont {Farr}()}]{time_marg}%
  \BibitemOpen
  \bibfield  {author} {\bibinfo {author} {\bibfnamefont {W.}~\bibnamefont
  {Farr}},\ }\href@noop {} {\enquote {\bibinfo {title} {Marginalisation of the
  time parameter in gravitational wave parameter estimation},}\ }\bibinfo
  {howpublished} {\url{https://dcc.ligo.org/T1400460-v2/public}}\BibitemShut
  {NoStop}%
\bibitem [{\citenamefont {Veitch}\ and\ \citenamefont
  {Del~Pozzo}()}]{phase_marg}%
  \BibitemOpen
  \bibfield  {author} {\bibinfo {author} {\bibfnamefont {J.}~\bibnamefont
  {Veitch}}\ and\ \bibinfo {author} {\bibfnamefont {W.}~\bibnamefont
  {Del~Pozzo}},\ }\href@noop {} {\enquote {\bibinfo {title} {Analytic
  marginalisation of phase parameter},}\ }\bibinfo {howpublished}
  {\url{https://dcc.ligo.org/LIGO-T1300326/public}}\BibitemShut {NoStop}%
\bibitem [{\citenamefont {Thrane}\ and\ \citenamefont
  {Talbot}(2019)}]{thrane_talbot_2019}%
  \BibitemOpen
  \bibfield  {author} {\bibinfo {author} {\bibfnamefont {E.}~\bibnamefont
  {Thrane}}\ and\ \bibinfo {author} {\bibfnamefont {C.}~\bibnamefont
  {Talbot}},\ }\href {\doibase 10.1017/pasa.2019.2} {\bibfield  {journal}
  {\bibinfo  {journal} {Publications of the Astronomical Society of Australia}\
  }\textbf {\bibinfo {volume} {36}},\ \bibinfo {pages} {e010} (\bibinfo {year}
  {2019})}\BibitemShut {NoStop}%
\bibitem [{\citenamefont {Bohé}\ \emph {et~al.}(2017)\citenamefont {Bohé},
  \citenamefont {Shao}, \citenamefont {Taracchini}, \citenamefont {Buonanno},
  \citenamefont {Babak}, \citenamefont {Harry}, \citenamefont {Hinder},
  \citenamefont {Ossokine}, \citenamefont {Pürrer}, \citenamefont {Raymond},\
  and\ \citenamefont {et~al.}}]{seobnrv4_rom}%
  \BibitemOpen
  \bibfield  {author} {\bibinfo {author} {\bibfnamefont {A.}~\bibnamefont
  {Bohé}}, \bibinfo {author} {\bibfnamefont {L.}~\bibnamefont {Shao}},
  \bibinfo {author} {\bibfnamefont {A.}~\bibnamefont {Taracchini}}, \bibinfo
  {author} {\bibfnamefont {A.}~\bibnamefont {Buonanno}}, \bibinfo {author}
  {\bibfnamefont {S.}~\bibnamefont {Babak}}, \bibinfo {author} {\bibfnamefont
  {I.~W.}\ \bibnamefont {Harry}}, \bibinfo {author} {\bibfnamefont
  {I.}~\bibnamefont {Hinder}}, \bibinfo {author} {\bibfnamefont
  {S.}~\bibnamefont {Ossokine}}, \bibinfo {author} {\bibfnamefont
  {M.}~\bibnamefont {Pürrer}}, \bibinfo {author} {\bibfnamefont
  {V.}~\bibnamefont {Raymond}}, \ and\ \bibinfo {author} {\bibnamefont
  {et~al.}},\ }\href {\doibase 10.1103/physrevd.95.044028} {\bibfield
  {journal} {\bibinfo  {journal} {Phys. Rev. D}\ }\textbf {\bibinfo {volume}
  {95}} (\bibinfo {year} {2017}),\ 10.1103/physrevd.95.044028}\BibitemShut
  {NoStop}%
\bibitem [{\citenamefont {Husa}\ \emph {et~al.}(2016)\citenamefont {Husa},
  \citenamefont {Khan}, \citenamefont {Hannam}, \citenamefont {P\"urrer},
  \citenamefont {Ohme}, \citenamefont {Forteza},\ and\ \citenamefont
  {Boh\'e}}]{imrphenom_1}%
  \BibitemOpen
  \bibfield  {author} {\bibinfo {author} {\bibfnamefont {S.}~\bibnamefont
  {Husa}}, \bibinfo {author} {\bibfnamefont {S.}~\bibnamefont {Khan}}, \bibinfo
  {author} {\bibfnamefont {M.}~\bibnamefont {Hannam}}, \bibinfo {author}
  {\bibfnamefont {M.}~\bibnamefont {P\"urrer}}, \bibinfo {author}
  {\bibfnamefont {F.}~\bibnamefont {Ohme}}, \bibinfo {author} {\bibfnamefont
  {X.~J.}\ \bibnamefont {Forteza}}, \ and\ \bibinfo {author} {\bibfnamefont
  {A.}~\bibnamefont {Boh\'e}},\ }\href {\doibase 10.1103/PhysRevD.93.044006}
  {\bibfield  {journal} {\bibinfo  {journal} {Phys. Rev. D}\ }\textbf {\bibinfo
  {volume} {93}},\ \bibinfo {pages} {044006} (\bibinfo {year}
  {2016})}\BibitemShut {NoStop}%
\bibitem [{\citenamefont {Khan}\ \emph {et~al.}(2016)\citenamefont {Khan},
  \citenamefont {Husa}, \citenamefont {Hannam}, \citenamefont {Ohme},
  \citenamefont {P\"urrer}, \citenamefont {Forteza},\ and\ \citenamefont
  {Boh\'e}}]{imrphenom_2}%
  \BibitemOpen
  \bibfield  {author} {\bibinfo {author} {\bibfnamefont {S.}~\bibnamefont
  {Khan}}, \bibinfo {author} {\bibfnamefont {S.}~\bibnamefont {Husa}}, \bibinfo
  {author} {\bibfnamefont {M.}~\bibnamefont {Hannam}}, \bibinfo {author}
  {\bibfnamefont {F.}~\bibnamefont {Ohme}}, \bibinfo {author} {\bibfnamefont
  {M.}~\bibnamefont {P\"urrer}}, \bibinfo {author} {\bibfnamefont {X.~J.}\
  \bibnamefont {Forteza}}, \ and\ \bibinfo {author} {\bibfnamefont
  {A.}~\bibnamefont {Boh\'e}},\ }\href {\doibase 10.1103/PhysRevD.93.044007}
  {\bibfield  {journal} {\bibinfo  {journal} {Phys. Rev. D}\ }\textbf {\bibinfo
  {volume} {93}},\ \bibinfo {pages} {044007} (\bibinfo {year}
  {2016})}\BibitemShut {NoStop}%
\bibitem [{\citenamefont {Pratten}\ \emph {et~al.}(2021)\citenamefont {Pratten}
  \emph {et~al.}}]{Pratten:2020ceb}%
  \BibitemOpen
  \bibfield  {author} {\bibinfo {author} {\bibfnamefont {G.}~\bibnamefont
  {Pratten}} \emph {et~al.},\ }\href {\doibase 10.1103/PhysRevD.103.104056}
  {\bibfield  {journal} {\bibinfo  {journal} {Phys. Rev. D}\ }\textbf {\bibinfo
  {volume} {103}},\ \bibinfo {pages} {104056} (\bibinfo {year} {2021})},\
  \Eprint {http://arxiv.org/abs/2004.06503} {arXiv:2004.06503 [gr-qc]}
  \BibitemShut {NoStop}%
\bibitem [{\citenamefont {Ossokine}\ \emph {et~al.}(2020)\citenamefont
  {Ossokine} \emph {et~al.}}]{Ossokine:2020kjp}%
  \BibitemOpen
  \bibfield  {author} {\bibinfo {author} {\bibfnamefont {S.}~\bibnamefont
  {Ossokine}} \emph {et~al.},\ }\href {\doibase 10.1103/PhysRevD.102.044055}
  {\bibfield  {journal} {\bibinfo  {journal} {Phys. Rev. D}\ }\textbf {\bibinfo
  {volume} {102}},\ \bibinfo {pages} {044055} (\bibinfo {year} {2020})},\
  \Eprint {http://arxiv.org/abs/2004.09442} {arXiv:2004.09442 [gr-qc]}
  \BibitemShut {NoStop}%
\bibitem [{\citenamefont {Rodriguez}\ \emph {et~al.}(2014)\citenamefont
  {Rodriguez}, \citenamefont {Farr}, \citenamefont {Raymond}, \citenamefont
  {Farr}, \citenamefont {Littenberg}, \citenamefont {Fazi},\ and\ \citenamefont
  {Kalogera}}]{Rodriguez_2014}%
  \BibitemOpen
  \bibfield  {author} {\bibinfo {author} {\bibfnamefont {C.~L.}\ \bibnamefont
  {Rodriguez}}, \bibinfo {author} {\bibfnamefont {B.}~\bibnamefont {Farr}},
  \bibinfo {author} {\bibfnamefont {V.}~\bibnamefont {Raymond}}, \bibinfo
  {author} {\bibfnamefont {W.~M.}\ \bibnamefont {Farr}}, \bibinfo {author}
  {\bibfnamefont {T.~B.}\ \bibnamefont {Littenberg}}, \bibinfo {author}
  {\bibfnamefont {D.}~\bibnamefont {Fazi}}, \ and\ \bibinfo {author}
  {\bibfnamefont {V.}~\bibnamefont {Kalogera}},\ }\href {\doibase
  10.1088/0004-637x/784/2/119} {\bibfield  {journal} {\bibinfo  {journal} {The
  Astrophysical Journal}\ }\textbf {\bibinfo {volume} {784}},\ \bibinfo {pages}
  {119} (\bibinfo {year} {2014})}\BibitemShut {NoStop}%
\bibitem [{gwt()}]{gwtc_data}%
  \BibitemOpen
  \href@noop {} {\enquote {\bibinfo {title} {Parameter estimation sample
  release for gwtc-1},}\ }\bibinfo {howpublished}
  {\url{https://dcc.ligo.org/LIGO-P1800370/public}}\BibitemShut {NoStop}%
\bibitem [{\citenamefont {Wu}\ \emph {et~al.}(2020{\natexlab{b}})\citenamefont
  {Wu}, \citenamefont {Cao},\ and\ \citenamefont {Zhu}}]{Wu2020}%
  \BibitemOpen
  \bibfield  {author} {\bibinfo {author} {\bibfnamefont {S.}~\bibnamefont
  {Wu}}, \bibinfo {author} {\bibfnamefont {Z.}~\bibnamefont {Cao}}, \ and\
  \bibinfo {author} {\bibfnamefont {Z.-H.}\ \bibnamefont {Zhu}},\ }\href
  {\doibase 10.1093/mnras/staa1176} {\bibfield  {journal} {\bibinfo  {journal}
  {Monthly Notices of the Royal Astronomical Society}\ }\textbf {\bibinfo
  {volume} {495}},\ \bibinfo {pages} {466} (\bibinfo {year}
  {2020}{\natexlab{b}})},\ \Eprint {http://arxiv.org/abs/2002.05528}
  {arXiv:2002.05528} \BibitemShut {NoStop}%
\bibitem [{\citenamefont {Cutler}\ and\ \citenamefont
  {Flanagan}(1994)}]{Cutler_1994}%
  \BibitemOpen
  \bibfield  {author} {\bibinfo {author} {\bibfnamefont {C.}~\bibnamefont
  {Cutler}}\ and\ \bibinfo {author} {\bibfnamefont {E.~E.}\ \bibnamefont
  {Flanagan}},\ }\href {\doibase 10.1103/physrevd.49.2658} {\bibfield
  {journal} {\bibinfo  {journal} {Phys. Rev. D}\ }\textbf {\bibinfo {volume}
  {49}},\ \bibinfo {pages} {2658–2697} (\bibinfo {year} {1994})}\BibitemShut
  {NoStop}%
\bibitem [{\citenamefont {Lange}\ \emph {et~al.}(2018)\citenamefont {Lange},
  \citenamefont {O'Shaughnessy},\ and\ \citenamefont {Rizzo}}]{lange2018rapid}%
  \BibitemOpen
  \bibfield  {author} {\bibinfo {author} {\bibfnamefont {J.}~\bibnamefont
  {Lange}}, \bibinfo {author} {\bibfnamefont {R.}~\bibnamefont
  {O'Shaughnessy}}, \ and\ \bibinfo {author} {\bibfnamefont {M.}~\bibnamefont
  {Rizzo}},\ }\href@noop {} {\enquote {\bibinfo {title} {Rapid and accurate
  parameter inference for coalescing, precessing compact binaries},}\ }
  (\bibinfo {year} {2018}),\ \Eprint {http://arxiv.org/abs/1805.10457}
  {arXiv:1805.10457 [gr-qc]} \BibitemShut {NoStop}%
\bibitem [{\citenamefont {Abbott}\ \emph
  {et~al.}(2016{\natexlab{b}})\citenamefont {Abbott}, \citenamefont {Abbott},
  \citenamefont {Abbott}, \citenamefont {Abernathy}, \citenamefont {Acernese},
  \citenamefont {Ackley}, \citenamefont {Adams}, \citenamefont {Adams},
  \citenamefont {Addesso}, \citenamefont {Adhikari},\ and\ \citenamefont
  {et~al.}}]{Abbott_2016}%
  \BibitemOpen
  \bibfield  {author} {\bibinfo {author} {\bibfnamefont {B.}~\bibnamefont
  {Abbott}}, \bibinfo {author} {\bibfnamefont {R.}~\bibnamefont {Abbott}},
  \bibinfo {author} {\bibfnamefont {T.}~\bibnamefont {Abbott}}, \bibinfo
  {author} {\bibfnamefont {M.}~\bibnamefont {Abernathy}}, \bibinfo {author}
  {\bibfnamefont {F.}~\bibnamefont {Acernese}}, \bibinfo {author}
  {\bibfnamefont {K.}~\bibnamefont {Ackley}}, \bibinfo {author} {\bibfnamefont
  {C.}~\bibnamefont {Adams}}, \bibinfo {author} {\bibfnamefont
  {T.}~\bibnamefont {Adams}}, \bibinfo {author} {\bibfnamefont
  {P.}~\bibnamefont {Addesso}}, \bibinfo {author} {\bibfnamefont
  {R.}~\bibnamefont {Adhikari}}, \ and\ \bibinfo {author} {\bibnamefont
  {et~al.}},\ }\href {\doibase 10.1103/physrevx.6.041015} {\bibfield  {journal}
  {\bibinfo  {journal} {Phys. Rev. X}\ }\textbf {\bibinfo {volume} {6}}
  (\bibinfo {year} {2016}{\natexlab{b}}),\
  10.1103/physrevx.6.041015}\BibitemShut {NoStop}%
\bibitem [{\citenamefont {Dickey}(1971)}]{dickey}%
  \BibitemOpen
  \bibfield  {author} {\bibinfo {author} {\bibfnamefont {J.~M.}\ \bibnamefont
  {Dickey}},\ }\href {\doibase 10.1214/aoms/1177693507} {\bibfield  {journal}
  {\bibinfo  {journal} {The Annals of Mathematical Statistics}\ }\textbf
  {\bibinfo {volume} {42}},\ \bibinfo {pages} {204 } (\bibinfo {year}
  {1971})}\BibitemShut {NoStop}%
\bibitem [{Rom({\natexlab{a}})}]{Romero_Shaw_2019_github_mchirp_gw151226}%
  \BibitemOpen
  \href@noop {} {\enquote {\bibinfo {title} {Supplementary figure for monthly
  notices of the royal astronomical society, volume 490, issue 4, december
  2019, pages 5210–5216},}\ }\bibinfo {howpublished}
  {\url{https://github.com/IsobelMarguarethe/eccentric-GWTC-1/blob/master/events/GW151226/corner_all_corner.pdf}}
  ({\natexlab{a}})\BibitemShut {NoStop}%
\bibitem [{Rom({\natexlab{b}})}]{Romero_Shaw_2019_github_mchirp_gw170608}%
  \BibitemOpen
  \href@noop {} {\enquote {\bibinfo {title} {Supplementary figure for monthly
  notices of the royal astronomical society, volume 490, issue 4, december
  2019, pages 5210–5216},}\ }\bibinfo {howpublished}
  {\url{https://github.com/IsobelMarguarethe/eccentric-GWTC-1/blob/master/events/GW170608/corner_all_corner.pdf}}
  ({\natexlab{b}})\BibitemShut {NoStop}%
\bibitem [{EOB(2021{\natexlab{b}})}]{EOBEBitbucket}%
  \BibitemOpen
  \href@noop {} {\enquote {\bibinfo {title} {\texttt{TEOBResumE}:
  Effective-one-body model with spin and tidal interactions on eccentric
  orbits},}\ }\bibinfo {howpublished}
  {\url{https://bitbucket.org/eob_ihes/teobresums/src/eccentric/}} (\bibinfo
  {year} {2021}{\natexlab{b}})\BibitemShut {NoStop}%
\bibitem [{\citenamefont {Hinder}\ \emph {et~al.}(2008)\citenamefont {Hinder},
  \citenamefont {Vaishnav}, \citenamefont {Herrmann}, \citenamefont
  {Shoemaker},\ and\ \citenamefont {Laguna}}]{Hinder_2008}%
  \BibitemOpen
  \bibfield  {author} {\bibinfo {author} {\bibfnamefont {I.}~\bibnamefont
  {Hinder}}, \bibinfo {author} {\bibfnamefont {B.}~\bibnamefont {Vaishnav}},
  \bibinfo {author} {\bibfnamefont {F.}~\bibnamefont {Herrmann}}, \bibinfo
  {author} {\bibfnamefont {D.~M.}\ \bibnamefont {Shoemaker}}, \ and\ \bibinfo
  {author} {\bibfnamefont {P.}~\bibnamefont {Laguna}},\ }\href {\doibase
  10.1103/physrevd.77.081502} {\bibfield  {journal} {\bibinfo  {journal} {Phys.
  Rev. D}\ }\textbf {\bibinfo {volume} {77}} (\bibinfo {year} {2008}),\
  10.1103/physrevd.77.081502}\BibitemShut {NoStop}%
\end{thebibliography}%
\bibliographystyle{apsrev4-1}

\end{document}